\documentclass[aps,prd,showpacs,
notitlepage,
reprint,
onecolumn,
superscriptaddress,nofootinbib,preprintnumbers]{revtex4-1} 

\makeatletter
\def\p@subsection{}
\makeatother
\usepackage{color,graphicx,amsmath,amssymb,lipsum}
\usepackage[colorlinks=true,citecolor=blue,linkcolor=blue,urlcolor=blue, backref=false,pdfborder={0 0 0}]{hyperref}

\usepackage[normalem]{ulem}
\usepackage[utf8]{inputenc} 
\usepackage{mathtools}
\usepackage{wrapfig}

\usepackage{float}

\usepackage{ulem}
\usepackage{diagbox}

\usepackage[dvipsnames]{xcolor}

\newcommand{\be}{\begin{equation}}
\newcommand{\ee}{\end{equation}}
\newcommand{\beqa}{\begin{eqnarray}}
\newcommand{\eeqa}{\end{eqnarray}}

\renewcommand\k{ {\bf k}}
\newcommand\p{ {\bf p}}
\newcommand\x{\textbf x}

\newcommand{\fnl}{f_{\rm NL}}

\newcommand{\kmax}{k_{\rm max}}
\newcommand{\kmin}{k_{\rm min}}
\newcommand{\hMpc}{h \text{Mpc}^{-1}}
\newcommand{\q}{{\bf q}}
\newcommand{\z}{\hat{\textbf{z}}}

\def\d{\partial}
\newcommand{\bseq}{\begin{subequations}}
\newcommand{\eseq}{\end{subequations}}

\DeclarePairedDelimiter{\abs}{\lvert}{\rvert}

\makeatletter
\newlength{\apb@width}
\newcommand{\autoparbox}[2][c]{\settowidth{\apb@width}{#2}\parbox[#1]{\apb@width}{#2}}

\makeatother

\usepackage{amsfonts}

\newcommand{\dif}{\mathrm{d}}
\renewcommand{\vec}{\bm}

\def\gsim{\raise0.3ex\hbox{$\;>$\kern-0.75em\raise-1.1ex\hbox{$\sim\;$}}}
\def\lsim{\raise0.3ex\hbox{$\;<$\kern-0.75em\raise-1.1ex\hbox{$\sim\;$}}}

\def\beqn#1{\begin{equation}\label{#1}}
\def\eeqn{\end{equation}}

\def\beqa#1{\begin{eqnarray}\label{#1}}
\def\eeqa{\end{eqnarray}}

\def\Z2{$\mathcal{Z_2}$}

\newcommand {\ignore}[1]{}

\begin{document}


\preprint{CERN-TH-2022-055}

\title{Constraints on Multi-Field Inflation from the BOSS Galaxy Survey}

\author{Giovanni Cabass}
\email{gcabass@ias.edu}
\affiliation{School of Natural Sciences, Institute for Advanced Study, 1 Einstein Drive, Princeton, NJ 08540, USA}

\author{Mikhail M. Ivanov}
\email{ivanov@ias.edu}\thanks{Einstein Fellow}
\affiliation{School of Natural Sciences, Institute for Advanced Study, 1 Einstein Drive, Princeton, NJ 08540, USA}

\author{Oliver H.\,E. Philcox}
\affiliation{Department of Astrophysical Sciences, Princeton University, Princeton, NJ 08540, USA}
\affiliation{School of Natural Sciences, Institute for Advanced Study, 1 Einstein Drive, Princeton, NJ 08540, USA}

\author{Marko Simonovi\'c}
\affiliation{Theoretical Physics Department, CERN, 1 Esplanade des Particules, Geneva 23, CH-1211, Switzerland}
\author{Matias Zaldarriaga}
\affiliation{School of Natural Sciences, Institute for Advanced Study, 1 Einstein Drive, Princeton, NJ 08540, USA}

\begin{abstract} 
We use redshift-space galaxy clustering data from the
BOSS survey to constrain
local primordial non-Gaussianity (LPNG). 
This 
is of particular importance due to the consistency relations, which imply that a detection of 
LPNG would 
rule out all single-field inflationary models. 
Our constraints are based on the consistently analyzed redshift-space galaxy power spectra and bispectra,
extracted from the public BOSS data with optimal window-free estimators. 
We use a complete perturbation 
theory model 
including all one-loop power spectrum corrections generated by LPNG.
Our 
constraint on the amplitude of the local non-Gaussian shape is
$f_{\rm NL}^{\rm local}=-33\pm 28$ at 68\%\,CL, yielding no evidence for primordial non-Gaussianity.
The addition of the bispectrum tightens the 
$f_{\rm NL}^{\rm local}$ constraints from BOSS by $20\%$, and allows breaking of degeneracies with non-Gaussian galaxy bias. These results set the stage for the analysis of future surveys, whose larger volumes will yield significantly tighter constraints on LPNG.
\end{abstract}

\maketitle

\section{Introduction}

Inflation provides a mechanism to seed 
density fluctuations that we observe in the late Universe. 
The physics responsible for it, which may have operated at energies as high as $10^{16}\,{\rm GeV}$, has left observable imprints in these density fluctuations. 
The observations of cosmic microwave background (CMB) anisotropies and the distribution of galaxies in the large-scale structure (LSS) present particularly appealing 
ways to probe the inflationary epoch, and thus the physics of this high-energy regime. 

There is a special class of inflationary models, in which inflation is driven by a medium whose quantum fluctuations are the only source of the observable overdensities. Assuming the attractor solution and the Bunch-Davies vacuum, these {\em single-field} (or single-clock) models generically predict purely adiabatic fluctuations with vanishing physical coupling of long-wavelength and short-wavelength modes. This result is formalized in the well-known consistency relations~\cite{Maldacena:2002vr, Creminelli:2004yq}. Given the bispectrum $B_\phi(k_1,k_2,k_3)$ of the primordial Bardeen potential~$\phi$, they have the following form
\be
\label{eq:consistency_relation}
B_\phi(k_1,k_2,k_3) \Big|_{k_3\ll k_1,k_2} =
-{\frac{5}{3}}P_\phi(k_3) \left[3+ k_1\frac{\partial}{\partial k_1}\right]P_\phi(k_1)  \;.
\ee
This limit, when one of the wavenumbers is much smaller than the other two, is called the squeezed limit. Eq.~\eqref{eq:consistency_relation} implies that in single-field models the only effect of the long-wavelength modes of $\phi$ on the short-scale modes is a simple rescaling of coordinates, which is locally unobservable.
Therefore, any detection of local primordial non-Gaussianity (LPNG), i.e.~a detection of a non-vanishing amplitude in the squeezed limit of the initial bispectrum due to physical interactions of long and short modes, would rule out single-field inflation~\cite{Creminelli:2004yq}.

Whilst the simplicity of single-field inflation is very appealing (and so far supported by observations), having more than one relevant fluctuating degree of freedom besides the inflaton is easy to achieve. Some well-known examples are the curvaton scenario~\cite{Enqvist:2001zp,Lyth:2001nq,Moroi:2001ct} and modulated reheating~\cite{Zaldarriaga:2003my}. In contrast to single-field inflation, these {\em multi-field}
models can produce large and observable LPNG.

This important distinction between the two classes of inflationary models makes the amplitude of the initial bispectrum containing LPNG, called $f_{\rm NL}^{\rm local}$, the key observable that we can use to make quantitative, model-independent statements about the primordial Universe. It is defined in the following way
\begin{equation} 
\label{eq:B_primordial_local_definition}
B_\phi(k_1,k_2,k_3) = 6f_{\rm NL}^{\rm local} \Delta^4_\phi\frac{{\cal S}_{\rm local} (k_1,k_2,k_3)}{k_1^2k_2^2k_3^2} \,\,, 
\end{equation}
where $\Delta^2_\phi$ is the amplitude 
of the primordial power spectrum $k^3P_\phi(k)=\Delta^2_\phi (k/k_\ast)^{n_s-1}$ and $n_s$ is the spectral index.\footnote{\emph{Planck} 2018~\cite{Aghanim:2018eyx} gives precise measurements for both of these quantities: $\Delta^2_\phi\approx 1.5\times 10^{-8}$ and $n_s\approx 0.96$, for the pivot scale $k_\ast=0.05$ Mpc$^{-1}$.}
The local template is given by
\be
\label{eq:local_template}
{\cal S}_{\rm local}(k_1,k_2,k_3) = \frac 13 \frac{k_1^2}{k_2 k_3} + 2~{\rm perms.}\,\,.
\ee
This template is such that the squeezed limit bispectrum generated by the LPNG takes the following form
\begin{equation} 
\label{eq:B_primordial_local_squeezed}
B_\phi(k_1,k_2,k_3) \Big|_{k_3\ll k_1,k_2} = 4f_{\rm NL}^{\rm local} P_\phi(k_1) P_\phi(k_3)  \,\,,
\end{equation}
which is very different from the single-field result given by Eq.~\eqref{eq:consistency_relation}. 
Generic values of $f_{\rm NL}^{\rm local}$ in multifield models are of order one or higher (for some counterexamples, see~\cite{Vernizzi:2006ve,Senatore:2010wk}), making~$f_{\rm NL}^{\rm local} \approx 1$ an interesting and well-motivated observational target. 

Significant efforts are aimed at measuring~$f_{\rm NL}^{\rm local}$, with the tightest current constraints coming from CMB observations. In particular, the \emph{Planck} 2018 data yields~$f_{\rm NL}^{\rm local} = -0.9 \pm 5.1$~\cite{Planck:2019kim}. Measurements from galaxy clustering are currently somewhat weaker, but are expected to improve significantly with upcoming galaxy surveys. These surveys will eventually reach the target of~$f_{\rm NL}^{\rm local} \approx 1$ (see for example~\cite{Dore:2014cca,Ferraro:2019uce,Castorina:2020zhz}). Almost all LSS analyses done so far use the fact that LPNG produces the so-called scale-dependent galaxy bias~\cite{Dalal:2007cu,Matarrese:2008nc}, and therefore can be constrained by observations of galaxy power spectra on large scales~\cite{Slosar:2008hx,Xia:2010yu,2013MNRAS.428.1116R,Castorina:2019wmr,Mueller:2021tqa}. Whilst measuring the galaxy power spectrum and looking for scale-dependent bias has the advantage of being relatively straightforward, this may not be an optimal way to constrain LPNG from LSS data. Indeed, as many Fisher forecasts 
and full likelihood 
analyses indicate, the dominant source of information on~$f_{\rm NL}^{\rm local}$ 
for the shot-noise limited samples 
is the galaxy bispectrum~\cite{Scoccimarro:2003wn,Baldauf:2010vn,Dore:2014cca,Ferraro:2019uce,MoradinezhadDizgah:2020whw,Barreira:2020ekm}. 
Developing consistent and robust pipelines to harvest this information is one of the major milestones on the way towards achieving the tightest possible bounds on LPNG.

Performing an optimal search for LPNG in the observed galaxy bispectrum is not a trivial task for a number of reasons. One of the main difficulties is the survey geometry, which mixes the Fourier modes on large scales. In order to circumvent this problem, in this paper we use recently developed optimal window-free power spectrum and bispectrum estimators~\cite{Philcox:2020vbm,Philcox:2021ukg}. In principle, such an approach guarantees that the results are unbiased, close-to-optimal and that all effects related to the window function are consistently taken into account. This is particularly important for constraints on~$f_{\rm NL}^{\rm local}$, since most of the signal comes from the largest scales in the survey, either through the scale-dependent bias or through the squeezed triangles. An alternative is to model the effects of window convolution when calculating the observed bispectrum. Doing this exactly is very challenging numerically, and novel methods to tackle this problem were developed only very recently~\cite{Pardede:2022udo}. On the other hand, if the effects of the window function are modelled using approximate treatments available in the literature \citep[e.g.,][]{Gil-Marin:2014sta,Gil-Marin:2016wya}, this can lead to biases at low $k$, leading to such bins needing to be dropped from the analysis. This approach was used recently in~\cite{DAmico:2022gki} to measure LPNG, and it remains unclear to what extent the results are impacted by the approximate treatment of the window function. 

Another non-trivial task 
is the modeling of the galaxy bispectrum signal.
This includes all aspects of the nonlinear evolution such as the backreaction of short-scale physics on large-scale modes, the nonlinear evolution of the BAO signature (IR resummation),
as well as a robust control over projection and 
binning effects. 
Many years of intense theoretical efforts~\cite{Scoccimarro:1999ed,Scoccimarro:2000sn,Sefusatti:2006pa,Sefusatti:2007ih,Sefusatti:2009qh,Baldauf:2014qfa,Angulo:2014tfa,Blas:2016sfa,Nadler:2017qto,Eggemeier:2018qae,Eggemeier:2021cam,Desjacques:2016bnm,Desjacques:2018pfv,Ivanov:2018gjr,Oddo:2019run,Oddo:2021iwq,Ivanov:2021kcd} 
have recently made the incorporation of 
these effects possible,
so that the bispectrum data can be routinely used 
in cosmological parameter analyses~\cite{Philcox:2021kcw}.

In this work, we present a 
search of LPNG using galaxy power spectrum and 
bispectrum from the publicly available BOSS data~\cite{Alam:2016hwk}.
This is a natural continuation of our previous analysis of PNG in single-field inflation~\cite{Cabass:2022wjy}, based on the same tree-level bispectrum model and the data cuts that were extensively tested in~\cite{Ivanov:2021kcd}. 
Importantly, the galaxy bispectrum
treatment presented in~\cite{Ivanov:2021kcd}
is fully systematic, i.e.~there is a 
way to control the precision 
of various effects
so it can be 
applied to future high-precision 
galaxy survey
data.

The remainder of this paper is structured as follows. In Section~\ref{sec:theory}
we present key theoretical ingredients needed to 
extract LPNG from the galaxy clustering data. 
They include the complete calculation of the one-loop
galaxy power spectrum in the presence of LPNG. 
Section~\ref{sec:data} describes the data and analysis details. 
We validate our pipeline on the mock galaxy clustering data 
in Section~\ref{sec:mocks}, and then apply it to the BOSS
data in Section~\ref{sec:boss}.
We present limits on non-Gaussian 
bias parameters from the BOSS data in Sec.~\ref{sec:biases}.
Section~\ref{sec:conclusions}
draws conclusions. Additional details of the theory model are given in the Appendix~\ref{sec:fnlsq}.

\section{Structure Formation in the presence of LPNG}
\label{sec:theory}

In this section we present our theoretical model,
which includes all necessary terms generated by LPNG. 
We work in the framework of 
the effective field theory of large scale structure (hereafter EFT of LSS), as described in \cite{Perko:2016puo,Ivanov:2019pdj,DAmico:2019fhj,Ivanov:2019hqk,Nishimichi:2020tvu,Cabass:2022avo} and references therein. 
Since the perturbative model for structure formation has been discussed in detail 
in the works cited above, we will provide only a brief overview in what follows, focusing on the ingredients that will be necessary 
to carry out the calculation of the one-loop LPNG contributions.
For dark matter and biased tracers in real space, these contributions 
have already been studied in the literature,~\cite{Taruya:2008pg,Assassi:2015jqa,Assassi:2015fma,MoradinezhadDizgah:2020whw}. Here, we extend these results to the case of galaxies in redshift-space needed for comparison to observations.

To simplify the text and formulas, we omit the superscript ``local'' 
in our notation of the $\fnl$ parameter in this section.
We stress, however, that all results obtained here, 
strictly speaking, apply only to the case
of LPNG. The analogous theory model for non-local primordial non-Gaussianity is presented in Ref.~\citep{Cabass:2022wjy}.

\subsection{Gaussian part}

For Gaussian initial conditions, the statistical properties 
of the first-order density field, $\delta^{(1)}$, are completely determined 
by its power spectrum $P_{11}$:
\be
\langle  \delta^{(1)}(\k) \delta^{(1)}(\k') \rangle  = (2\pi)^3\delta^{(3)}_D(\k'+\k)P_{11}(k)\,,
\ee
where we have suppressed the explicit time dependence for
brevity.
In this work, we restrict our analysis of the galaxy power spectrum to one-loop order in the EFT of LSS, where the usual Gaussian part reads
\be
P_{\text{Gauss}} = P_{\rm tree}+ P_{\rm 1-loop}+P_{\rm ctr}+P_{\rm stoch}\,\,,
\ee
where $P_{\rm tree}$ is the linear term, 
$P_{\rm 1-loop}$ is the one-loop correction (with Gaussian initial conditions), 
$P_{\rm ctr}$ is the higher derivative term (counterterm),
and $P_{\rm stoch}$ is the term that captures galaxy stochasticity.

We now present our model for the galaxy bias, referring the interested reader to~\cite{Desjacques:2016bnm} for an extensive review. 
For the statistics considered herein, it is sufficient to consider 
the galaxy density field at cubic order. We use the following set of bias operators
\be
\begin{split}
\label{eq:real_space_gaussian} 
\delta_g^{(r)} =  b_1\delta + \frac{b_2}{2}\delta^2 + b_{{\cal G}_2}{\cal G}_2 + \frac{b_3}{6}\delta^3 
+ b_{\delta{\cal G}_2}\delta{\cal G}_2 + b_{{\cal G}_3}{\cal G}_3 + b_{\Gamma_3}\Gamma_3 + R^2_\ast\partial^2\delta\,\,. 
\end{split}
\ee 
The Galileon operator ${\cal G}_2$ is defined as $(\partial_i\partial_j\Phi_g)^2 - (\partial^2\Phi_g)^2$, 
where $\Phi_g$ is the gravitational potential. The $\Gamma_3$ operator instead 
is defined as ${\cal G}_2[\Phi_g] - {\cal G}_2[\Phi_v]$, where $\Phi_v$ is the velocity potential. 
The cubic operators $\delta^3$, $\delta{\cal G}_2$, ${\cal G}_3$ do not contribute 
to the one-loop power spectrum after renormalization. 

The redshift-space mapping for fixed 
line of sight $\z$ at order $(\delta^{(1)})^3$ is given by
\be
\label{eq:RS_mapping} 
\delta_g^{(s)}=\delta_g^{(r)} - \partial_z\big(u_z(1+\delta^{(r)}_g)\big) + \frac{1}{2}\partial^2_z\big(u^2_z(1+\delta^{(r)}_g)\big) 
- \frac{1}{6}\partial^3_z\big(u^3_z\big)\,\,, 
\ee 
(from expanding the usual real-space to redshift-space relation), where $u_z \equiv \z\cdot {\bf v}/{\cal H}$, ${\bf v}$ is the 
peculiar velocity field and ${\cal H}$
is the conformal Hubble parameter. These can be written in terms of the Fourier-space kernels
\begin{eqnarray}
Z_1(\k)&=&b_1 + f\mu^2\,,\\\nonumber
Z_2(\k_1,\k_2)
&=& \frac{b_2}{2}+b_{\mathcal{G}_2}\left(\frac{(\k_1\cdot \k_2)^2}{k_1^2 k_2^2}-1\right) + b_1 F_2(\k_1,\k_2)+f\mu^2 G_2(\k_1,\k_2)+\frac{f\mu k_{12}}{2}\left(
\frac{\mu_1}{k_1}(b_1+f\mu^2_2)+
\frac{\mu_2}{k_2}(b_1+f\mu^2_1)
\right)\,,
\end{eqnarray}
and $Z_3$, whose expression can be found in Section~2 of \cite{Chudaykin:2020aoj},
where $f$ is the logarithmic growth rate 
and $\mu = \hat\k \cdot  \hat{\textbf{z}}$. In this notation, the deterministic part of the redshift-space galaxy density field can be written as 
 \begin{equation}\label{eq:deltazs}
\delta_g(\k) = Z_1(\k)\delta^{(1)}(\k)
+\,\int_{\p_{12}=\k}
Z_2(\p_1,\p_2)\delta^{(1)}(\p_1)\delta^{(1)}(\p_2) +\int_{\p_{123}=\k}
Z_3(\p_1,\p_2,\p_3)\delta^{(1)}(\p_1)\delta^{(1)}(\p_2)
\delta^{(1)}(\p_3)  \,\,,
\end{equation}
where we have introduced the following notation $\int_{\p_{1\ldots n}=\k} \equiv \int \frac{d^3\p_1}{(2\pi)^3} \cdots \frac{d^3\p_n}{(2\pi)^3} (2\pi)^3 \delta^D (\k-\p_{1\ldots n}) $ and  $\p_{1\ldots n}\equiv \p_1+ \cdots + \p_n$.
We also supplement these kernels with the appropriate redshift-space counterterms
that are omitted in \eqref{eq:RS_mapping} for clarity; these are discussed in Ref.\,\citep{Ivanov:2021kcd}.

\subsection{LPNG-related non-linearity} 
\label{subsec:PNG_bias}

\noindent LPNG affects the statistics of the galaxy overdensity in two ways. First, we have a non-zero bispectrum 
for the linear matter overdensity $\delta^{(1)}$. This generates a connected bispectrum contribution $B_{111}$, 
\be\label{eq:b111}
\langle \delta^{(1)}(\k_1)\delta^{(1)}(\k_2)\delta^{(1)}(\k_3)\rangle   \equiv(2\pi)^3\delta_D^{(3)}(\k_{123})
B_{111}(k_1,k_2,k_3)=(2\pi)^3\delta_D^{(3)}(\k_{123})
\prod_{a=1}^3\mathcal{M}(k_a)
B_{\phi}(k_1,k_2,k_3)\,,
\ee
where we have introduced the transfer functions 
\be 
\delta^{(1)}(\k) = {\cal M}(k)\phi(\k)\quad \Rightarrow 
\quad 
{\cal M}(k)=\sqrt{\frac{P_{11}(k)}{P_\phi(k)}}\,\,.
\ee
The initial
bispectrum 
\eqref{eq:b111}
also generates
an additional loop correction to the matter power spectrum dubbed $P_{12}$.
We will discuss this term shortly.

Second,
LPNG
modulates the correlation between the long and short modes, 
which ultimately alters the probability of galaxy formation (inducing scale-dependent bias~\cite{Dalal:2007cu,Matarrese:2008nc}). In order to 
reproduce this effect in the perturbative galaxy bias expansion, one needs to include
new operators with the appropriate bias coefficients analogous to the Gaussian case \eqref{eq:real_space_gaussian}. At linear order in $f_{\rm NL} \Delta_\phi$ and cubic order in $\delta^{(1)}$ these operators 
are given by
\be\label{eq: bias-lpng}
\begin{split}
& \delta_g^{\rm LPNG}(\x) = b_\phi f_{\rm NL}\phi(\q) + b_{\phi\delta} f_{\rm NL}\phi(\q)\delta(\x)  + b_{\phi\delta^2} f_{\rm NL}\phi(\q)\delta^2(\x) 
+ b_{\phi{\cal G}_2} f_{\rm NL}\phi(\q){\cal G}_2(\x)\,\,.
\end{split} 
\ee
Note that this expansion is valid only for 
LPNG. For non-local primordial non-Gaussianity
the squeezed bispectrum is typically
proportional to derivatives of $\phi$,
and hence $\phi$ in the above expansion
must be replaced by appropriate 
higher derivative operators like $\partial^2\phi$~\cite{Schmidt:2018bkr}.
These operators appear to be higher order 
and hence their effect can be neglected 
at the one-loop order in the EFT of LSS~\cite{Cabass:2022wjy}.

In contrast to Eq.~\eqref{eq:real_space_gaussian}, here we have made the argument of all relevant 
fields explicit. 
More precisely, the Bardeen potential appearing on the right-hand side is evaluated at the Lagrangian 
position $\q$ corresponding to the Eulerian position $\x$~\cite{Assassi:2015jqa,Assassi:2015fma,MoradinezhadDizgah:2020whw}. In order to evaluate all fields
at the Eulerian coordinates we need to Taylor expand the primordial gravitational potential. If we want to keep all terms up to cubic order we can write
\be
\phi(\q) = \phi(\x-{\boldsymbol{ \psi}}(\q)) = \phi\big(\x-{\boldsymbol \psi}(\x-{\boldsymbol \psi}(\x))\big) \,\,.
\ee
Expanding perturbatively in the displacement field ${\boldsymbol \psi}$ we get
\be
\label{eq:displacement} 
\phi(\q) = \phi - {\bf \psi}^i\partial_i\phi + \psi^k(\partial_k\psi^i)\partial_i\phi 
+ \frac{1}{2}\psi^i\psi^j\partial_i\partial_j\phi\,\,, 
\ee 
where the fields on the right-hand side are all evaluated at the Eulerian position $\x$, 
and we emphasize that the displacement ${\bf \psi}$ contains both the linear and the second-order contribution. We keep terms up to cubic order in the expansion
\eqref{eq:displacement} since they are needed for the consistent calculation of the one-loop power spectrum. Before we move on, let us comment on the omission of higher derivative 
terms of the form $\partial^2_{\q}\phi(\q)$ in \eqref{eq: bias-lpng}. These corrections 
can be straightforwardly included, see e.g.~\cite{Assassi:2015jqa},
but for realistic values of $\fnl$ they are always suppressed
compared to the two-loop Gaussian contributions that 
we neglect here. Therefore, we neglect the higher derivative 
LPNG terms in what follows. 

Let us now shift our attention to redshift-space. 
In this case LPNG 
generates additional counterterms in $\delta_{g}$ involving the matter velocity field ${\bf v}$. 
However, as we have just discussed, these terms can be neglected 
in our analysis because they have the same order of magnitude as the 
higher derivative LPNG operators.
Hence, it is enough to use \eqref{eq:RS_mapping} to map the rest-frame galaxy overdensity in presence of LPNG to redshift-space. 

All in all, the Taylor expansion of $\delta_{\rm NL}$ though $\delta^{(1)}$
in the presence of LPNG will take a form identical to Eq.~\eqref{eq:deltazs},
but with the new kernels $Z^{\rm tot}_n =Z_n +Z^{\rm NG}_n$ ($n=1,2,3$), 
where $Z^{\rm NG}_n$ are the additional PNG kernel contributions. 
The linear kernel is given by
\be
Z^{\rm NG}_1(k) = b_\phi f_{\rm NL} \,,
\ee 
with the second kernel taking the form
\be
\begin{split} Z^{\rm NG}_2(\p_1,\p_2) &=  b_\phi f_{\rm NL}\,\frac{\p_1\cdot\p_2}{2p_1p_2}\bigg(\frac{p_2}{p_1}\frac{1}{{\cal M}(p_2)} + \frac{p_1}{p_2}\frac{1}{{\cal M}(p_1)}\bigg) + b_\phi f_{\rm NL}\, \frac{f\mu k}{2}\bigg(\frac{\mu_1}{p_1}\frac{1}{{\cal M}(p_2)} + \frac{\mu_2}{p_2}\frac{1}{{\cal M}(p_1)}\bigg) \\ 
&\;\;\;\; + b_{\phi\delta}f_{\rm NL}\,\frac{1}{2}\bigg(\frac{1}{{\cal M}(p_1)} + \frac{1}{{\cal M}(p_2)}\bigg) \,\,,
\end{split} 
\ee 
where we have introduced
\be 
\mu_i = \z\cdot\hat{\bf p}_i\,\,, \quad 
\mu_{ij} = \z\cdot({\bf p}_i+{\bf p}_j)/
\abs{{\bf p}_i+{\bf p}_j}\,\,.
\ee
For the cubic fields we find \be
\label{eq:Z3NG}
\begin{split}
& Z^{\rm NG}_3(\p_1,\p_2,\p_3) = b_\phi f_{\rm NL} \,\bigg({-\frac{1}{14}}{\cal G}_2(\p_1,\p_2)\frac{(\p_1+\p_2)\cdot\p_3}{\abs{\p_1+\p_2}^2}\frac{1}{{\cal M}(p_3)} + \text{$2$ perms.}\bigg) \\
&\;\;\;\; + b_\phi f_{\rm NL}\, \bigg(\frac{1}{6}\frac{\p_1\cdot\p_2}{p^2_1p^2_2}\frac{\p_2\cdot\p_3}{{\cal M}(p_3)} + \text{$5$ perms.}\bigg) + b_\phi f_{\rm NL}\, \bigg(\frac{1}{6}\frac{\p_1\cdot\p_3}{p^2_1p^2_2}\frac{\p_2\cdot\p_3}{{\cal M}(p_3)} + \text{$2$ perms.}\bigg) \\
&\;\;\;\; + b_\phi f_{\rm NL}\, f\mu p_{123}\bigg(\frac{1}{3}G_2(\p_1,\p_2)\frac{\mu_{12}}{\abs{\p_1+\p_2}}\frac{1}{{\cal M}(p_3)} + \text{$2$ perms.}\bigg) + b_\phi f_{\rm NL}\, (f\mu p_{123})^2\bigg(\frac{1}{6}\frac{\mu_1\mu_2}{p_1p_2}\frac{1}{{\cal M}(p_3)} + \text{$2$ perms.}\bigg)\,\,. 
\end{split} 
\ee 
Note that $b_{\phi\delta^2}$ and $b_{\phi{\cal G}_2}$ do not appear in $Z^{\rm NG}_3$: 
the reason for this is that they are removed after renormalization of $b_1$ and $b_\phi$.
Moreover, the contributions from $b_{\phi\delta}$ where either $\phi$ or $\delta$ are expanded 
at second order in perturbations are also absorbed by renormalization of these two parameters, was first proved in Ref.~\cite{MoradinezhadDizgah:2020whw}
in the context of the real space perturbation theory. Finally, let us note that compared to the analysis of~\cite{DAmico:2022gki}, we include the cubic non-Gaussian kernel in the model, which is needed to calculate
corrections to the one-loop galaxy power spectrum induced by LPNG.

\subsection{Stochasticity}

So far we have focused on the deterministic part of the galaxy overdensity. PNG leads to 
additional contributions to the stochastic part of $\delta_g$ as we discuss below.

Refs.~\cite{Desjacques:2016bnm,Ivanov:2019pdj,Ivanov:2021kcd} contain a detailed description of stochastic terms in 
the case of Gaussian initial conditions. 
As far as the tree-level bispectrum and one-loop power spectrum in the presence
of LPNG are concerned, 
the the full stochastic contribution is given by~\cite{Perko:2016puo,Ivanov:2021kcd}
\be\label{eq:NG_stochasticity} 
\delta_g^{\rm stoch.}(\k) =   \epsilon + \frac{d_2}{2} b_1 [\delta \epsilon]_\k 
-f \left[\d_z(\epsilon u_z )\right]_\k 
+\fnl \frac{b_{\epsilon \phi}}{2} [\epsilon \phi]_\k + a_0 R_*^2 k^2   \epsilon + a_2 k_z \hat z^i  (\epsilon^i + k^i \epsilon) 
+a_4 k_z^2  \hat z_i \hat z_j\epsilon^{ij}\,,
\ee 
where $\epsilon,~\epsilon^i,~\epsilon^{ij}$ 
are the stochastic density, velocity and tidal fields. 
The final three terms in \eqref{eq:NG_stochasticity} are higher derivative stochastic contributions that are 
important only for the Gaussian part. The only new LPNG contribution 
here is $\epsilon \phi$, where 
we emphasize again
that $\phi$ is evaluated at the Lagrangian position $\q$.

\subsection{Summary of the power spectrum and bispectrum models}

Once the new kernels in the presence of LPNG are obtained,
it is straightforward to compute the 
one-loop power spectrum and the tree-level bispectrum.  
Modulo the counterterms, the deterministic part is given by 
\be
\label{eq:pfull}
\begin{split}
P_{\text{ $1$-loop}}^{\rm tot} &= 2\int_\p [Z^{\rm tot}_2(\p,\k-\p)]^2 P_{11}(p)P_{11}(|\k-\p|)
+6 Z^{\rm tot}_1(\k)P_{11}(k)\int_\p Z^{\rm tot}_3(\k,-\p,\p)P_{11}(p) \\
&\;\;\;\;+2Z^{\rm tot}_1(\k)\int_\p Z^{\rm tot}_2(\p,\k-\p)B_{111}(k,p,|\k-\p|)\,\,,\\
B^{\rm tot}_{\rm tree}&= Z^{\rm tot}_1(\k_1)Z^{\rm tot}_1(\k_2)
Z^{\rm tot}_1(\k_3)B_{111}(k_1,k_2,k_3)
+ 2Z_2^{\rm tot}(\k_1,\k_2) Z^{\rm tot}_1(\k_1)Z^{\rm tot}_2(\k_2)
P_{11}(k_1)
P_{11}(k_2)
+\text{perms.}\,\,. 
\end{split}
\ee
It is instructive to simplify this expression 
and break it down into separate pieces. 
In this section we give explicit expressions for different terms 
in Eq.~\eqref{eq:pfull}.
We focus on contributions that are linear in $\fnl^{\rm local}\Delta_\phi$.
The contributions $\mathcal{O}((\fnl^{\rm local}\Delta_\phi)^2)$
can be straightforwardly computed, but turn out to be irrelevant for our analysis (as shown in Section \ref{subsec:scaling_universe}) except for the linear $\fnl^2$ scale-dependent bias term.
We briefly discuss other $\mathcal{O}((\fnl^{\rm local}\Delta_\phi)^2)$
corrections in Appendix~\ref{sec:fnlsq}. 

\subsubsection{Power spectrum}

\noindent The power spectrum has three kinds of additional contributions proportional to $f_{\rm NL}$. 
At tree level we have
\be
P^{\rm LPNG}_{\text{tree-level}} = P_{11}^{f_{\rm NL}} + P_{11}^{f_{\rm NL}^2} \,\,,
\ee
while at one-loop order the total contribution linear in $\fnl^{\rm local}$ is given by 
\be
P^{\rm LPNG}_{\text{$1$-loop}} = P_{22}^{f_{\rm NL}} + P_{13}^{f_{\rm NL}} + P_{12}\,\,. 
\ee 
The first contributions are the scale-dependent bias 
\be
\begin{split}
P_{11}^{f_{\rm NL}}(k,\mu) = 2(b_1+f\mu^2)b_{\phi}f_{\rm NL}\frac{P_{11}(k)}{{\cal M}(k)}\,\,,\quad P_{11}^{f_{\rm NL}^2}(k,\mu) =b^2_{\phi}f^2_{\rm NL}\frac{P_{11}(k)}{{\cal M}^2(k)} \,\,.
\end{split}
\ee 
Defining $P_{1\phi}(k)\equiv P_{11}(k)/{\cal M}(k)$, the $P_{22}^{f_{\rm NL}}$ contribution can be written as 
\be
\begin{split}
P_{22}^{f_{\rm NL}}(\k) = 4 f_{\rm NL}  \int_{\p} \tilde{Z}^{\rm NG}_2(\p,\k-\p)Z_2(\p,\k-\p) P_{11}(p)P_{1\phi}(\abs{\k-\p})\,\,, 
\end{split}
\ee 
where we have introduced the new kernel
\be
\begin{split} 
\tilde{Z}^{\rm NG}_2(\p_1,\p_2) &= b_\phi \,\frac{\p_1\cdot\p_2}{p_1^2} 
+ b_\phi \, {f\mu k} \frac{\mu_1}{p_1}   + b_{\phi\delta}\,\,,
\end{split} 
\ee 
which is just a simplified version of $Z_2^{\rm NG}$.
The use of $P_{1\phi}$ and 
$\tilde{Z}^{\rm NG}_2$ 
is particularly convenient for the FFTLog evaluation of $P_{22}^{\fnl}$,
which we perform in this work following the approach of~\cite{Simonovic:2017mhp}.
$P_{13}^{f_{\rm NL}}$ is given by the sum of three contributions: 
\be 
P_{13}^{f_{\rm NL}} =  
P_{13}^{f_{\rm NL}} {}^{(1)} +P_{13}^{f_{\rm NL}} {}^{(2)}+P_{13}^{f_{\rm NL}} {}^{(3)}\,\,.
\ee
The first is simply
\be
P_{13}^{f_{\rm NL}} {}^{(1)}=\frac{b_\phi f_{\rm NL}}{{\cal M}(k)}\,\int_{\p}6Z_3(\k,\p,{-\p})P_{11}(p)\,\,, 
\ee 
with the second being
\be
P_{13}^{f_{\rm NL}} {}^{(2)}={-Z_1(\k)}b_{\phi}f_{\rm NL}k^2(1+f^2\mu^2)\sigma^2_v {P_{1\phi}(k)}\,\,, 
\ee 
where $\sigma^2_v\equiv \int_0^{\infty}\dif q\,P_{11}(q)/6\pi^2$. 
This comes from the terms $\psi^i\psi^j\partial_i\partial_j\phi/2$ in \eqref{eq:displacement} 
and $\partial^2_z(u^2_z\phi)/2$ in the redshift-space mapping of $b_\phi f_{\rm NL}\phi$,
i.e.~the third and fifth terms in \eqref{eq:Z3NG}. 
The contribution $P_{13}^{f_{\rm NL}} {}^{(2)}$ exactly 
cancels with the IR limit of the $P_{22}^{\fnl}$ integral 
just like the IR limits of the $P_{13}$
and $P_{22}$ contributions in the Gaussian case.
Finally, there is a term of the form 
\be
 P_{13}^{f_{\rm NL}} {}^{(3)}=Z_1(\k)P_{11}(k)\int_{\p}6\tilde{Z}^{\rm NG}_3(\k,\p,{-\p})P_{11}(p)\,\,,
\ee 
where 
\be
\begin{split}
\tilde Z^{\rm NG}_3(\p_1,\p_2,\p_3) &= b_\phi f_{\rm NL} \,\bigg({-\frac{1}{14}}{\cal G}_2(\vec{p}_1,\vec{p}_2)\frac{(\vec{p}_1+\vec{p}_2)\cdot\vec{p}_3}{\abs{\vec{p}_1+\vec{p}_2}^2}\frac{1}{{\cal M}(p_3)} + \text{$2$ perms.}\bigg)  + b_\phi f_{\rm NL}\, \bigg(\frac{1}{6}\frac{\vec{p}_1\cdot\vec{p}_2}{p^2_1p^2_2}\frac{\vec{p}_2\cdot\vec{p}_3}{{\cal M}(p_3)} + \text{$5$ perms.}\bigg) \\
&\;\;\;\; + b_\phi f_{\rm NL}\, f\mu p_{123}\bigg(\frac{1}{3}G_2(\p_1,\p_2)\frac{\mu_{12}}{\abs{\p_1+\p_2}}\frac{1}{{\cal M}(p_3)} + \text{$2$ perms.}\bigg)\,\,. 
\end{split} 
\ee 
 Notice that the only permutations surviving are those for which the transfer function remains inside the loop integral, i.e.~it is $P_{1\phi}$ that is integrated in $P_{13}^{f_{\rm NL}} {}^{(3)}$.
The last term $P_{12}$ is given by 
\be
\begin{split}
& P_{12}(\k) = 12 f_{\rm NL} Z_1(\k)
\Delta_\phi \, {\cal T}(k) \int_{\p}\big[{\cal S}(k,p,\abs{\k-\p})Z_2(\p,\k-\p)\big] {\cal T}(p){\cal T}(\abs{\k-\p})\,\,, 
\end{split}
\ee
where we have defined ${\cal T}(k) \equiv \Delta_\phi{\cal M}(k)/k^2$.

Finally, we note that the contribution from 
the stochastic term $\epsilon \phi$ 
to the one-loop power spectrum 
is degenerate with the stochastic shot noise contributions we have in the zero-$f_{\rm NL}$ case. 
For this reason we do not include it in the model.

\begin{figure*}[htb!]
\centering
\includegraphics[width=0.49\textwidth]{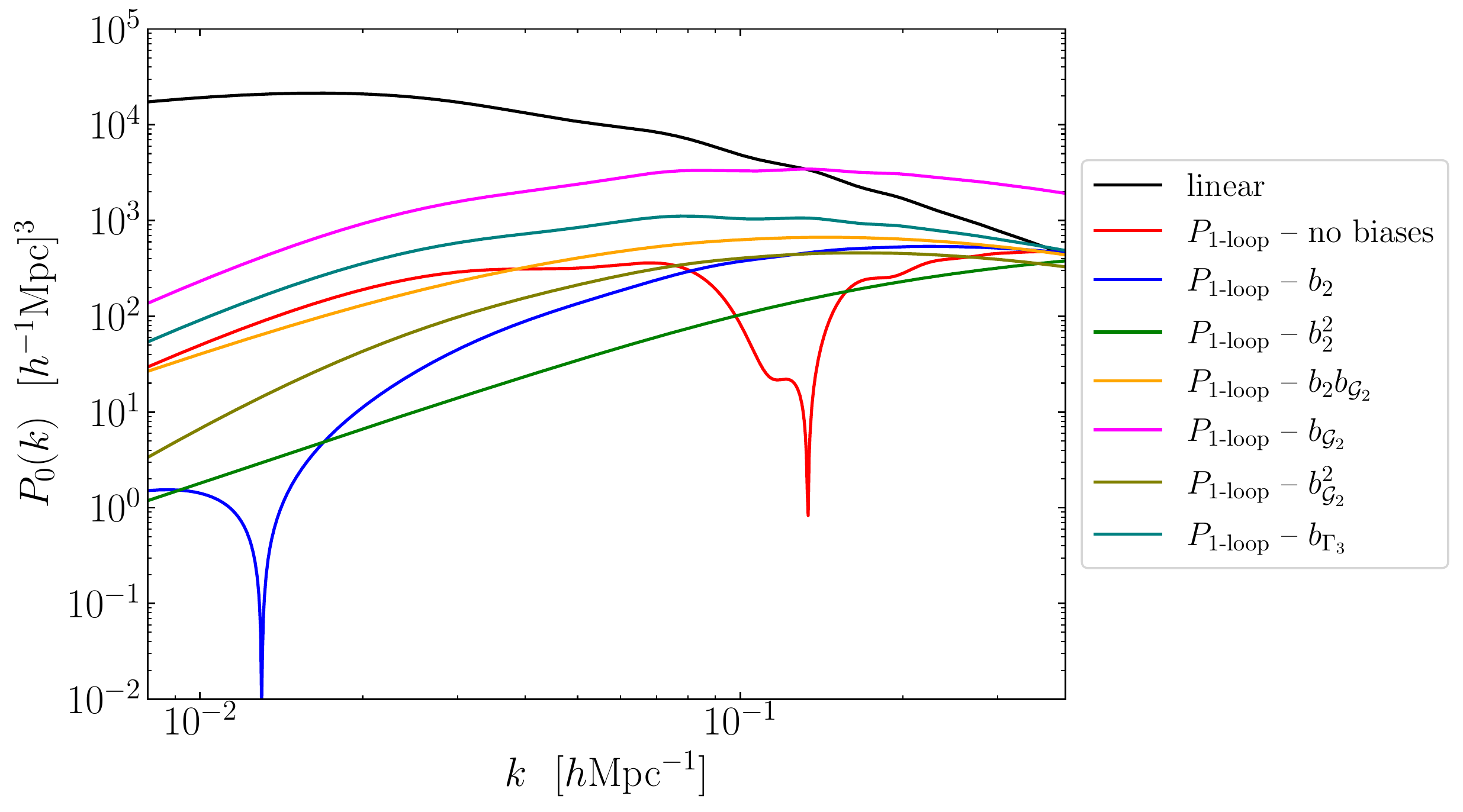}
\includegraphics[width=0.50\textwidth]{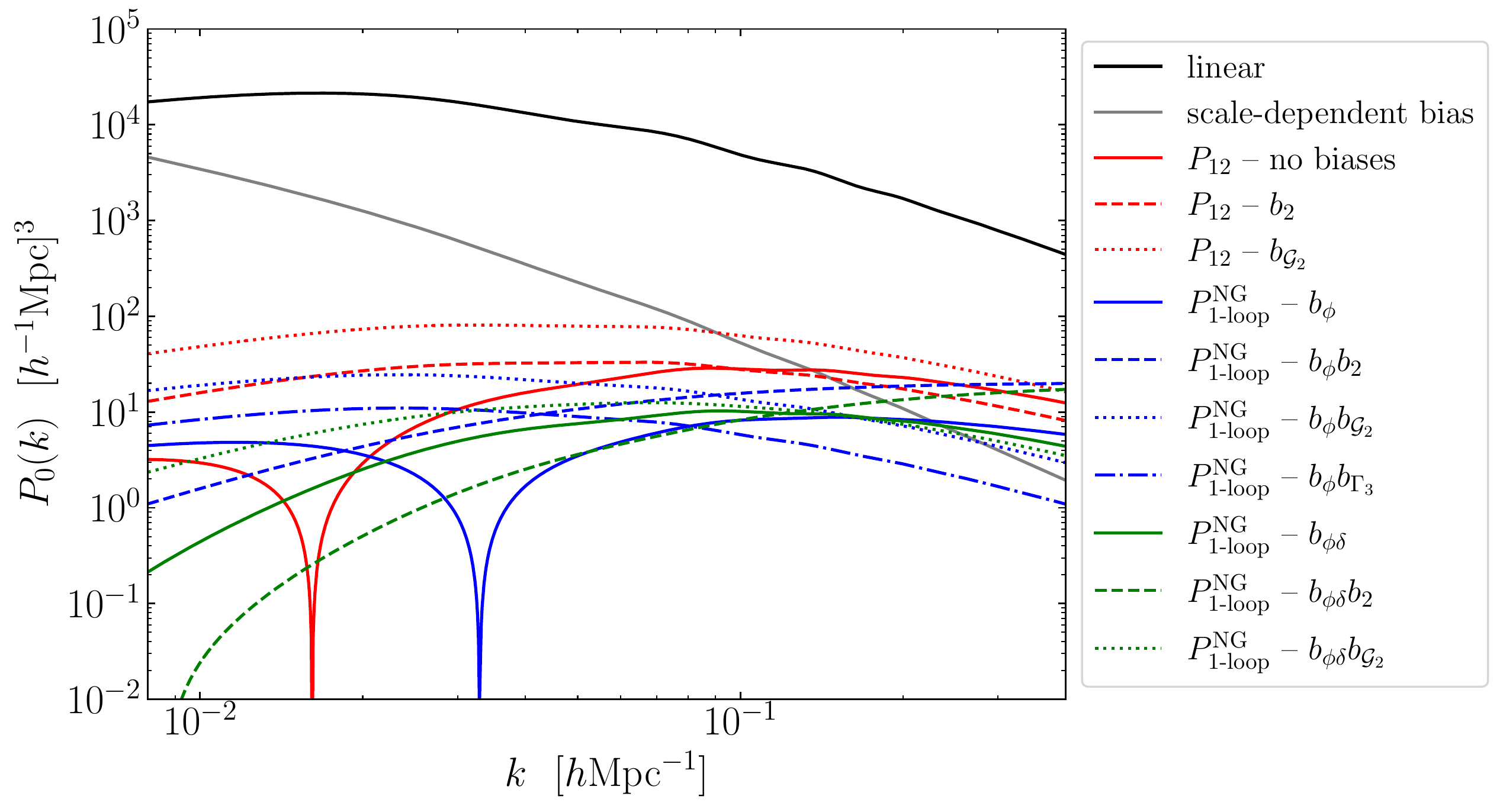}
\caption{
\textbf{Left panel} -- ``Gaussian'' one-loop contributions to the power spectrum monopole at $z=0.61$ compared with linear theory. We take $b_1 = 1$, and the different curves have the corresponding bias parameters set to unity.  
\textbf{Right panel} -- PNG contributions to the power spectrum monopole at $z=0.61$ compared with linear theory. We take $f_{\rm NL} = 100$ and $b_1=1$. The grey curve shows the scale-dependent bias contribution for $b_\phi=1$. The remaining curves show the different contributions ($P_{12}$ and $P_{22}^{f_{\rm NL}} + P_{13}^{f_{\rm NL}}$) to $P^{\rm NG}_{\text{$1$-loop}}$ for unit values of the corresponding bias parameters.
\label{fig:oneloops} } 
\end{figure*}

The Gaussian and non-Gaussian one-loop corrections to the galaxy 
power spectrum monopole are shown 
in the left and right panels of Fig.~\ref{fig:oneloops}, respectively, for $f_{\rm NL}=100$. We show all separate shapes without multiplying them 
by the nuisance parameters to clearly illustrate the size of these terms 
regardless of particular galaxy samples. 
The terms labeled ``no biases'' correspond to pure matter contributions (i.e.~they have $b_1=1$ and all other biases set to zero, which corresponds a contribution from pure matter). 
We see that some of the  LPNG loops are actually as large as the matter loops 
for $k\lesssim 0.1~\hMpc$. Thus, these terms must be included for consistency.

\subsubsection{Bispectrum}

\noindent Working at tree level in perturbations 
and at the linear order in $f_{\rm NL}\Delta_\phi$, 
the PNG contributions to the redshift-space bispectrum are 
\be
B^{\rm NG}_{\text{tree-level}} =f_{\rm NL} B_{211}^{f_{\rm NL}} + B^{(s)}_{111}\,\,. 
\ee 
$B_{211}^{f_{\rm NL}}$ arises from scale-dependent bias and is given by 
\begin{widetext}
\be
\begin{split} 
B_{211}^{f_{\rm NL}}(\k_1,\k_2,\k_3) &= 
Z_1(\k_1)Z_1(\k_2) b_{\delta \phi}   \Big[P_{11}(k_1)P_{1\phi}(k_2) + P_{11}(k_2)P_{1\phi}(k_1)\Big] \\
&\;\;\;\; + Z_1(\k_1)Z_1(\k_2) b_{\phi}   \frac{\k_1\cdot\k_2}{k_1k_2}\bigg(\frac{k_2}{k_1}\frac{1}{{\cal M}(k_2)} + \frac{k_1}{k_2}\frac{1}{{\cal M}(k_1)}\bigg)P_{11}(k_1)P_{11}(k_2) \\
&\;\;\;\; + Z_1(\k_1)Z_1(\k_2) b_{\phi}  \, 
f\mu_{12} k_{12} \bigg(\frac{\mu_1}{k_1}\frac{1}{{\cal M}(k_2)} + \frac{\mu_2}{k_2}\frac{1}{{\cal M}(k_1)}\bigg)P_{11}(k_1)P_{11}(k_2) \\
&\;\;\;\; + 2 b_{\phi} Z_2(\k_1,\k_2) \Big[Z_1(\k_1) P_{11}(k_1) P_{1\phi}(k_2) + Z_1(\k_2)P_{11}(k_2)P_{1\phi}(k_1)\Big] + \text{2 permutations}\,\,. 
\end{split} 
\ee 
\end{widetext}
$B^{(s)}_{111}$ is the standard tree-level redshift-space PNG contribution, 
\be
B^{(s)}_{111}(\k_1,\k_2,\k_3) =  \prod_{a=1}^3Z_1(\k_a) \mathcal{M}(k_a) B_\phi (k_1,k_2,k_3) = Z_1(\k_1)Z_1(\k_2)Z_1(\k_3)\,f_{\rm NL}\Delta_\phi ~ 6~{\cal S}(k_1,k_2,k_3)\, {\cal T}(k_1){\cal T}(k_1){\cal T}(k_1)\,\,. 
\ee
So far we have discussed the deterministic contributions. In contrast to the power spectrum case, 
there is an additional stochastic contribution that is not degenerate with the ones 
present also in the zero-$f_{\rm NL}$ case. This contribution comes 
from \eqref{eq:NG_stochasticity}, and takes the form
\be
\begin{split}
& B^{f_{\rm NL}}_{\rm stoch} = \fnl b_{\phi}\mathcal{M}^{-1}(k_1) Z_1(\k_1)P_{11}(k_1)P_\epsilon(k_2) +  \fnl b_{\epsilon \phi} \mathcal{M}^{-1}(k_1) Z_1(\k_1)P_{11}(k_1)P_\epsilon(k_2) 
+\text{$5$ perms.}\,\,,
\end{split}
\ee
where $P_\epsilon(k)$ is the power spectrum of $\epsilon$.
At leading order it is proportional to the constant 
shot noise value $\bar n^{-1}$.
As we will argue shortly, these terms turn out to be irrelevant for our analysis.

Before closing this section, we also note
that we have implemented IR resummation for all
the LPNG terms entering the power
spectra and bispectra models, following 
the formalism of time-sliced perturbation theory~\cite{Blas:2015qsi,Blas:2016sfa,Ivanov:2018gjr,Vasudevan:2019ewf}. 
After implementing both IR resummation and the Alcock-Pazcynski 
projection effects~\cite{Alcock:1979mp}
in our models for the tree-level bispectrum
and the one-loop power spectra, we numerically compute the Legendre multipoles
of the power spectrum and the bispectrum monopole, allowing for robust comparison to data.

\subsection{Behavior in a scaling universe}
\label{subsec:scaling_universe}

\noindent Let us estimate the relative importance of the different $\fnl$ contributions.
This can be done using the scaling universe approach~\cite{Pajer:2013jj,Assassi:2015jqa}. 
It is based on the fact that the 
linear power spectrum in our Universe can be well 
approximated by a power law: $P_{11}\propto (k/k_{\rm NL})^{n}k_{\rm NL}^{-3}$ with $n\approx -1.5$ for 
quasi-linear wavenumbers $k\simeq 0.1\,h{\rm Mpc}^{-1}$. We also introduced
the nonlinear scale $k_{\rm NL}=0.5~\hMpc$
at $z=0.5$.

We choose to focus on this particular range for the following reason. Given that the leading LPNG contribution is a linear scale-dependent bias enhanced on large scales, and the LPNG loop corrections dominate the usual Gaussian loops at low-$k$, large scales should be crucial for our analysis. The relative contributions of these terms diminish compared to the Gaussian loops at small scales, but the errorbars also get smaller. This suggests that the relative importance of the LPNG corrections should
be maximal at some intermediate wavenumber scale, 
which we choose we to be $k_{\rm ref}=0.1~\hMpc$, roughly 
in the center of the wavenumber range that we use in the data analysis.
In what follows, all estimates will be presented for $k=k_{\rm ref}$.

Assuming that there is a single non-linear scale in the problem, 
the estimates for the total dimensionless galaxy power spectrum $\Delta^2(k)\equiv k^3 P(k)$ for purely Gaussian initial conditions give 
\be 
\begin{split}
\Delta^2(k)& =
{\underbrace{\left(\frac{k}{k_{\rm NL}}\right)^{1.5}}_{P_{\text{tree}}}}+
\underbrace{\left(\frac{k}{k_{\rm NL}}\right)^{3}}_{P_{\text{$1$-loop}}} 
+ \underbrace{\left(\frac{k}{k_{\rm NL}}\right)^{3.5}}_{\rm ctr} 
+ \underbrace{\left(\frac{k}{k_{\rm NL}}\right)^{3}}_{\rm stoch} \,\,.
\end{split} 
\ee 
Recalling that the Bardeen potential has a nearly scale-invariant spectrum, 
we get the following expressions for the LPNG terms: 
\be 
\begin{split}
\Delta^2_{\rm NG}(k)& = \underbrace{\fnl \Delta_\phi\left(\frac{k}{k_{\rm NL}}\right)^{0.75}}_{P_{\text{tree-level}}^{\text{NG,}~ \fnl}}
+ \underbrace{(\fnl \Delta_\phi)^2}_{P_{\text{tree-level}}^{\text{NG,}~ \fnl^2}}+ \underbrace{\fnl \Delta_\phi\left(\frac{k}{k_{\rm NL}}\right)^{2.25}}_{P_{\text{$1$-loop}}^{\rm NG}} \,\,.
\end{split} 
\ee 
Evaluating these corrections at the reference scale $k_{\rm ref}=0.1~\hMpc$, we get
\be 
\begin{split}
&\Delta^2_{P_{\text{tree}}}\simeq 0.089\,\,,\\
&\Delta^2_{P_{\text{tree-level}}^{\text{NG,}~ \fnl}}\simeq 1.1\cdot 10^{-2} \times  \frac{\fnl}{300}\,\,, \\
\end{split}
\qquad
\begin{split}
&\Delta^2_{P_{\text{1-loop}}}=\Delta^2_{P_{\text{stoch}}}\simeq 8\cdot 10^{-3}\,\,,\\
&\Delta^2_{P_{\text{tree-level}}^{\text{NG,}~ \fnl^2}} \simeq 
1.3\cdot 10^{-3}\times \frac{\fnl}{300}\,\,, \\
\end{split}
\qquad
\begin{split}
&\Delta^2_{P_{\text{ctr}}}\simeq 3.6\cdot 10^{-3}\,\,, \\
&\Delta^2_{P^{\rm NG}_{\text{1-loop}}}\simeq 9.6\cdot 10^{-4} \times \frac{\fnl}{300} \,\,.
\end{split} 
\ee
As expected, we see that the scale-dependent bias contribution $P_{\text{tree-level}}^{\rm LPNG}$ always dominates over 
$P_{\text{$1$-loop}}^{\rm LPNG}$
, and it is the main source of constraining power in the power spectrum data.
For $f_{\rm NL}\lesssim 300$ typical for our analyses 
we also see that the one-loop PNG contributions are a 
small fraction of the ``Gaussian'' $P_{\text{$1$-loop}}$. 

The leading correction to the above result is given by the 
Gaussian two-loop contribution, which can be estimates as
\be
 \Delta^2_{\rm 2-loop} = \left(\frac{k_{\rm ref}}{k_{\rm NL}}\right)^{4.5}  \simeq 7.2\cdot 10^{-4}\,\,.
\ee
This can be contrasted with the terms that we have dropped. 
Higher derivative $\fnl$ contributions stemming down from terms like $\partial^2\phi$ 
would be suppressed compared to other 1-loop
LPNG terms that we retain in the theory model,
\be 
\Delta^2_{\langle \partial^2\phi\delta \rangle }=\fnl \Delta_\phi
 \left(\frac{k_{\rm ref}}{k_{\rm NL}}\right)^{2.75}\simeq 4.3\cdot 10^{-4}
 \times \left(\frac{\fnl}{300}\right)
 \,\,.
\ee
This justifies our choice of  dropping these terms in Section~\ref{subsec:PNG_bias}. 

We can make a similar argument for the loop terms $\mathcal{O}(\fnl^2)$,
which are also suppressed,
\be 
\Delta^2_{\langle \phi \delta^2 \rangle }=(\fnl \Delta_\phi)^2 
 \left(\frac{k_{\rm ref}}{k_{\rm NL}}\right)^{1.5}
\simeq 1.1\cdot 10^{-4} \times \left(\frac{\fnl}{300}\right)^2\,\,.
\ee
All in all, our scaling universe estimates confirm 
that the one-loop LPNG corrections can be important in the 
data analysis~\cite{MoradinezhadDizgah:2020whw}.
In addition, we also need to retain the leading $\fnl^2$
tree-level power spectrum contribution.

For the squeezed-limit tree-level bispectrum, where the shape of LPNG plays the most important role, 
it is straightforward to see that $B_{211}^{f_{\rm NL}}$ and $B_{111}$ scale in the same way. Their relative importance 
with respect to the ``Gaussian'' $B^{(s)}_{211}$ is the same as that of the scale-dependent bias piece versus $P_{11}$ in 
the power spectrum. The contribution from $B^{f_{\rm NL}}_{\rm stoch}$
is suppressed in the squeezed limit, and thus we do not include it in the analysis.

\subsection{LPNG bias parameters}\label{subsec:lpng-bias}

In the context of the EFT of LSS, $b_\phi$ and $b_{\phi\delta}$ (as well as 
usual bias parameters like $b_1$, $b_2$, etc.)
should be treated as free nuisance parameters and marginalized 
over in data analysis.
However, there are certain phenomenological models of 
dark matter halo formation, which predict $b_\phi$ and $b_{\phi\delta}$
parameters as a function of the linear bias $b_1$. These are known as ``universality relations''~\cite{Desjacques:2016bnm}. For the relevant LPNG bias coefficients they predict 
\be
\label{eq:bphi0}
b_\phi = 2\delta_c (b_1-1)\,,\quad b_{\phi \delta} = b_\phi -(b_1-1) +\delta_c\left[b_2-\frac{8}{21}(b_1-1)\right]\,,
\ee
where $\delta_c=1.686$. A typical approach then is to assume that the same
relationship holds true even for galaxies. 
The universality relation is routinely used 
in most of $\fnl$ constraints from galaxy surveys.
However, the relationships~\eqref{eq:bphi0}
fail for galaxies~\cite{Barreira:2020kvh}
from realistic hydrodynamical simulations. 
The most accurate analysis to date gives the following fits
based on the state-of-the art galaxy formation simulations~\cite{Barreira:2020ekm,Barreira:2021ueb}:
\be 
\label{eq:bphi}
b_\phi = 2\delta_c (b_1-0.55)\,,\quad
b_{\delta \phi}=3.85 - 9.49b_1 +3.44 b_1^2\,\,.
\ee
We adopt relationship \eqref{eq:bphi} in
our baseline analysis.
As a cross check, we also repeat our analysis for the 
vanilla universality relations~\eqref{eq:bphi0}.
In Section~\ref{sec:biases}
we go beyond any assumptions on the LPNG parameters and 
fit $b_{\delta \phi}\fnl^{\rm local}$
and $b_{\phi}\fnl^{\rm local}$
directly from the data for the first time.

\section{Data and Analysis details}
\label{sec:data}

Our analysis is based on the twelfth data release  (DR12)~\cite{Alam:2016hwk} of  the  Baryon  Oscillation  Spectroscopic  Survey  (BOSS).   
The galaxy clustering data covers two redshift bins with effective centers $z=0.38,\,0.61$,  
for the Northern  and Southern galactic caps, resulting in 
four independent slices. 
The BOSS DR12 release contains a total of 
$\sim 1.2\times 10^6$ galaxies  in a total  volume  of  $6\,(h^{-1}\text{Gpc})^3$.
From each data chunk, we extract  the redshift-space 
power  spectrum multipole moments $P_{\ell}$ ($\ell=0,2,4$),
the real space power spectrum proxy $Q_0$ \cite{Ivanov:2021haa}, the redshift-space bispectrum monopoles
for triangle configurations within the range of $k_i\in [0.01,0.08) \,\hMpc$ 
(a total of 62 bispectrum data points per data chunk),
and the BAO parameters $\alpha_\parallel, \alpha_\perp$ measured 
from the post-reconstructed power spectrum data using the method of~\cite{Philcox:2020vvt}. Both power spectra and bispectra are measured using window-free estimators \citep{Philcox:2020vbm,Philcox:2021ukg}, thus we do not need to include the survey window function in our theoretical model.

We use the data cuts\footnote{Note that we use $\kmax$
for $P_\ell$ that is
slightly larger than that adopted 
in Ref.~\cite{Philcox:2021bwo}.
This is because the particular 
choice of the data cut
in that paper $\kmax=0.2~\hMpc$ 
was based on detecting biases
in the cosmological parameter posteriors. However, in contrast to~\cite{Philcox:2021bwo},
here 
we fix all cosmological parameters, 
in which case
the fit to $\fnl^{\rm local}$ is unbiased
up to somewhat larger $\kmax$.
We stress that this choice is not 
essential for the purposes of our work, as the 
$\fnl^{\rm local}$ constraints 
are dominated 
by the linear LPNG bias and hence
are saturated at large scales. } $\kmax=0.25~\hMpc$,
$\kmin=0.01~\hMpc$ for $P_{\ell}$ and $\kmin=0.25~\hMpc$, $\kmax=0.27~\hMpc$ for $Q_0$,
so that the two statistics are largely independent. 
We use lower $\kmax$ for $Q_0$ because the two-loop
corrections can be non-negligible compared to the PNG contributions for $\kmax>0.3~\hMpc$.
Note that our choice 
$\kmin=0.01~\hMpc$ for both the power spectra and bispectra is conservative. 
We remove the first bin in order to limit systematic effects related to 
stellar contamination and residual radial and atmospheric systematics, as well as integral constraints.\footnote{ As shown in Fig.\,19 of \citep{Kalus:2018qsy}, weight-based approaches to removing large-scale systematics (such as those applied in \citep{BOSS:2016apd} and herein) produce comparable results to more sophisticated methods on comparatively large scales. For $k\lesssim0.01~\hMpc$, the differences between different approaches become significant (and the systematics become larger than the statistical errors, if uncorrected), thus these modes are excised from our analysis.}

The power spectra and bispectra used in this work 
are extracted using the window-free estimators~\cite{Philcox:2020vbm,Philcox:2021ukg}.
The covariances for our total datavector $\{P_0,P_2,P_4,Q_0,B_0,\alpha_\parallel,\alpha_\perp\}$ for each data chunk
are extracted from a suite of 2048 MultiDark-Patchy  mocks~\cite{Kitaura:2015uqa},
using the standard empirical covariance matrix estimator.\footnote{See \cite{Wadekar:2019rdu,Wadekar:2020hax,Philcox:2020zyp,Byun:2020rgl,Biagetti:2021tua} for alternative covariance
matrix estimation techniques.} 

\begin{figure}
\centering
\begin{minipage}{.46\textwidth}
  \centering
  \includegraphics[width=\textwidth]{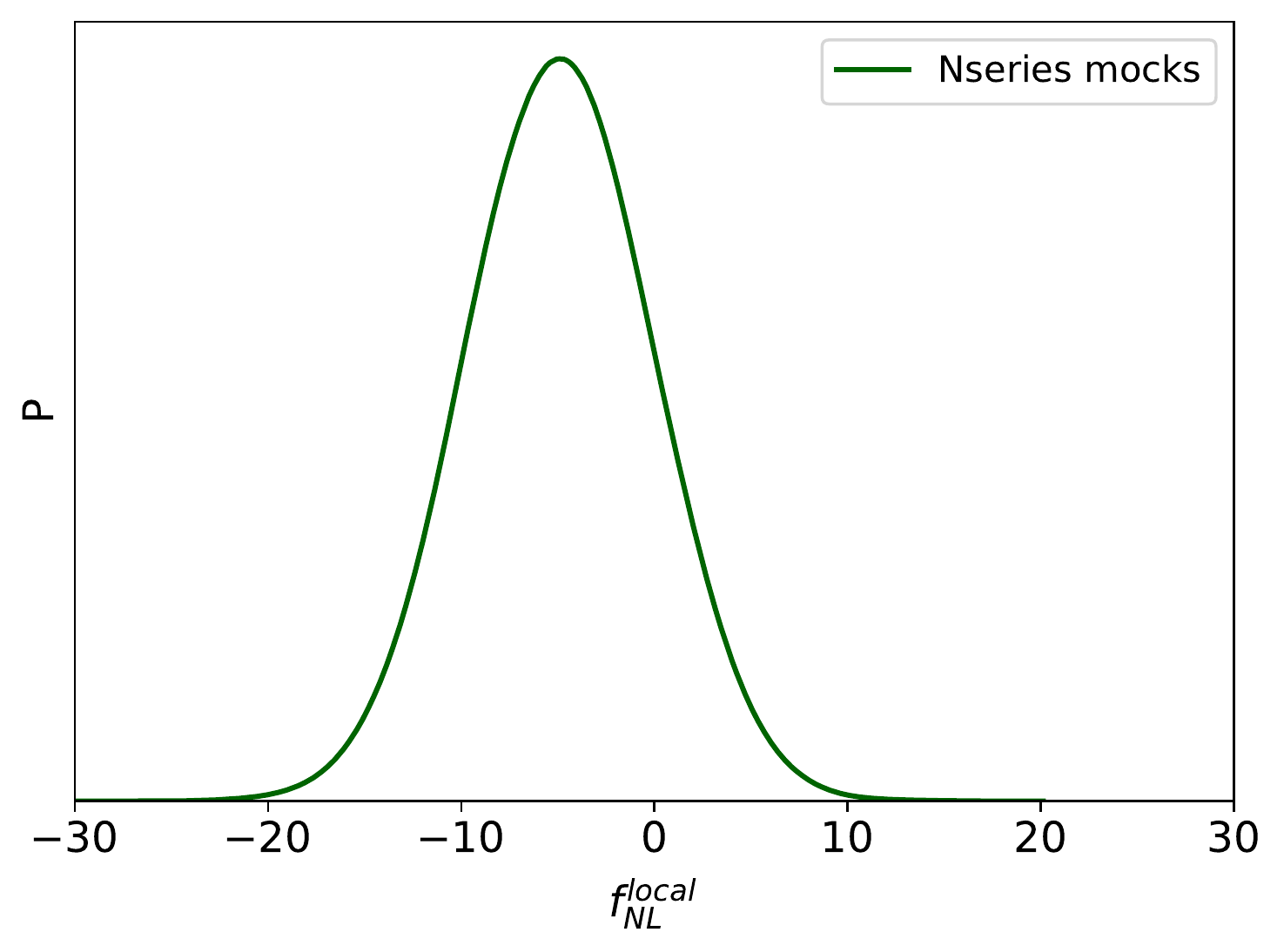}
\caption{
Marginalized constraint on $\fnl^{\rm local}$ from the mean of $84$ \textsc{Nseries} simulations, with a total volume approximately $40\times$ larger than that of BOSS. Note that we do \textit{not} rescale the covariance to the BOSS volume, but use that appropriate for the entire \textsc{Nseries} volume, allowing a robust probe of theoretical systematics. Here, we find $f_{\rm NL}^{\rm local} = -4.9 \pm 5.0$ at $68\%$\,CL.
\label{fig:fnl}}
\end{minipage}%
\hfill
\begin{minipage}{.52\textwidth}
  \centering
\includegraphics[width=\textwidth]{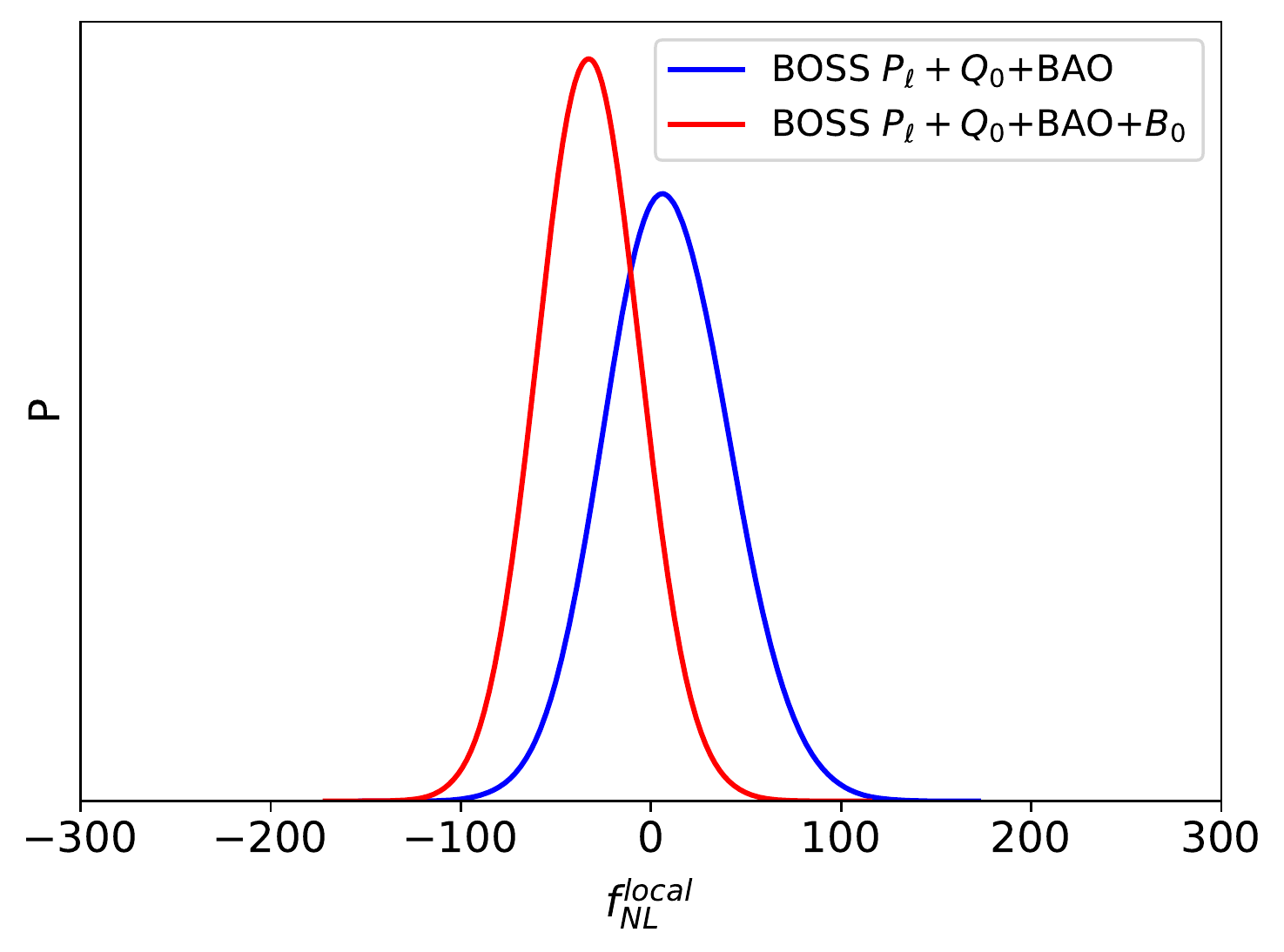}
\caption{
Marginalized constraints on local-type primordial non-Gaussianity from the BOSS power spectrum (blue) and power spectrum plus bispectrum (red). We find $f_{\rm NL}^{\rm local} = 9^{+33}_{-35}$ and $-33\pm 28$ in the two cases respectively at $68\%$\,CL, with the bispectrum tightening the constraints by $\approx 20\%$. These are the main results of this work.
\label{fig:fnlboss}  }
\end{minipage}
\end{figure}

Our full-shape analysis matches the ones of~\cite{Ivanov:2019pdj,Philcox:2020vvt,Philcox:2021kcw,Cabass:2022wjy}.
Unlike these works, we explicitly fix all cosmological parameters to the 
Planck 2018 priors~\cite{Aghanim:2018eyx}. This is done because in this work
we are interested only in the constraints on $\fnl^{\rm local}$ from the BOSS survey.
Formally, this corresponds to a combination of the CMB power 
spectra and BOSS power+bispectrum data. Thus, in our MCMC analysis we 
only fit $\fnl^{\rm local}$ with an infinitely large flat prior,
plus the Gaussian EFT nuisance parameters (encompassing biases, stochasticity, and counterterms). 

As discussed in Section~\ref{subsec:lpng-bias}, our fiducial analysis fixes the PNG
bias coefficients to values predicted by the dark matter halo 
relations as functions of the corresponding linear bias $b_1$
for each data chunk. This choice is optional. In principle, we
can fit both $b_{\phi}$ and $b_{\phi \delta}$ directly from the data,
but the current limits on these parameters are not very constraining, as shown in Section~\ref{sec:biases}.
Therefore, they are fully consistent with the fits from 
simulations~\eqref{eq:bphi}, making it reasonable to fix them for the primary purposes of this paper.

Our analysis is based on the publicly available \textsc{class-pt} code~\cite{Chudaykin:2020aoj}.
Since we do not vary cosmology in this study, we compute the full one-loop power spectrum corrections including the LPNG terms only once, utilizing the Planck cosmology, and only vary the bias parameters and $f_{\rm NL}^{\rm local}$ in the likelihood. For the \textsc{Nseries} mocks we recompute the relevant templates to match the \textsc{Nseries} fiducial cosmology. 
We plan to implement the full cosmology-dependent LPNG calculation in a future update of \textsc{class-pt}, using which we will systematically study the sensitivity of the $\fnl^{\rm local}$ constraints to uncertainties in cosmological parameters. Our Markov Chain Monte Carlo (MCMC) analysis is run with the Montepython code  \citep{Brinckmann:2018cvx} and is based on the previously-used public likelihoods.\footnote{Available at \href{https://github.com/oliverphilcox/full_shape_likelihoods}{github.com/oliverphilcox/full\_shape\_likelihoods}.}

\section{Validation on Mocks}
\label{sec:mocks}

As a validation test, we apply our pipeline to \textsc{Nseries} mock catalogs. 
These catalogs were used by the BOSS collaboration 
for internal validation tests~\cite{Alam:2016hwk}.
The suite consists of 84 semi-independent simulation boxes. 
The \textsc{Nseries} mocks are designed to reproduce the 
clustering signal of the high-z NGC BOSS sample. 
Each box has a similar effective volume
and mean effective redshift $z=0.56$. 
We fit the mean of 84 \textsc{Nseries} boxes with the covariance
of one box, divided by 84. Effectively, this is equivalent 
to fitting a dataset which is $\approx 40$ times larger 
than the BOSS survey. Just like in the actual BOSS data analysis, 
we fix all the cosmological parameters (to the true values used in the simulations),
and vary only $\fnl^{\rm local}$ and nuisance parameters in the fit.

The \textsc{Nseries} mocks were produced
for Gaussian initial conditions, 
which we can recover with our pipeline.
Indeed, we find $\fnl^{\rm local}$ consistent with 
zero, 
\be 
\fnl^{\rm local}=-4.9 \pm 5.0  \quad \text{at 68\%\,CL}\,\,,
\ee 
with the 1d marginalized posterior shown in Fig.~\ref{fig:fnl}. 
Note that the mean is expected to differ from zero
to be different from zero by $(1-2)~\sigma$ due to random fluctuations.
This also gives us an estimate of the theory systematic error, 
$\Delta\fnl^{\rm local}|_{\rm syst}\lesssim 5$,
which is less than $0.2\sigma$ of the actual BOSS 1d marginalized statistical error (rescaling by the square-root of the volume ratio) .

We stress that the main goal of our validation test is to estimate the bias 
due to the theory systematic error. An alternative approach for validation is 
to fit the mean of the mocks with the covariance that matches the overall
volume of the BOSS survey. This test, however, does not allow one to assess the theory bias because the posterior distribution 
in that case is affected by prior volume effects (arising from the priors necessarily imposed on nuisance parameters), which can be 
as large as the actual theory bias. This obscures the estimation
of theory bias and can lead to wrong conclusions on the validity of the fitting pipeline. 
For example, the prior volume effects exactly cancel the theory bias 
on $\sigma_8$ for BOSS-like mocks~\cite{Ivanov:2019pdj,Chudaykin:2020ghx,Philcox:2021kcw}. Thus, 
if one fits the mean of the mocks with the covariance matching the BOSS survey volume, 
one can erroneously conclude that the theory model is valid even for $\kmax=0.3~\hMpc$, 
whereas fitting the same data with the actual covariance of the simulation 
suggests that the theory systematic bias on $\sigma_8$
becomes sizable in 
the analysis of multipoles 
$P_\ell$ 
for $\kmax > 0.20~\hMpc$.

\section{Results for BOSS}
\label{sec:boss}

 \begin{figure*}[ht!]
    \centering
         \includegraphics[width=0.49\textwidth]{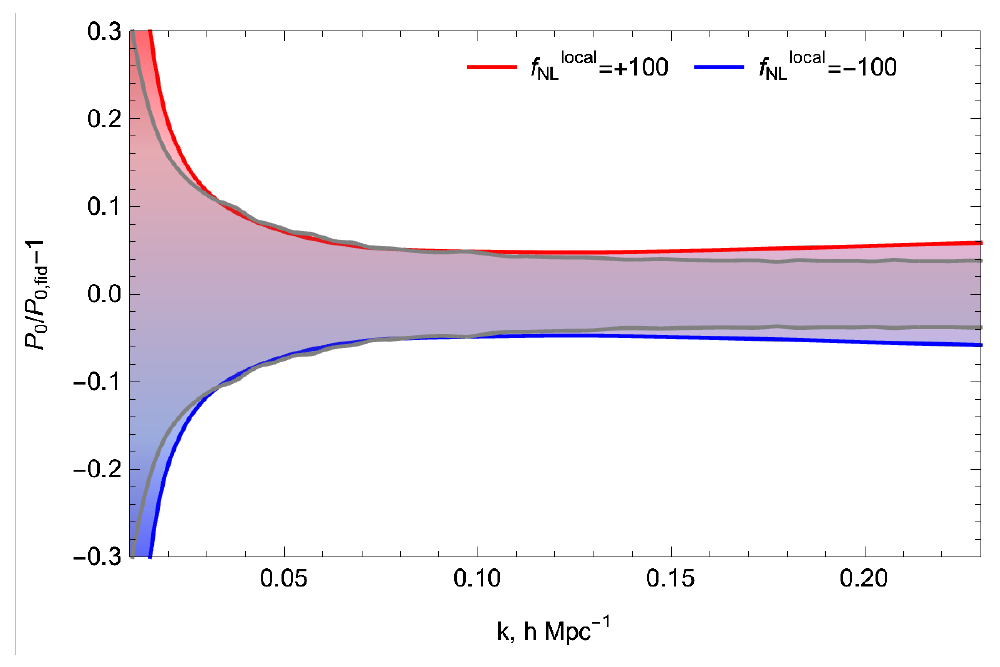}
    \includegraphics[width=0.49\textwidth]{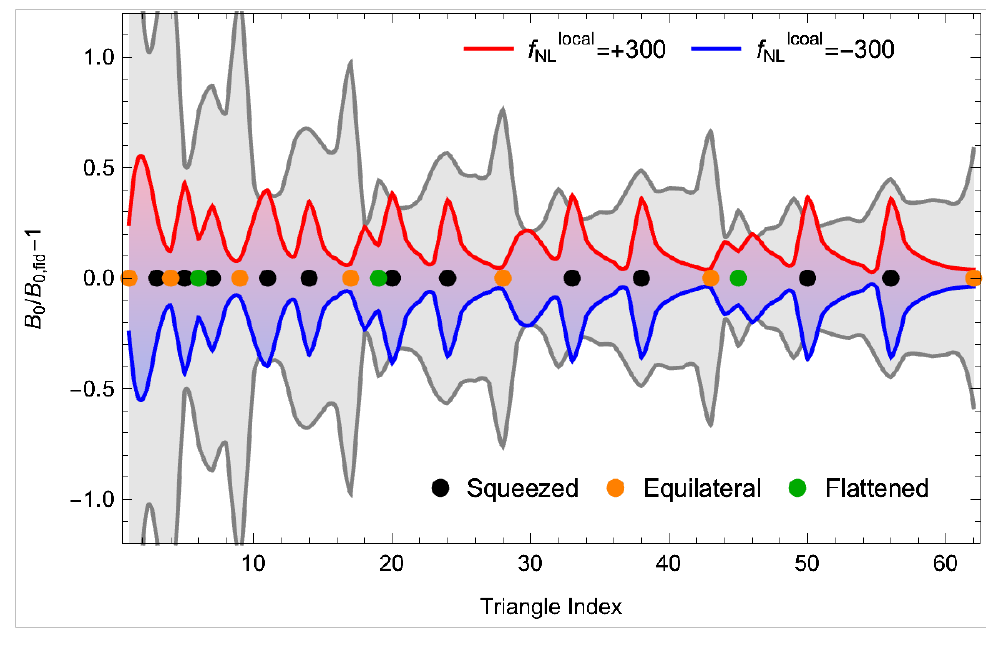}
    \caption{
        \textbf{Left panel}: residual variations of the galaxy power spectrum monopole $P_0$
    w.r.t.~variations of $\fnl^{\rm local}$.
    These variations 
    have distinctive shape dependence 
    and therefore can be constrained 
    by the data. 
\textbf{Right panel}: residual variations of the galaxy bispectrum monopole
    w.r.t.~variations of $\fnl^{\rm local}$. Black, orange and green
    dots denote the squeezed ($k_3=0.015~\hMpc$, $k_{1},k_2>k_3$), equilateral ($k_1=k_2=k_3$), and flattened ($k_2=k_3$, $2k_2=k_1+0.015~\hMpc$)
    triangle configurations, respectively. }
    \label{fig:residual}
\end{figure*}

Here, we present the main results from the combined analysis of four BOSS
data chunks. We start with the power spectra + BAO data alone, i.e.~without the bispectrum. In this case we obtain
\be
\begin{split}
\text{$P_\ell$+$Q_0$+BAO:}
\quad
\fnl^{\rm local}=9_{-35}^{+33}\,\,,\quad \text{(68\%~CL)}\,\,,\quad -57<\fnl^{\rm local} < 78\,\,,\quad \text{(95\%~CL)}\,\,.
 \end{split}
\ee
The marginalized posterior distribution is shown in Fig.~\ref{fig:fnlboss}.
These results are similar, but somewhat 
stronger than those obtained previously 
from BOSS DR9, $-45<\fnl^{\rm local}<195$ (95\%\,CL)~\cite{2013MNRAS.428.1116R}. This suggests that the constraints in
the power spectrum are dominated by 
the linear scale-dependent bias from large scales.
Indeed, repeating our analysis for $\kmin=0.05~\hMpc$ we find constraints that are worse 
by a factor of four, 
\be 
\text{ $\kmin=0.05~\hMpc$~~:}\quad 
\fnl^{\rm local}=-120_{-140}^{+100}\,\,,\quad \text{(68\%~CL)}\,\,,\quad -353<\fnl^{\rm local} < 140\,\,,\quad \text{(95\%~CL)}\,\,.
\ee
These results are consistent with the expectation 
that LPNG constraints are dominated by the scale-dependent bias, which is sensitive to the lowest available bin in the survey (see also~\cite{MoradinezhadDizgah:2020whw}). 

Note that there are important differences between 
our analysis and that of~\cite{2013MNRAS.428.1116R}.
That work was based on the monopole
power spectrum moment of the BOSS DR9 CMASS
sample and had
a lower scale cut.
In contrast to that, we use all three 
power spectrum multipole moments 
plus the real space proxy 
of the complete BOSS DR12 data sample,
but impose a conservative scale cut 
$\kmin = 0.01~\hMpc$, significantly reducing any systematics caused by observational effects, such as galactic foregrounds. Nevertheless, 
the results of the two analyses are pleasingly consistent.

The addition of the bispectrum monopole
shrinks the errorbar on $\fnl^{\rm local}$ by $\simeq 20\%$,
\be
\text{$P_\ell$+$Q_0$+BAO+$B_0$:}\quad 
\fnl^{\rm local}=-33\pm 28\,\,,\quad \text{(68\%~CL)}\,,\quad 
-88<\fnl^{\rm local} < 23 \,\,,\quad \text{(95\%~CL)},.
\ee
We do not find any evidence for LPNG: the 95\%~CL limits are consistent with zero.
The final posterior distribution is presented in Fig.~\ref{fig:fnlboss}.

To estimate the dependence 
of our results on the 
LPNG priors, 
we have repeated our
analysis 
assuming the universality
relations \eqref{eq:bphi0}
instead of the more accurate simulation-calibrated fits 
\eqref{eq:bphi}. We found 
somewhat weaker bounds, 
\be 
\text{Prior~\eqref{eq:bphi0}:}\quad 
\fnl^{\rm local}=-50\pm 40\,\,,\quad \text{(68\%~CL)}\,,\quad 
-130<\fnl^{\rm local} < 30\,\quad \text{(95\%CL)}\,.
\ee 
This weakening of the constraints 
is expected, since the 
universality relations 
underpredict the actual values
of $b_\phi$ by $\sim 30\%$
compared to~\eqref{eq:bphi}
for $b_1\approx 2$.
Note that when the universality 
bias relations~\eqref{eq:bphi0}
are used, the relative impact of the bispectrum
is somewhat stronger: it tightens 
the constraints by $\simeq 30\%$.
The result
without the bispectrum 
in this case is $\fnl^{\rm local}=64.7_{-60}^{+52}$.

\begin{figure}[htb!]
\centering
\includegraphics[width=0.99\textwidth]{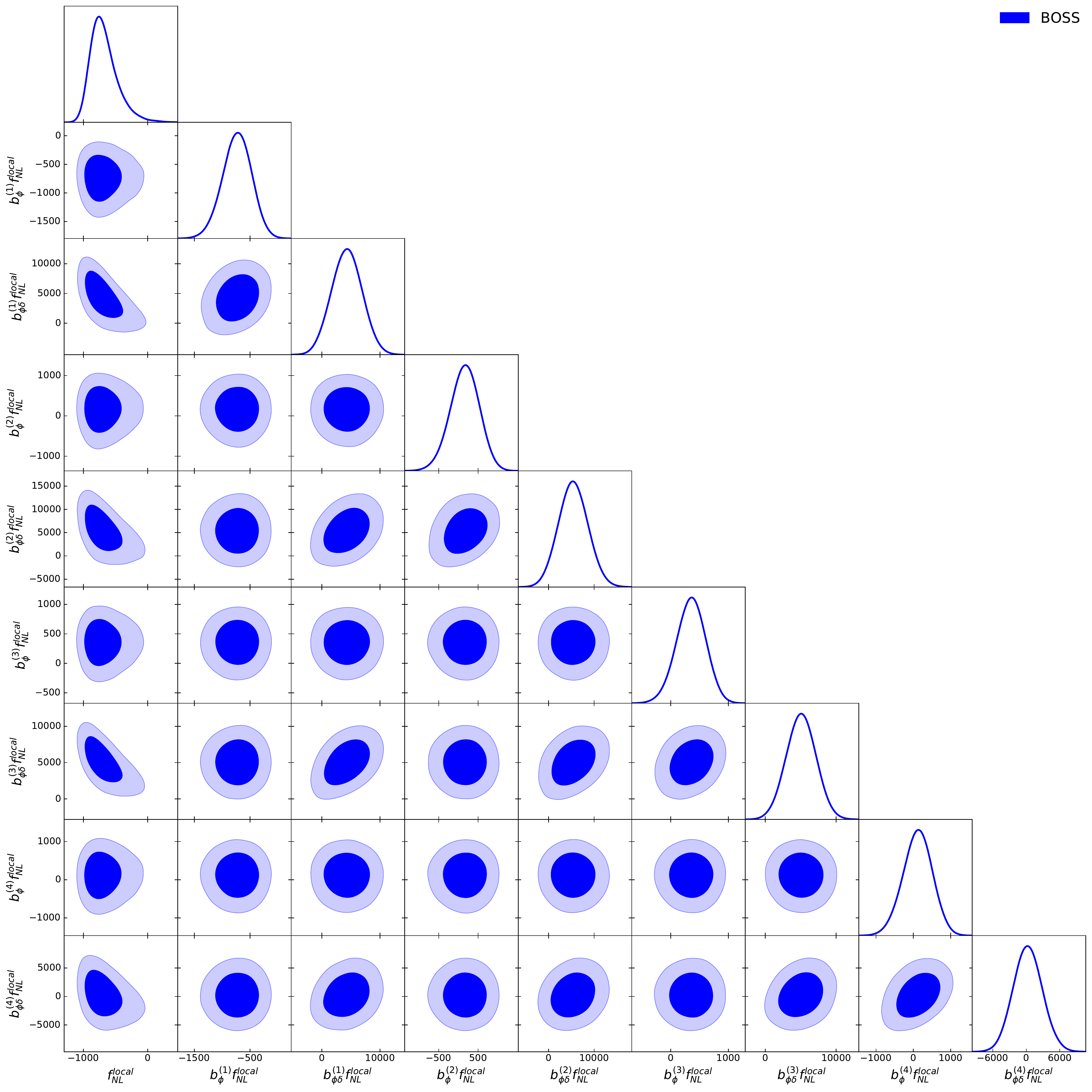}
\caption{
1d and 2d marginalized posteriors for $\fnl^{\rm local}$ 
and the normalized LPNG bias parameters $b_\phi^{(i)}\fnl^{\rm local}$,  
$b_{\phi\delta}^{(i)}\fnl^{\rm local}$, extracted from the BOSS galaxy power spectra 
and bispectra data. $i=1,2,3,4$ corresponds to the NGCz3, SGCz3, NGCz1, and SGCz1 BOSS
data slices, respectively. 
\label{fig:all_bs}}
\end{figure}

It is instructive to study where our constraints originate from. To this end we 
illustrate the effect of the variation of $\fnl^{\rm local}$ on the galaxy power 
spectrum and bispectrum monopole, showing the corresponding residuals in Fig.~\ref{fig:residual}.
We focus on the NGC z3 data chunk, and show the residuals w.r.t.~the best-fit model 
with $\fnl^{\rm local}=0$.
The bispectrum residuals are shown as a function of the triangle index 
which is defined by the bin center $(k_1,k_2,k_3)$. The bin centers here 
satisfy $k_i\in [0.015,0.075]~\hMpc$, 
encompassing  all wavenumbers with 
$k\in (0.01,0.08)~\hMpc$. We additionally mark the squeezed, equilateral, 
and flattened triangles with black, orange, and green dots on the x-axis (note that our notation differs somewhat from~\cite{Cabass:2022wjy}).
The shaded region corresponds to data errors.

Let us first focus on the power spectrum. We observe that the largest deviations take place 
both on large scales (due to the linear scale-dependent bias), and on short scales due to the 
one-loop LPNG corrections. This is another indication 
that the one-loop LPNG corrections should be included in data analysis.
Of course, their effect is washed out, to some extent, by marginalization over the standard EFT nuisance parameters, which are important at short scales. Nevertheless, 
the LPNG nonlinear corrections must to be included for the sake of consistency, and additionally, the degeneracies are greatly reduced in practice by the inclusion of higher-order statistics. 

Let us now move on to the bispectrum. We see that the PNG 
contribution has a very significant configuration-dependence. 
The LPNG terms peak at the squeezed triangles. This behavior is very different 
from the typical effect of the standard non-linear galaxy bias~\cite{Cabass:2022wjy},  which illustrates 
that this particular type of non-Gaussianity does not strongly suffer from 
degeneracy with the unknown galaxy formation details.

Let us compare our results with the $\fnl^{\rm local}$
measurements available in the literature. Our 
limit is a factor of ten worse than the Planck 2018 constraint, $\fnl^{\rm local}=-0.9 \pm 5.1$ (68\%\,CL)~\cite{Planck:2019kim}.
As we stressed before, our limit is better than the one obtained 
from BOSS DR9~\cite{2013MNRAS.428.1116R},
which is equivalent to $\sigma_{\fnl^{\rm local}}\approx 60$.
The main reasons for this improvement 
are
new data, the complete theory model
for the power spectrum,
more accurate priors for LPNG 
bias parameters, and 
the large-scale 
galaxy bispectrum,
which is quite sensitive to the scale-dependent bias signal. 
Note that our measurement has a precision 
somewhat worse but
comparable with the eBOSS quasars
$\fnl^{\rm local}=-12 \pm 21$ (68\%\,CL)~\cite{Castorina:2019wmr,Mueller:2021tqa} (which boast a much longer redshift baseline, and thus a substantially lower $k_{\rm min}$), and with WMAP,
$\fnl^{\rm local}=37 \pm 20$ (68\%\,CL)~\cite{2013ApJS..208...20B}.  We find comparable results to those from an independent analysis of the BOSS power spectrum and bispectrum (using a partial one-loop theory model for the latter statistic) \citep{DAmico:2022gki}: $\fnl^{\rm local} = -30\pm29$, though, as noted above, our analysis differs due to the use of a fully consistent theory model and complete treatment of the survey window, allowing larger-scale information to be robustly included.
Finally, our measurements are somewhat better than the ones
coming from the UV luminosity function $\fnl^{\rm local}=71^{+426}_{-237}$ (95\%\,CL)~\cite{Sabti:2020ser,Sabti:2021unj,Sabti:2021xvh}, although they include information from scales with $k>0.3~\hMpc$, which we do not consider in our study.

\section{LPNG bias parameters}
\label{sec:biases}

The galaxy bispectrum
allows us to 
measure $\fnl^{\rm local}$
separately from 
$b_\phi \fnl^{\rm local}$. 
At the power 
spectrum level this is, essentially 
impossible,
because the constraints are dominated by scale-dependent bias controlled by combination $b_\phi \fnl^{\rm local}$~\cite{MoradinezhadDizgah:2020whw}.
However, the bispectrum allows 
us to extract $\fnl^{\rm local}$
directly from the $B_{111}^{(s)}$
shape that is generated by
the matter clustering 
and not the LPNG bias. 
In this section we present the 
constraints on $b_\phi \fnl^{\rm local}$ and $\fnl^{\rm local}$
independently. Moreover, we also present
constraints on the 
quadratic scale-dependent 
bias term $b_{\phi \delta} \fnl^{\rm local}$, 
which shows up in the tree-level 
galaxy bispectrum with LPNG. 
We found that $\fnl^{\rm local}$
can be quite large 
in our chains if 
$b_\phi$ and $b_{\phi\delta}$
are not fixed, and therefore 
we have included the $(\fnl^{\rm local})^2$
corrections to the tree-level galaxy bispectrum.
For these pieces, we keep the additional 
$b_{\phi^2}$ bias fixed to the  
prediction of the universality relation, as discussed in Appendix~\ref{sec:fnlsq}.

We fit $\fnl^{\rm local}$ plus parameters
$b_\phi \fnl^{\rm local}$,  
$b_{\phi\delta}\fnl^{\rm local}$
for each independent data chunk. 
Our results are displayed in Fig.~\ref{fig:all_bs} and 
Table~\ref{tab:phi}. We observe that 
the BOSS data can constrain 
the LPNG bias parameter only at the level $\sigma_{b_{\phi}\fnl^{\rm local}}\sim 3\times 10^2$ and $\sigma_{b_{\phi \delta}\fnl^{\rm local}}\sim 3\times 10^3 $.
We see that most of the posteriors are compatible with 
zero values of corresponding parameters within $95\%$\,CL. 
However, the parameters 
$\fnl^{\rm local}, b_\phi^{(1)}\fnl^{\rm local},b_{\phi\delta}^{(3)}\fnl^{\rm local}$, overlap with zero only within 
$99\%$\,CL of the marginalized posterior. Inspecting the 2d marginalized contours (shown in Fig.~\ref{fig:all_bs}) suggests that this is a result of degeneracies 
between $\fnl^{\rm local}$
and the LPNG bias parameter combinations. 
Note that the resulting posteriors are also significantly 
non-Gaussian, which implies that having non-zero $\fnl^{\rm local}$
at $95\%$\,CL does not actually imply a detection
at a significance level 
equivalent to that of a Gaussian-distributed parameter at $2\sigma$, as evidenced by the lack of detection of LPNG in the fiducial analysis.

\begin{table}
\centering 
\begin{tabular}{|l|c|c|c|c|}
 \hline
Param & best-fit & mean$\pm\sigma$ & 95\% lower & 95\% upper \\ \hline
$\fnl^{\rm local}$ &$-720$ & $-676_{-250}^{+150}$ & $-1080$ & $-210$ \\
$b_{\phi }^{(1)}\fnl^{\rm local}$ 
&$-878$ & $-740_{-250}^{+280}$ & $-1280$ & $-210$ \\
$b_{\phi\delta}^{(1)}\fnl^{\rm local}$ 
&$1900$ & $4300_{-2700}^{+2700}$ & $-950$ & $9500$ \\
$b_\phi^{(2)} \fnl^{\rm local} $ 
&$170$ & $160_{-360}^{+390}$ & $-600$ & $900$ \\
$b_{\phi\delta}^{(2)}\fnl^{\rm local}$ 
&$5400$ & $5400_{-3400}^{+3200}$ & $-1100$ & $12000$ \\
$b_\phi^{(3)}\fnl^{\rm local}$ 
&$180$ & $350_{-250}^{+270}$ & $-170$ & $860$ \\
$b_{\phi\delta}^{(3)} \fnl^{\rm local}$ 
&$4400$ & $5000_{-2200}^{+2100}$ & $800$ & $9300$ \\
$b_\phi^{(4)}\fnl^{\rm local}$ 
&$26$ & $120_{-390}^{+420}$ & $-700$ & $900$ \\
$b_{\phi\delta}^{(4)}\fnl^{\rm local}$ 
&$-22$ & $290_{-2700}^{+2600}$ & $-5000$ & $5700$ \\
\hline
 \end{tabular}
  \caption{1d marginalized limits for $\fnl^{\rm local}$ 
and the normalized LPNG bias parameters $b_\phi^{(i)}\fnl^{\rm local}$,  
$b_{\phi\delta}^{(i)}\fnl^{\rm local}$, extracted from the BOSS galaxy power spectra 
and bispectra data. $i=1,2,3,4$ corresponds to the NGCz3, SGCz3, NGCz1, and SGCz1 BOSS
data slices, respectively. 
 \label{tab:phi}
 }
 \end{table}

\section{Conclusions}
\label{sec:conclusions}

We have presented constraints on 
local primordial non-Gaussianity from the BOSS 
full-shape galaxy clustering data. The two main novelties 
of our analysis are (a) we use the full one-loop power spectrum
model that includes all necessary non-linear one-loop corrections generated by 
LPNG, and (b) we include the consistently analyzed galaxy bispectrum, incorporating a full treatment of all relevant theoretical and observational effects.
We have found that the latter improves $\fnl^{\rm local}$
constraints by $20\%$ compared to the power spectrum analysis.
Thus, our paper extends and complements previous works
on LPNG from the galaxy clustering data.

There are many ways in which our analysis can be improved. First, the $k$-range can be expanded, including additional information from both small and large scales. The maximum wavenumber used in the analysis $k_{\rm max}$ can be significantly enhanced by the addition of the redshift-space galaxy two-loop power spectrum, one-loop bispectrum, as well as the tree-level trispectrum. Partial calculations of these observables already exist in the literature, e.g.~\cite{Bertolini:2016bmt,Eggemeier:2018qae,Konstandin:2019bay}, and we plan to incorporate them in our future analyses. Going beyond perturbative analysis, additional constraints on $f_{\rm NL}^{\rm local}$ can be obtained from the nonlinear regime using consistency relations for LSS~\cite{Creminelli:2013mca}. They guarantee that the local shape of the bispectrum in the squeezed limit is protected by the equivalence principle and it can be though of as a feature which is very distinct from anything that can be produced by the astrophysical processes. Such feature can be extracted even when the short modes are in the nonlinear regime, marginalizing over the standard nonlinear physics using theoretical error~\cite{Baldauf:2016sjb,Chudaykin:2020hbf}, similarly to what was done in extracting the BAO feature from the broadband in~\cite{Philcox:2020vvt}.
In addition, we plan to increase 
the $k$-range also on the lower end,
by including modes with $k<0.01~\hMpc$
that are omitted in the present analysis. This is particularly important in order to enhance constraints from the scale-dependent bias. A simple Fisher forecast indicates that the errors on $f_{\rm NL}^{\rm local}$ from BOSS can improve by a factor of two by including all low-$k$ modes. Whilst straightforward from a theoretical point of view, this will require a detailed study of large-scale
systematics, such as the integral 
constraint (both global and radial), foreground stars, atmospheric effects, seeing, and galactic extinction, which can produce large-scale radial and angular distortions, e.g.~\citep{BOSS:2016apd,Kalus:2018qsy}. Such work will be of particular importance as the survey volume increases, and the range of $\fnl$ parameters allowed by data tightens. Another important point to keep in mind is that the Gaussian
power spectra and bispectra 
likelihood, which is used in our analysis, 
is not valid for the lowest $k$ bins and it can skew the constraints on $f_{\rm NL}^{\rm local}$. For some recent reflections on how to deal with this problem, see~\cite{Wang:2018xuy}. 

Second, it would be interesting to study the dependence of the result 
on the priors on EFT nuisance parameters. Previous works~\cite{Wadekar:2020hax,Cabass:2022wjy}
have found that marginalization over 
Gaussian nuisance parameters 
leads to a very significant degradation of parameter errorbars. 
It is important to understand to what extent this can be avoided with priors 
on nuisance parameters extracted from high 
fidelity simulations. 
We have also shown that our current 
constraints significantly rely 
on using universality-like relationships 
for the LPNG bias parameters $b_\phi,b_{\phi\delta}$. We will study if using relationships is accurate enough
for
simulated data with injected LPNG.
A similar analysis was 
done in Ref.~\cite{MoradinezhadDizgah:2020whw}
for the case of real space halo clustering. 

Third, one may include redshift-space multipoles
of the galaxy bispectrum beyond the monopole moment,
which we considered in this work~\cite{Scoccimarro:1999ed}.
Fourth, one should perform a systematic 
sensitivity forecast for future surveys like DESI~\cite{Aghamousa:2016zmz}, Euclid~\cite{Amendola:2016saw}, 
and MegaMapper~\cite{Schlegel:2019eqc}. 
In particular, just based on the ratio of 
BOSS and DESI volumes, one may expect 
improvements by a factor of three, i.e.~reaching
$\sigma_{\fnl^{\rm local}}\approx 10$.
Some forecasts have already been performed e.g.~\cite{Karagiannis:2018jdt,Sailer:2021yzm},
but most of them have been based on simplistic assumptions
about the theoretical modeling of the power spectrum and the bispectrum, and it is rare for forecasts to include both the power spectrum and bispectrum in combination. It will be interesting to see
if the inclusion of all necessary non-linear corrections can impact 
the conclusions of these works, i.e.~to perform a fully consistent 
forecast similar to~\cite{Chudaykin:2019ock,Sailer:2021yzm}. 
Fifth, another important ingredient 
is a systematic study of the properties of galaxy samples 
that will be targeted by future surveys, e.g.~emission line galaxies 
admit higher $\kmax$ and therefore better LPNG 
measurements can be obtained from this sample~\cite{Ivanov:2021zmi}. 
Finally, it would be interesting to extend our analysis 
to the case of projected statistics, which is motivated by 
photometric surveys like SPHEREx~\cite{Dore:2014cca} and the Vera Rubin observatory~\cite{LSST:2008ijt}.
This analysis will naturally  
require including relativistic and full-sky corrections, which can impact constraints on LPNG~\cite{Camera:2014bwa,Alonso:2015sfa,DiDio:2016gpd,Castorina:2021xzs}, and exploring to what extent the analysis based on correlation functions
is optimal and how does it compare to recent results obtained using forward modeling~\cite{Andrews:2022nvv}.

\section*{Acknowledgments}

We are grateful to Kazuyuki Akitsu and Azadeh Moradinezhad Dizgah for valuable discussions. 
We are especially grateful to Alexandre Barreira for useful discussions about priors on non-Gaussian biases and careful reading of the manuscript. 
GC acknowledges support from the Institute for Advanced Study. The work of 
MMI has been supported by NASA through the NASA Hubble Fellowship grant \#HST-HF2-51483.001-A awarded by the Space Telescope Science Institute, which is operated by the Association of Universities for Research in Astronomy, Incorporated, under NASA contract NAS5-26555. 
OHEP thanks Will Coulton and Kendrick Smith for useful discussions and acknowledges support from the Simons Foundation. Parameter estimates presented in this paper have been obtained with the \textsc{class-pt} 
Boltzmann code \cite{Chudaykin:2020aoj}
(see also \cite{Blas:2011rf}) interfaced with the \textsc{Montepython} MCMC sampler \cite{Audren:2012wb,Brinckmann:2018cvx}. 
The triangle plots are generated with the \textsc{getdist} package\footnote{\href{https://getdist.readthedocs.io/en/latest/}{
\textcolor{blue}{https://getdist.readthedocs.io/en/latest/}}
}~\cite{Lewis:2019xzd}.


\appendix 

\section{Higher order PNG corrections to the galaxy bispectrum}
\label{sec:fnlsq}

In this section we present, for completeness, the bispectrum 
corrections at $\mathcal{O}(\fnl^2)$. To obtain them, we need first include a new bias operator 
\be 
\delta_g^{\text{real space}}\supset  \frac{1}{2}\fnl^2 b_{\phi^2}\phi^2\,.
\ee
The full $\mathcal{O}(\fnl^2)$ kernel is then given by 
\begin{widetext}
\be 
\begin{split}
& B_{NG}^{f_{\rm NL}^2} = Z_1(\k_1)Z_1(\k_2)Z_1(\k_3)b_\phi \left(\frac{1}{Z_1(\k_1)\mathcal{M}(k_1)}+\frac{1}{Z_1(\k_2)\mathcal{M}(k_2)}+\frac{1}{Z_1(\k_3)\mathcal{M}(k_3)}\right)B_{111}(k_1,k_2,k_3) \\
& + \Bigg[ b_\phi^2 \left(\frac{Z_1(\k_1)}{\mathcal{M}(k_2)}+\frac{Z_1(\k_2)}{\mathcal{M}(k_1)}\right)
\frac{(\k_1\cdot \k_2)}{k_1 k_2}\left(\frac{k_1}{k_2 \mathcal{M}(k_1)}+\frac{k_2}{k_1 \mathcal{M}(k_2)}\right)
+ fb^2_\phi \mu k 
\left(\frac{\mu_1}{k_1 \mathcal{M}(k_2)}+\frac{\mu_2}{k_2 \mathcal{M}(k_1)}\right)
\left(\frac{Z_1(\k_1)}{\mathcal{M}(k_2)}+\frac{Z_1(\k_2)}{\mathcal{M}(k_1)}\right)
\\
& + 2b_\phi^2 \frac{Z_2(\k_1,\k_2)}{\mathcal{M}(k_1)\mathcal{M}(k_2)} 
+ b_{\phi\delta} b_\phi  \left(\frac{Z_1(\k_1)}{M(k_2)}+\frac{Z_1(\k_2)}{\mathcal{M}(k_1)}\right)
\left(\frac{1}{\mathcal{M}(k_1)}+\frac{1}{\mathcal{M}(k_2)}\right) + b_{\phi^2} \frac{Z_1(\k_1)Z_1(\k_2)}{2\mathcal{M}(k_1)\mathcal{M}(k_2)} 
\Bigg]P_{11}(k_1)P_{11}(k_2)\,\,.
\end{split}
\ee
\end{widetext}
Finally, the universality relations for the quadratic operators dictate 
\be 
\begin{split}
 & b_{\phi^2}=4\delta_c(b_2^L\delta_c - 2b_1^L)\,\,,\quad b_{\phi\delta} = b_\phi+(-b_1^L+\delta_c b_2^L)  = b_\phi -(b_1-1) +\delta_c\left[b_2-\frac{8}{21}(b_1-1)\right]\,\,,
\end{split}
\ee 
where $b_1^L\equiv b_1-1$ and $b_2^L$
are Lagrangian bias coefficients.

\bibliography{short.bib}

\begin{thebibliography}{107}%
\makeatletter
\providecommand \@ifxundefined [1]{%
 \@ifx{#1\undefined}
}%
\providecommand \@ifnum [1]{%
 \ifnum #1\expandafter \@firstoftwo
 \else \expandafter \@secondoftwo
 \fi
}%
\providecommand \@ifx [1]{%
 \ifx #1\expandafter \@firstoftwo
 \else \expandafter \@secondoftwo
 \fi
}%
\providecommand \natexlab [1]{#1}%
\providecommand \enquote  [1]{``#1''}%
\providecommand \bibnamefont  [1]{#1}%
\providecommand \bibfnamefont [1]{#1}%
\providecommand \citenamefont [1]{#1}%
\providecommand \href@noop [0]{\@secondoftwo}%
\providecommand \href [0]{\begingroup \@sanitize@url \@href}%
\providecommand \@href[1]{\@@startlink{#1}\@@href}%
\providecommand \@@href[1]{\endgroup#1\@@endlink}%
\providecommand \@sanitize@url [0]{\catcode `\\12\catcode `\$12\catcode
  `\&12\catcode `\#12\catcode `\^12\catcode `\_12\catcode `\%12\relax}%
\providecommand \@@startlink[1]{}%
\providecommand \@@endlink[0]{}%
\providecommand \url  [0]{\begingroup\@sanitize@url \@url }%
\providecommand \@url [1]{\endgroup\@href {#1}{\urlprefix }}%
\providecommand \urlprefix  [0]{URL }%
\providecommand \Eprint [0]{\href }%
\providecommand \doibase [0]{http://dx.doi.org/}%
\providecommand \selectlanguage [0]{\@gobble}%
\providecommand \bibinfo  [0]{\@secondoftwo}%
\providecommand \bibfield  [0]{\@secondoftwo}%
\providecommand \translation [1]{[#1]}%
\providecommand \BibitemOpen [0]{}%
\providecommand \bibitemStop [0]{}%
\providecommand \bibitemNoStop [0]{.\EOS\space}%
\providecommand \EOS [0]{\spacefactor3000\relax}%
\providecommand \BibitemShut  [1]{\csname bibitem#1\endcsname}%
\let\auto@bib@innerbib\@empty
\bibitem [{\citenamefont {Maldacena}(2003)}]{Maldacena:2002vr}%
  \BibitemOpen
  \bibfield  {author} {\bibinfo {author} {\bibfnamefont {J.~M.}\ \bibnamefont
  {Maldacena}},\ }\href {\doibase 10.1088/1126-6708/2003/05/013} {\bibfield
  {journal} {\bibinfo  {journal} {JHEP}\ }\textbf {\bibinfo {volume} {05}},\
  \bibinfo {pages} {013} (\bibinfo {year} {2003})},\ \Eprint
  {http://arxiv.org/abs/astro-ph/0210603} {arXiv:astro-ph/0210603} \BibitemShut
  {NoStop}%
\bibitem [{\citenamefont {Creminelli}\ and\ \citenamefont
  {Zaldarriaga}(2004)}]{Creminelli:2004yq}%
  \BibitemOpen
  \bibfield  {author} {\bibinfo {author} {\bibfnamefont {P.}~\bibnamefont
  {Creminelli}}\ and\ \bibinfo {author} {\bibfnamefont {M.}~\bibnamefont
  {Zaldarriaga}},\ }\href {\doibase 10.1088/1475-7516/2004/10/006} {\bibfield
  {journal} {\bibinfo  {journal} {JCAP}\ }\textbf {\bibinfo {volume} {10}},\
  \bibinfo {pages} {006} (\bibinfo {year} {2004})},\ \Eprint
  {http://arxiv.org/abs/astro-ph/0407059} {arXiv:astro-ph/0407059} \BibitemShut
  {NoStop}%
\bibitem [{\citenamefont {Enqvist}\ and\ \citenamefont
  {Sloth}(2002)}]{Enqvist:2001zp}%
  \BibitemOpen
  \bibfield  {author} {\bibinfo {author} {\bibfnamefont {K.}~\bibnamefont
  {Enqvist}}\ and\ \bibinfo {author} {\bibfnamefont {M.~S.}\ \bibnamefont
  {Sloth}},\ }\href {\doibase 10.1016/S0550-3213(02)00043-3} {\bibfield
  {journal} {\bibinfo  {journal} {Nucl. Phys. B}\ }\textbf {\bibinfo {volume}
  {626}},\ \bibinfo {pages} {395} (\bibinfo {year} {2002})},\ \Eprint
  {http://arxiv.org/abs/hep-ph/0109214} {arXiv:hep-ph/0109214} \BibitemShut
  {NoStop}%
\bibitem [{\citenamefont {Lyth}\ and\ \citenamefont
  {Wands}(2002)}]{Lyth:2001nq}%
  \BibitemOpen
  \bibfield  {author} {\bibinfo {author} {\bibfnamefont {D.~H.}\ \bibnamefont
  {Lyth}}\ and\ \bibinfo {author} {\bibfnamefont {D.}~\bibnamefont {Wands}},\
  }\href {\doibase 10.1016/S0370-2693(01)01366-1} {\bibfield  {journal}
  {\bibinfo  {journal} {Phys. Lett. B}\ }\textbf {\bibinfo {volume} {524}},\
  \bibinfo {pages} {5} (\bibinfo {year} {2002})},\ \Eprint
  {http://arxiv.org/abs/hep-ph/0110002} {arXiv:hep-ph/0110002} \BibitemShut
  {NoStop}%
\bibitem [{\citenamefont {Moroi}\ and\ \citenamefont
  {Takahashi}(2001)}]{Moroi:2001ct}%
  \BibitemOpen
  \bibfield  {author} {\bibinfo {author} {\bibfnamefont {T.}~\bibnamefont
  {Moroi}}\ and\ \bibinfo {author} {\bibfnamefont {T.}~\bibnamefont
  {Takahashi}},\ }\href {\doibase 10.1016/S0370-2693(01)01295-3} {\bibfield
  {journal} {\bibinfo  {journal} {Phys. Lett. B}\ }\textbf {\bibinfo {volume}
  {522}},\ \bibinfo {pages} {215} (\bibinfo {year} {2001})},\ \bibinfo {note}
  {[Erratum: Phys.Lett.B 539, 303--303 (2002)]},\ \Eprint
  {http://arxiv.org/abs/hep-ph/0110096} {arXiv:hep-ph/0110096} \BibitemShut
  {NoStop}%
\bibitem [{\citenamefont {Zaldarriaga}(2004)}]{Zaldarriaga:2003my}%
  \BibitemOpen
  \bibfield  {author} {\bibinfo {author} {\bibfnamefont {M.}~\bibnamefont
  {Zaldarriaga}},\ }\href {\doibase 10.1103/PhysRevD.69.043508} {\bibfield
  {journal} {\bibinfo  {journal} {Phys. Rev. D}\ }\textbf {\bibinfo {volume}
  {69}},\ \bibinfo {pages} {043508} (\bibinfo {year} {2004})},\ \Eprint
  {http://arxiv.org/abs/astro-ph/0306006} {arXiv:astro-ph/0306006} \BibitemShut
  {NoStop}%
\bibitem [{\citenamefont {Aghanim}\ \emph {et~al.}(2018)\citenamefont {Aghanim}
  \emph {et~al.}}]{Aghanim:2018eyx}%
  \BibitemOpen
  \bibfield  {author} {\bibinfo {author} {\bibfnamefont {N.}~\bibnamefont
  {Aghanim}} \emph {et~al.} (\bibinfo {collaboration} {Planck}),\ }\href@noop
  {} {\  (\bibinfo {year} {2018})},\ \Eprint {http://arxiv.org/abs/1807.06209}
  {arXiv:1807.06209 [astro-ph.CO]} \BibitemShut {NoStop}%
\bibitem [{\citenamefont {Vernizzi}\ and\ \citenamefont
  {Wands}(2006)}]{Vernizzi:2006ve}%
  \BibitemOpen
  \bibfield  {author} {\bibinfo {author} {\bibfnamefont {F.}~\bibnamefont
  {Vernizzi}}\ and\ \bibinfo {author} {\bibfnamefont {D.}~\bibnamefont
  {Wands}},\ }\href {\doibase 10.1088/1475-7516/2006/05/019} {\bibfield
  {journal} {\bibinfo  {journal} {JCAP}\ }\textbf {\bibinfo {volume} {05}},\
  \bibinfo {pages} {019} (\bibinfo {year} {2006})},\ \Eprint
  {http://arxiv.org/abs/astro-ph/0603799} {arXiv:astro-ph/0603799} \BibitemShut
  {NoStop}%
\bibitem [{\citenamefont {Senatore}\ and\ \citenamefont
  {Zaldarriaga}(2012)}]{Senatore:2010wk}%
  \BibitemOpen
  \bibfield  {author} {\bibinfo {author} {\bibfnamefont {L.}~\bibnamefont
  {Senatore}}\ and\ \bibinfo {author} {\bibfnamefont {M.}~\bibnamefont
  {Zaldarriaga}},\ }\href {\doibase 10.1007/JHEP04(2012)024} {\bibfield
  {journal} {\bibinfo  {journal} {JHEP}\ }\textbf {\bibinfo {volume} {04}},\
  \bibinfo {pages} {024} (\bibinfo {year} {2012})},\ \Eprint
  {http://arxiv.org/abs/1009.2093} {arXiv:1009.2093 [hep-th]} \BibitemShut
  {NoStop}%
\bibitem [{\citenamefont {Akrami}\ \emph {et~al.}(2020)\citenamefont {Akrami}
  \emph {et~al.}}]{Planck:2019kim}%
  \BibitemOpen
  \bibfield  {author} {\bibinfo {author} {\bibfnamefont {Y.}~\bibnamefont
  {Akrami}} \emph {et~al.} (\bibinfo {collaboration} {Planck}),\ }\href
  {\doibase 10.1051/0004-6361/201935891} {\bibfield  {journal} {\bibinfo
  {journal} {Astron. Astrophys.}\ }\textbf {\bibinfo {volume} {641}},\ \bibinfo
  {pages} {A9} (\bibinfo {year} {2020})},\ \Eprint
  {http://arxiv.org/abs/1905.05697} {arXiv:1905.05697 [astro-ph.CO]}
  \BibitemShut {NoStop}%
\bibitem [{\citenamefont {Dor\'e}\ \emph {et~al.}(2014)\citenamefont {Dor\'e}
  \emph {et~al.}}]{Dore:2014cca}%
  \BibitemOpen
  \bibfield  {author} {\bibinfo {author} {\bibfnamefont {O.}~\bibnamefont
  {Dor\'e}} \emph {et~al.},\ }\href@noop {} {\  (\bibinfo {year} {2014})},\
  \Eprint {http://arxiv.org/abs/1412.4872} {arXiv:1412.4872 [astro-ph.CO]}
  \BibitemShut {NoStop}%
\bibitem [{\citenamefont {Ferraro}\ \emph {et~al.}(2019)\citenamefont {Ferraro}
  \emph {et~al.}}]{Ferraro:2019uce}%
  \BibitemOpen
  \bibfield  {author} {\bibinfo {author} {\bibfnamefont {S.}~\bibnamefont
  {Ferraro}} \emph {et~al.},\ }\href@noop {} {\  (\bibinfo {year} {2019})},\
  \Eprint {http://arxiv.org/abs/1903.09208} {arXiv:1903.09208 [astro-ph.CO]}
  \BibitemShut {NoStop}%
\bibitem [{\citenamefont {Castorina}\ \emph {et~al.}(2020)\citenamefont
  {Castorina} \emph {et~al.}}]{Castorina:2020zhz}%
  \BibitemOpen
  \bibfield  {author} {\bibinfo {author} {\bibfnamefont {E.}~\bibnamefont
  {Castorina}} \emph {et~al.},\ }\href@noop {} {\  (\bibinfo {year} {2020})},\
  \Eprint {http://arxiv.org/abs/2002.05072} {arXiv:2002.05072 [astro-ph.IM]}
  \BibitemShut {NoStop}%
\bibitem [{\citenamefont {Dalal}\ \emph {et~al.}(2008)\citenamefont {Dalal},
  \citenamefont {Dore}, \citenamefont {Huterer},\ and\ \citenamefont
  {Shirokov}}]{Dalal:2007cu}%
  \BibitemOpen
  \bibfield  {author} {\bibinfo {author} {\bibfnamefont {N.}~\bibnamefont
  {Dalal}}, \bibinfo {author} {\bibfnamefont {O.}~\bibnamefont {Dore}},
  \bibinfo {author} {\bibfnamefont {D.}~\bibnamefont {Huterer}}, \ and\
  \bibinfo {author} {\bibfnamefont {A.}~\bibnamefont {Shirokov}},\ }\href
  {\doibase 10.1103/PhysRevD.77.123514} {\bibfield  {journal} {\bibinfo
  {journal} {Phys. Rev. D}\ }\textbf {\bibinfo {volume} {77}},\ \bibinfo
  {pages} {123514} (\bibinfo {year} {2008})},\ \Eprint
  {http://arxiv.org/abs/0710.4560} {arXiv:0710.4560 [astro-ph]} \BibitemShut
  {NoStop}%
\bibitem [{\citenamefont {Matarrese}\ and\ \citenamefont
  {Verde}(2008)}]{Matarrese:2008nc}%
  \BibitemOpen
  \bibfield  {author} {\bibinfo {author} {\bibfnamefont {S.}~\bibnamefont
  {Matarrese}}\ and\ \bibinfo {author} {\bibfnamefont {L.}~\bibnamefont
  {Verde}},\ }\href {\doibase 10.1086/587840} {\bibfield  {journal} {\bibinfo
  {journal} {Astrophys. J. Lett.}\ }\textbf {\bibinfo {volume} {677}},\
  \bibinfo {pages} {L77} (\bibinfo {year} {2008})},\ \Eprint
  {http://arxiv.org/abs/0801.4826} {arXiv:0801.4826 [astro-ph]} \BibitemShut
  {NoStop}%
\bibitem [{\citenamefont {Slosar}\ \emph {et~al.}(2008)\citenamefont {Slosar},
  \citenamefont {Hirata}, \citenamefont {Seljak}, \citenamefont {Ho},\ and\
  \citenamefont {Padmanabhan}}]{Slosar:2008hx}%
  \BibitemOpen
  \bibfield  {author} {\bibinfo {author} {\bibfnamefont {A.}~\bibnamefont
  {Slosar}}, \bibinfo {author} {\bibfnamefont {C.}~\bibnamefont {Hirata}},
  \bibinfo {author} {\bibfnamefont {U.}~\bibnamefont {Seljak}}, \bibinfo
  {author} {\bibfnamefont {S.}~\bibnamefont {Ho}}, \ and\ \bibinfo {author}
  {\bibfnamefont {N.}~\bibnamefont {Padmanabhan}},\ }\href {\doibase
  10.1088/1475-7516/2008/08/031} {\bibfield  {journal} {\bibinfo  {journal}
  {JCAP}\ }\textbf {\bibinfo {volume} {08}},\ \bibinfo {pages} {031} (\bibinfo
  {year} {2008})},\ \Eprint {http://arxiv.org/abs/0805.3580} {arXiv:0805.3580
  [astro-ph]} \BibitemShut {NoStop}%
\bibitem [{\citenamefont {Xia}\ \emph {et~al.}(2010)\citenamefont {Xia},
  \citenamefont {Viel}, \citenamefont {Baccigalupi}, \citenamefont {De~Zotti},
  \citenamefont {Matarrese},\ and\ \citenamefont {Verde}}]{Xia:2010yu}%
  \BibitemOpen
  \bibfield  {author} {\bibinfo {author} {\bibfnamefont {J.-Q.}\ \bibnamefont
  {Xia}}, \bibinfo {author} {\bibfnamefont {M.}~\bibnamefont {Viel}}, \bibinfo
  {author} {\bibfnamefont {C.}~\bibnamefont {Baccigalupi}}, \bibinfo {author}
  {\bibfnamefont {G.}~\bibnamefont {De~Zotti}}, \bibinfo {author}
  {\bibfnamefont {S.}~\bibnamefont {Matarrese}}, \ and\ \bibinfo {author}
  {\bibfnamefont {L.}~\bibnamefont {Verde}},\ }\href {\doibase
  10.1088/2041-8205/717/1/L17} {\bibfield  {journal} {\bibinfo  {journal}
  {Astrophys. J. Lett.}\ }\textbf {\bibinfo {volume} {717}},\ \bibinfo {pages}
  {L17} (\bibinfo {year} {2010})},\ \Eprint {http://arxiv.org/abs/1003.3451}
  {arXiv:1003.3451 [astro-ph.CO]} \BibitemShut {NoStop}%
\bibitem [{\citenamefont {{Ross}}\ \emph {et~al.}(2013)\citenamefont {{Ross}},
  \citenamefont {{Percival}}, \citenamefont {{Carnero}}, \citenamefont
  {{Zhao}}, \citenamefont {{Manera}}, \citenamefont {{Raccanelli}},
  \citenamefont {{Aubourg}}, \citenamefont {{Bizyaev}}, \citenamefont
  {{Brewington}}, \citenamefont {{Brinkmann}}, \citenamefont {{Brownstein}},
  \citenamefont {{Cuesta}}, \citenamefont {{da Costa}}, \citenamefont
  {{Eisenstein}}, \citenamefont {{Ebelke}}, \citenamefont {{Guo}},
  \citenamefont {{Hamilton}}, \citenamefont {{Maga{\~n}a}}, \citenamefont
  {{Malanushenko}}, \citenamefont {{Malanushenko}}, \citenamefont {{Maraston}},
  \citenamefont {{Montesano}}, \citenamefont {{Nichol}}, \citenamefont
  {{Oravetz}}, \citenamefont {{Pan}}, \citenamefont {{Prada}}, \citenamefont
  {{S{\'a}nchez}}, \citenamefont {{Samushia}}, \citenamefont {{Schlegel}},
  \citenamefont {{Schneider}}, \citenamefont {{Seo}}, \citenamefont
  {{Sheldon}}, \citenamefont {{Simmons}}, \citenamefont {{Snedden}},
  \citenamefont {{Swanson}}, \citenamefont {{Thomas}}, \citenamefont
  {{Tinker}}, \citenamefont {{Tojeiro}},\ and\ \citenamefont
  {{Zehavi}}}]{2013MNRAS.428.1116R}%
  \BibitemOpen
  \bibfield  {author} {\bibinfo {author} {\bibfnamefont {A.~J.}\ \bibnamefont
  {{Ross}}}, \bibinfo {author} {\bibfnamefont {W.~J.}\ \bibnamefont
  {{Percival}}}, \bibinfo {author} {\bibfnamefont {A.}~\bibnamefont
  {{Carnero}}}, \bibinfo {author} {\bibfnamefont {G.-b.}\ \bibnamefont
  {{Zhao}}}, \bibinfo {author} {\bibfnamefont {M.}~\bibnamefont {{Manera}}},
  \bibinfo {author} {\bibfnamefont {A.}~\bibnamefont {{Raccanelli}}}, \bibinfo
  {author} {\bibfnamefont {E.}~\bibnamefont {{Aubourg}}}, \bibinfo {author}
  {\bibfnamefont {D.}~\bibnamefont {{Bizyaev}}}, \bibinfo {author}
  {\bibfnamefont {H.}~\bibnamefont {{Brewington}}}, \bibinfo {author}
  {\bibfnamefont {J.}~\bibnamefont {{Brinkmann}}}, \bibinfo {author}
  {\bibfnamefont {J.~R.}\ \bibnamefont {{Brownstein}}}, \bibinfo {author}
  {\bibfnamefont {A.~J.}\ \bibnamefont {{Cuesta}}}, \bibinfo {author}
  {\bibfnamefont {L.~A.~N.}\ \bibnamefont {{da Costa}}}, \bibinfo {author}
  {\bibfnamefont {D.~J.}\ \bibnamefont {{Eisenstein}}}, \bibinfo {author}
  {\bibfnamefont {G.}~\bibnamefont {{Ebelke}}}, \bibinfo {author}
  {\bibfnamefont {H.}~\bibnamefont {{Guo}}}, \bibinfo {author} {\bibfnamefont
  {J.-C.}\ \bibnamefont {{Hamilton}}}, \bibinfo {author} {\bibfnamefont
  {M.~V.}\ \bibnamefont {{Maga{\~n}a}}}, \bibinfo {author} {\bibfnamefont
  {E.}~\bibnamefont {{Malanushenko}}}, \bibinfo {author} {\bibfnamefont
  {V.}~\bibnamefont {{Malanushenko}}}, \bibinfo {author} {\bibfnamefont
  {C.}~\bibnamefont {{Maraston}}}, \bibinfo {author} {\bibfnamefont
  {F.}~\bibnamefont {{Montesano}}}, \bibinfo {author} {\bibfnamefont {R.~C.}\
  \bibnamefont {{Nichol}}}, \bibinfo {author} {\bibfnamefont {D.}~\bibnamefont
  {{Oravetz}}}, \bibinfo {author} {\bibfnamefont {K.}~\bibnamefont {{Pan}}},
  \bibinfo {author} {\bibfnamefont {F.}~\bibnamefont {{Prada}}}, \bibinfo
  {author} {\bibfnamefont {A.~G.}\ \bibnamefont {{S{\'a}nchez}}}, \bibinfo
  {author} {\bibfnamefont {L.}~\bibnamefont {{Samushia}}}, \bibinfo {author}
  {\bibfnamefont {D.~J.}\ \bibnamefont {{Schlegel}}}, \bibinfo {author}
  {\bibfnamefont {D.~P.}\ \bibnamefont {{Schneider}}}, \bibinfo {author}
  {\bibfnamefont {H.-J.}\ \bibnamefont {{Seo}}}, \bibinfo {author}
  {\bibfnamefont {A.}~\bibnamefont {{Sheldon}}}, \bibinfo {author}
  {\bibfnamefont {A.}~\bibnamefont {{Simmons}}}, \bibinfo {author}
  {\bibfnamefont {S.}~\bibnamefont {{Snedden}}}, \bibinfo {author}
  {\bibfnamefont {M.~E.~C.}\ \bibnamefont {{Swanson}}}, \bibinfo {author}
  {\bibfnamefont {D.}~\bibnamefont {{Thomas}}}, \bibinfo {author}
  {\bibfnamefont {J.~L.}\ \bibnamefont {{Tinker}}}, \bibinfo {author}
  {\bibfnamefont {R.}~\bibnamefont {{Tojeiro}}}, \ and\ \bibinfo {author}
  {\bibfnamefont {I.}~\bibnamefont {{Zehavi}}},\ }\href {\doibase
  10.1093/mnras/sts094} {\bibfield  {journal} {\bibinfo  {journal} {MNRAS}\
  }\textbf {\bibinfo {volume} {428}},\ \bibinfo {pages} {1116} (\bibinfo {year}
  {2013})},\ \Eprint {http://arxiv.org/abs/1208.1491} {arXiv:1208.1491
  [astro-ph.CO]} \BibitemShut {NoStop}%
\bibitem [{\citenamefont {Castorina}\ \emph {et~al.}(2019)\citenamefont
  {Castorina} \emph {et~al.}}]{Castorina:2019wmr}%
  \BibitemOpen
  \bibfield  {author} {\bibinfo {author} {\bibfnamefont {E.}~\bibnamefont
  {Castorina}} \emph {et~al.},\ }\href {\doibase 10.1088/1475-7516/2019/09/010}
  {\bibfield  {journal} {\bibinfo  {journal} {JCAP}\ }\textbf {\bibinfo
  {volume} {09}},\ \bibinfo {pages} {010} (\bibinfo {year} {2019})},\ \Eprint
  {http://arxiv.org/abs/1904.08859} {arXiv:1904.08859 [astro-ph.CO]}
  \BibitemShut {NoStop}%
\bibitem [{\citenamefont {Mueller}\ \emph {et~al.}(2021)\citenamefont {Mueller}
  \emph {et~al.}}]{Mueller:2021tqa}%
  \BibitemOpen
  \bibfield  {author} {\bibinfo {author} {\bibfnamefont {E.-M.}\ \bibnamefont
  {Mueller}} \emph {et~al.},\ }\href@noop {} {\  (\bibinfo {year} {2021})},\
  \Eprint {http://arxiv.org/abs/2106.13725} {arXiv:2106.13725 [astro-ph.CO]}
  \BibitemShut {NoStop}%
\bibitem [{\citenamefont {Scoccimarro}\ \emph {et~al.}(2004)\citenamefont
  {Scoccimarro}, \citenamefont {Sefusatti},\ and\ \citenamefont
  {Zaldarriaga}}]{Scoccimarro:2003wn}%
  \BibitemOpen
  \bibfield  {author} {\bibinfo {author} {\bibfnamefont {R.}~\bibnamefont
  {Scoccimarro}}, \bibinfo {author} {\bibfnamefont {E.}~\bibnamefont
  {Sefusatti}}, \ and\ \bibinfo {author} {\bibfnamefont {M.}~\bibnamefont
  {Zaldarriaga}},\ }\href {\doibase 10.1103/PhysRevD.69.103513} {\bibfield
  {journal} {\bibinfo  {journal} {Phys. Rev. D}\ }\textbf {\bibinfo {volume}
  {69}},\ \bibinfo {pages} {103513} (\bibinfo {year} {2004})},\ \Eprint
  {http://arxiv.org/abs/astro-ph/0312286} {arXiv:astro-ph/0312286} \BibitemShut
  {NoStop}%
\bibitem [{\citenamefont {Baldauf}\ \emph {et~al.}(2011)\citenamefont
  {Baldauf}, \citenamefont {Seljak},\ and\ \citenamefont
  {Senatore}}]{Baldauf:2010vn}%
  \BibitemOpen
  \bibfield  {author} {\bibinfo {author} {\bibfnamefont {T.}~\bibnamefont
  {Baldauf}}, \bibinfo {author} {\bibfnamefont {U.}~\bibnamefont {Seljak}}, \
  and\ \bibinfo {author} {\bibfnamefont {L.}~\bibnamefont {Senatore}},\ }\href
  {\doibase 10.1088/1475-7516/2011/04/006} {\bibfield  {journal} {\bibinfo
  {journal} {JCAP}\ }\textbf {\bibinfo {volume} {04}},\ \bibinfo {pages} {006}
  (\bibinfo {year} {2011})},\ \Eprint {http://arxiv.org/abs/1011.1513}
  {arXiv:1011.1513 [astro-ph.CO]} \BibitemShut {NoStop}%
\bibitem [{\citenamefont {Moradinezhad~Dizgah}\ \emph
  {et~al.}(2021)\citenamefont {Moradinezhad~Dizgah}, \citenamefont {Biagetti},
  \citenamefont {Sefusatti}, \citenamefont {Desjacques},\ and\ \citenamefont
  {Nore\~na}}]{MoradinezhadDizgah:2020whw}%
  \BibitemOpen
  \bibfield  {author} {\bibinfo {author} {\bibfnamefont {A.}~\bibnamefont
  {Moradinezhad~Dizgah}}, \bibinfo {author} {\bibfnamefont {M.}~\bibnamefont
  {Biagetti}}, \bibinfo {author} {\bibfnamefont {E.}~\bibnamefont {Sefusatti}},
  \bibinfo {author} {\bibfnamefont {V.}~\bibnamefont {Desjacques}}, \ and\
  \bibinfo {author} {\bibfnamefont {J.}~\bibnamefont {Nore\~na}},\ }\href
  {\doibase 10.1088/1475-7516/2021/05/015} {\bibfield  {journal} {\bibinfo
  {journal} {JCAP}\ }\textbf {\bibinfo {volume} {05}},\ \bibinfo {pages} {015}
  (\bibinfo {year} {2021})},\ \Eprint {http://arxiv.org/abs/2010.14523}
  {arXiv:2010.14523 [astro-ph.CO]} \BibitemShut {NoStop}%
\bibitem [{\citenamefont {Barreira}(2020)}]{Barreira:2020ekm}%
  \BibitemOpen
  \bibfield  {author} {\bibinfo {author} {\bibfnamefont {A.}~\bibnamefont
  {Barreira}},\ }\href {\doibase 10.1088/1475-7516/2020/12/031} {\bibfield
  {journal} {\bibinfo  {journal} {JCAP}\ }\textbf {\bibinfo {volume} {12}},\
  \bibinfo {pages} {031} (\bibinfo {year} {2020})},\ \Eprint
  {http://arxiv.org/abs/2009.06622} {arXiv:2009.06622 [astro-ph.CO]}
  \BibitemShut {NoStop}%
\bibitem [{\citenamefont {Philcox}(2021{\natexlab{a}})}]{Philcox:2020vbm}%
  \BibitemOpen
  \bibfield  {author} {\bibinfo {author} {\bibfnamefont {O.~H.~E.}\
  \bibnamefont {Philcox}},\ }\href {\doibase 10.1103/PhysRevD.103.103504}
  {\bibfield  {journal} {\bibinfo  {journal} {Phys. Rev. D}\ }\textbf {\bibinfo
  {volume} {103}},\ \bibinfo {pages} {103504} (\bibinfo {year}
  {2021}{\natexlab{a}})},\ \Eprint {http://arxiv.org/abs/2012.09389}
  {arXiv:2012.09389 [astro-ph.CO]} \BibitemShut {NoStop}%
\bibitem [{\citenamefont {Philcox}(2021{\natexlab{b}})}]{Philcox:2021ukg}%
  \BibitemOpen
  \bibfield  {author} {\bibinfo {author} {\bibfnamefont {O.~H.~E.}\
  \bibnamefont {Philcox}},\ }\href {\doibase 10.1103/PhysRevD.104.123529}
  {\bibfield  {journal} {\bibinfo  {journal} {Phys. Rev. D}\ }\textbf {\bibinfo
  {volume} {104}},\ \bibinfo {pages} {123529} (\bibinfo {year}
  {2021}{\natexlab{b}})},\ \Eprint {http://arxiv.org/abs/2107.06287}
  {arXiv:2107.06287 [astro-ph.CO]} \BibitemShut {NoStop}%
\bibitem [{\citenamefont {Pardede}\ \emph {et~al.}(2022)\citenamefont
  {Pardede}, \citenamefont {Rizzo}, \citenamefont {Biagetti}, \citenamefont
  {Castorina}, \citenamefont {Sefusatti},\ and\ \citenamefont
  {Monaco}}]{Pardede:2022udo}%
  \BibitemOpen
  \bibfield  {author} {\bibinfo {author} {\bibfnamefont {K.}~\bibnamefont
  {Pardede}}, \bibinfo {author} {\bibfnamefont {F.}~\bibnamefont {Rizzo}},
  \bibinfo {author} {\bibfnamefont {M.}~\bibnamefont {Biagetti}}, \bibinfo
  {author} {\bibfnamefont {E.}~\bibnamefont {Castorina}}, \bibinfo {author}
  {\bibfnamefont {E.}~\bibnamefont {Sefusatti}}, \ and\ \bibinfo {author}
  {\bibfnamefont {P.}~\bibnamefont {Monaco}},\ }\href@noop {} {\  (\bibinfo
  {year} {2022})},\ \Eprint {http://arxiv.org/abs/2203.04174} {arXiv:2203.04174
  [astro-ph.CO]} \BibitemShut {NoStop}%
\bibitem [{\citenamefont {Gil-Marín}\ \emph {et~al.}(2015)\citenamefont
  {Gil-Marín}, \citenamefont {Noreña}, \citenamefont {Verde}, \citenamefont
  {Percival}, \citenamefont {Wagner}, \citenamefont {Manera},\ and\
  \citenamefont {Schneider}}]{Gil-Marin:2014sta}%
  \BibitemOpen
  \bibfield  {author} {\bibinfo {author} {\bibfnamefont {H.}~\bibnamefont
  {Gil-Marín}}, \bibinfo {author} {\bibfnamefont {J.}~\bibnamefont {Noreña}},
  \bibinfo {author} {\bibfnamefont {L.}~\bibnamefont {Verde}}, \bibinfo
  {author} {\bibfnamefont {W.~J.}\ \bibnamefont {Percival}}, \bibinfo {author}
  {\bibfnamefont {C.}~\bibnamefont {Wagner}}, \bibinfo {author} {\bibfnamefont
  {M.}~\bibnamefont {Manera}}, \ and\ \bibinfo {author} {\bibfnamefont {D.~P.}\
  \bibnamefont {Schneider}},\ }\href {\doibase 10.1093/mnras/stv961} {\bibfield
   {journal} {\bibinfo  {journal} {Mon. Not. Roy. Astron. Soc.}\ }\textbf
  {\bibinfo {volume} {451}},\ \bibinfo {pages} {539} (\bibinfo {year}
  {2015})},\ \Eprint {http://arxiv.org/abs/1407.5668} {arXiv:1407.5668
  [astro-ph.CO]} \BibitemShut {NoStop}%
\bibitem [{\citenamefont {Gil-Marín}\ \emph {et~al.}(2017)\citenamefont
  {Gil-Marín}, \citenamefont {Percival}, \citenamefont {Verde}, \citenamefont
  {Brownstein}, \citenamefont {Chuang}, \citenamefont {Kitaura}, \citenamefont
  {Rodríguez-Torres},\ and\ \citenamefont {Olmstead}}]{Gil-Marin:2016wya}%
  \BibitemOpen
  \bibfield  {author} {\bibinfo {author} {\bibfnamefont {H.}~\bibnamefont
  {Gil-Marín}}, \bibinfo {author} {\bibfnamefont {W.~J.}\ \bibnamefont
  {Percival}}, \bibinfo {author} {\bibfnamefont {L.}~\bibnamefont {Verde}},
  \bibinfo {author} {\bibfnamefont {J.~R.}\ \bibnamefont {Brownstein}},
  \bibinfo {author} {\bibfnamefont {C.-H.}\ \bibnamefont {Chuang}}, \bibinfo
  {author} {\bibfnamefont {F.-S.}\ \bibnamefont {Kitaura}}, \bibinfo {author}
  {\bibfnamefont {S.~A.}\ \bibnamefont {Rodríguez-Torres}}, \ and\ \bibinfo
  {author} {\bibfnamefont {M.~D.}\ \bibnamefont {Olmstead}},\ }\href {\doibase
  10.1093/mnras/stw2679} {\bibfield  {journal} {\bibinfo  {journal} {Mon. Not.
  Roy. Astron. Soc.}\ }\textbf {\bibinfo {volume} {465}},\ \bibinfo {pages}
  {1757} (\bibinfo {year} {2017})},\ \Eprint {http://arxiv.org/abs/1606.00439}
  {arXiv:1606.00439 [astro-ph.CO]} \BibitemShut {NoStop}%
\bibitem [{\citenamefont {D'Amico}\ \emph {et~al.}(2022)\citenamefont
  {D'Amico}, \citenamefont {Lewandowski}, \citenamefont {Senatore},\ and\
  \citenamefont {Zhang}}]{DAmico:2022gki}%
  \BibitemOpen
  \bibfield  {author} {\bibinfo {author} {\bibfnamefont {G.}~\bibnamefont
  {D'Amico}}, \bibinfo {author} {\bibfnamefont {M.}~\bibnamefont
  {Lewandowski}}, \bibinfo {author} {\bibfnamefont {L.}~\bibnamefont
  {Senatore}}, \ and\ \bibinfo {author} {\bibfnamefont {P.}~\bibnamefont
  {Zhang}},\ }\href@noop {} {\  (\bibinfo {year} {2022})},\ \Eprint
  {http://arxiv.org/abs/2201.11518} {arXiv:2201.11518 [astro-ph.CO]}
  \BibitemShut {NoStop}%
\bibitem [{\citenamefont {Scoccimarro}\ \emph {et~al.}(1999)\citenamefont
  {Scoccimarro}, \citenamefont {Couchman},\ and\ \citenamefont
  {Frieman}}]{Scoccimarro:1999ed}%
  \BibitemOpen
  \bibfield  {author} {\bibinfo {author} {\bibfnamefont {R.}~\bibnamefont
  {Scoccimarro}}, \bibinfo {author} {\bibfnamefont {H.~M.~P.}\ \bibnamefont
  {Couchman}}, \ and\ \bibinfo {author} {\bibfnamefont {J.~A.}\ \bibnamefont
  {Frieman}},\ }\href {\doibase 10.1086/307220} {\bibfield  {journal} {\bibinfo
   {journal} {Astrophys. J.}\ }\textbf {\bibinfo {volume} {517}},\ \bibinfo
  {pages} {531} (\bibinfo {year} {1999})},\ \Eprint
  {http://arxiv.org/abs/astro-ph/9808305} {arXiv:astro-ph/9808305 [astro-ph]}
  \BibitemShut {NoStop}%
\bibitem [{\citenamefont {Scoccimarro}(2000)}]{Scoccimarro:2000sn}%
  \BibitemOpen
  \bibfield  {author} {\bibinfo {author} {\bibfnamefont {R.}~\bibnamefont
  {Scoccimarro}},\ }\href {\doibase 10.1086/317248} {\bibfield  {journal}
  {\bibinfo  {journal} {Astrophys. J.}\ }\textbf {\bibinfo {volume} {544}},\
  \bibinfo {pages} {597} (\bibinfo {year} {2000})},\ \Eprint
  {http://arxiv.org/abs/astro-ph/0004086} {arXiv:astro-ph/0004086} \BibitemShut
  {NoStop}%
\bibitem [{\citenamefont {Sefusatti}\ \emph {et~al.}(2006)\citenamefont
  {Sefusatti}, \citenamefont {Crocce}, \citenamefont {Pueblas},\ and\
  \citenamefont {Scoccimarro}}]{Sefusatti:2006pa}%
  \BibitemOpen
  \bibfield  {author} {\bibinfo {author} {\bibfnamefont {E.}~\bibnamefont
  {Sefusatti}}, \bibinfo {author} {\bibfnamefont {M.}~\bibnamefont {Crocce}},
  \bibinfo {author} {\bibfnamefont {S.}~\bibnamefont {Pueblas}}, \ and\
  \bibinfo {author} {\bibfnamefont {R.}~\bibnamefont {Scoccimarro}},\ }\href
  {\doibase 10.1103/PhysRevD.74.023522} {\bibfield  {journal} {\bibinfo
  {journal} {Phys. Rev.}\ }\textbf {\bibinfo {volume} {D74}},\ \bibinfo {pages}
  {023522} (\bibinfo {year} {2006})},\ \Eprint
  {http://arxiv.org/abs/astro-ph/0604505} {arXiv:astro-ph/0604505 [astro-ph]}
  \BibitemShut {NoStop}%
\bibitem [{\citenamefont {Sefusatti}\ and\ \citenamefont
  {Komatsu}(2007)}]{Sefusatti:2007ih}%
  \BibitemOpen
  \bibfield  {author} {\bibinfo {author} {\bibfnamefont {E.}~\bibnamefont
  {Sefusatti}}\ and\ \bibinfo {author} {\bibfnamefont {E.}~\bibnamefont
  {Komatsu}},\ }\href {\doibase 10.1103/PhysRevD.76.083004} {\bibfield
  {journal} {\bibinfo  {journal} {Phys. Rev. D}\ }\textbf {\bibinfo {volume}
  {76}},\ \bibinfo {pages} {083004} (\bibinfo {year} {2007})},\ \Eprint
  {http://arxiv.org/abs/0705.0343} {arXiv:0705.0343 [astro-ph]} \BibitemShut
  {NoStop}%
\bibitem [{\citenamefont {Sefusatti}(2009)}]{Sefusatti:2009qh}%
  \BibitemOpen
  \bibfield  {author} {\bibinfo {author} {\bibfnamefont {E.}~\bibnamefont
  {Sefusatti}},\ }\href {\doibase 10.1103/PhysRevD.80.123002} {\bibfield
  {journal} {\bibinfo  {journal} {Phys. Rev. D}\ }\textbf {\bibinfo {volume}
  {80}},\ \bibinfo {pages} {123002} (\bibinfo {year} {2009})},\ \Eprint
  {http://arxiv.org/abs/0905.0717} {arXiv:0905.0717 [astro-ph.CO]} \BibitemShut
  {NoStop}%
\bibitem [{\citenamefont {Baldauf}\ \emph {et~al.}(2015)\citenamefont
  {Baldauf}, \citenamefont {Mercolli}, \citenamefont {Mirbabayi},\ and\
  \citenamefont {Pajer}}]{Baldauf:2014qfa}%
  \BibitemOpen
  \bibfield  {author} {\bibinfo {author} {\bibfnamefont {T.}~\bibnamefont
  {Baldauf}}, \bibinfo {author} {\bibfnamefont {L.}~\bibnamefont {Mercolli}},
  \bibinfo {author} {\bibfnamefont {M.}~\bibnamefont {Mirbabayi}}, \ and\
  \bibinfo {author} {\bibfnamefont {E.}~\bibnamefont {Pajer}},\ }\href
  {\doibase 10.1088/1475-7516/2015/05/007} {\bibfield  {journal} {\bibinfo
  {journal} {JCAP}\ }\textbf {\bibinfo {volume} {1505}},\ \bibinfo {pages}
  {007} (\bibinfo {year} {2015})},\ \Eprint {http://arxiv.org/abs/1406.4135}
  {arXiv:1406.4135 [astro-ph.CO]} \BibitemShut {NoStop}%
\bibitem [{\citenamefont {Angulo}\ \emph {et~al.}(2015)\citenamefont {Angulo},
  \citenamefont {Foreman}, \citenamefont {Schmittfull},\ and\ \citenamefont
  {Senatore}}]{Angulo:2014tfa}%
  \BibitemOpen
  \bibfield  {author} {\bibinfo {author} {\bibfnamefont {R.~E.}\ \bibnamefont
  {Angulo}}, \bibinfo {author} {\bibfnamefont {S.}~\bibnamefont {Foreman}},
  \bibinfo {author} {\bibfnamefont {M.}~\bibnamefont {Schmittfull}}, \ and\
  \bibinfo {author} {\bibfnamefont {L.}~\bibnamefont {Senatore}},\ }\href
  {\doibase 10.1088/1475-7516/2015/10/039} {\bibfield  {journal} {\bibinfo
  {journal} {JCAP}\ }\textbf {\bibinfo {volume} {1510}},\ \bibinfo {pages}
  {039} (\bibinfo {year} {2015})},\ \Eprint {http://arxiv.org/abs/1406.4143}
  {arXiv:1406.4143 [astro-ph.CO]} \BibitemShut {NoStop}%
\bibitem [{\citenamefont {Blas}\ \emph
  {et~al.}(2016{\natexlab{a}})\citenamefont {Blas}, \citenamefont {Garny},
  \citenamefont {Ivanov},\ and\ \citenamefont {Sibiryakov}}]{Blas:2016sfa}%
  \BibitemOpen
  \bibfield  {author} {\bibinfo {author} {\bibfnamefont {D.}~\bibnamefont
  {Blas}}, \bibinfo {author} {\bibfnamefont {M.}~\bibnamefont {Garny}},
  \bibinfo {author} {\bibfnamefont {M.~M.}\ \bibnamefont {Ivanov}}, \ and\
  \bibinfo {author} {\bibfnamefont {S.}~\bibnamefont {Sibiryakov}},\ }\href
  {\doibase 10.1088/1475-7516/2016/07/028} {\bibfield  {journal} {\bibinfo
  {journal} {JCAP}\ }\textbf {\bibinfo {volume} {1607}},\ \bibinfo {pages}
  {028} (\bibinfo {year} {2016}{\natexlab{a}})},\ \Eprint
  {http://arxiv.org/abs/1605.02149} {arXiv:1605.02149 [astro-ph.CO]}
  \BibitemShut {NoStop}%
\bibitem [{\citenamefont {Nadler}\ \emph {et~al.}(2018)\citenamefont {Nadler},
  \citenamefont {Perko},\ and\ \citenamefont {Senatore}}]{Nadler:2017qto}%
  \BibitemOpen
  \bibfield  {author} {\bibinfo {author} {\bibfnamefont {E.~O.}\ \bibnamefont
  {Nadler}}, \bibinfo {author} {\bibfnamefont {A.}~\bibnamefont {Perko}}, \
  and\ \bibinfo {author} {\bibfnamefont {L.}~\bibnamefont {Senatore}},\ }\href
  {\doibase 10.1088/1475-7516/2018/02/058} {\bibfield  {journal} {\bibinfo
  {journal} {JCAP}\ }\textbf {\bibinfo {volume} {02}},\ \bibinfo {pages} {058}
  (\bibinfo {year} {2018})},\ \Eprint {http://arxiv.org/abs/1710.10308}
  {arXiv:1710.10308 [astro-ph.CO]} \BibitemShut {NoStop}%
\bibitem [{\citenamefont {Eggemeier}\ \emph {et~al.}(2018)\citenamefont
  {Eggemeier}, \citenamefont {Scoccimarro},\ and\ \citenamefont
  {Smith}}]{Eggemeier:2018qae}%
  \BibitemOpen
  \bibfield  {author} {\bibinfo {author} {\bibfnamefont {A.}~\bibnamefont
  {Eggemeier}}, \bibinfo {author} {\bibfnamefont {R.}~\bibnamefont
  {Scoccimarro}}, \ and\ \bibinfo {author} {\bibfnamefont {R.~E.}\ \bibnamefont
  {Smith}},\ }\href@noop {} {\  (\bibinfo {year} {2018})},\ \Eprint
  {http://arxiv.org/abs/1812.03208} {arXiv:1812.03208 [astro-ph.CO]}
  \BibitemShut {NoStop}%
\bibitem [{\citenamefont {Eggemeier}\ \emph {et~al.}(2021)\citenamefont
  {Eggemeier}, \citenamefont {Scoccimarro}, \citenamefont {Smith},
  \citenamefont {Crocce}, \citenamefont {Pezzotta},\ and\ \citenamefont
  {S\'anchez}}]{Eggemeier:2021cam}%
  \BibitemOpen
  \bibfield  {author} {\bibinfo {author} {\bibfnamefont {A.}~\bibnamefont
  {Eggemeier}}, \bibinfo {author} {\bibfnamefont {R.}~\bibnamefont
  {Scoccimarro}}, \bibinfo {author} {\bibfnamefont {R.~E.}\ \bibnamefont
  {Smith}}, \bibinfo {author} {\bibfnamefont {M.}~\bibnamefont {Crocce}},
  \bibinfo {author} {\bibfnamefont {A.}~\bibnamefont {Pezzotta}}, \ and\
  \bibinfo {author} {\bibfnamefont {A.~G.}\ \bibnamefont {S\'anchez}},\
  }\href@noop {} {\  (\bibinfo {year} {2021})},\ \Eprint
  {http://arxiv.org/abs/2102.06902} {arXiv:2102.06902 [astro-ph.CO]}
  \BibitemShut {NoStop}%
\bibitem [{\citenamefont {Desjacques}\ \emph
  {et~al.}(2018{\natexlab{a}})\citenamefont {Desjacques}, \citenamefont
  {Jeong},\ and\ \citenamefont {Schmidt}}]{Desjacques:2016bnm}%
  \BibitemOpen
  \bibfield  {author} {\bibinfo {author} {\bibfnamefont {V.}~\bibnamefont
  {Desjacques}}, \bibinfo {author} {\bibfnamefont {D.}~\bibnamefont {Jeong}}, \
  and\ \bibinfo {author} {\bibfnamefont {F.}~\bibnamefont {Schmidt}},\ }\href
  {\doibase 10.1016/j.physrep.2017.12.002} {\bibfield  {journal} {\bibinfo
  {journal} {Phys. Rept.}\ }\textbf {\bibinfo {volume} {733}},\ \bibinfo
  {pages} {1} (\bibinfo {year} {2018}{\natexlab{a}})},\ \Eprint
  {http://arxiv.org/abs/1611.09787} {arXiv:1611.09787 [astro-ph.CO]}
  \BibitemShut {NoStop}%
\bibitem [{\citenamefont {Desjacques}\ \emph
  {et~al.}(2018{\natexlab{b}})\citenamefont {Desjacques}, \citenamefont
  {Jeong},\ and\ \citenamefont {Schmidt}}]{Desjacques:2018pfv}%
  \BibitemOpen
  \bibfield  {author} {\bibinfo {author} {\bibfnamefont {V.}~\bibnamefont
  {Desjacques}}, \bibinfo {author} {\bibfnamefont {D.}~\bibnamefont {Jeong}}, \
  and\ \bibinfo {author} {\bibfnamefont {F.}~\bibnamefont {Schmidt}},\ }\href
  {\doibase 10.1088/1475-7516/2018/12/035} {\bibfield  {journal} {\bibinfo
  {journal} {JCAP}\ }\textbf {\bibinfo {volume} {1812}},\ \bibinfo {pages}
  {035} (\bibinfo {year} {2018}{\natexlab{b}})},\ \Eprint
  {http://arxiv.org/abs/1806.04015} {arXiv:1806.04015 [astro-ph.CO]}
  \BibitemShut {NoStop}%
\bibitem [{\citenamefont {Ivanov}\ and\ \citenamefont
  {Sibiryakov}(2018)}]{Ivanov:2018gjr}%
  \BibitemOpen
  \bibfield  {author} {\bibinfo {author} {\bibfnamefont {M.~M.}\ \bibnamefont
  {Ivanov}}\ and\ \bibinfo {author} {\bibfnamefont {S.}~\bibnamefont
  {Sibiryakov}},\ }\href {\doibase 10.1088/1475-7516/2018/07/053} {\bibfield
  {journal} {\bibinfo  {journal} {JCAP}\ }\textbf {\bibinfo {volume} {1807}},\
  \bibinfo {pages} {053} (\bibinfo {year} {2018})},\ \Eprint
  {http://arxiv.org/abs/1804.05080} {arXiv:1804.05080 [astro-ph.CO]}
  \BibitemShut {NoStop}%
\bibitem [{\citenamefont {Oddo}\ \emph {et~al.}(2020)\citenamefont {Oddo},
  \citenamefont {Sefusatti}, \citenamefont {Porciani}, \citenamefont {Monaco},\
  and\ \citenamefont {S\'anchez}}]{Oddo:2019run}%
  \BibitemOpen
  \bibfield  {author} {\bibinfo {author} {\bibfnamefont {A.}~\bibnamefont
  {Oddo}}, \bibinfo {author} {\bibfnamefont {E.}~\bibnamefont {Sefusatti}},
  \bibinfo {author} {\bibfnamefont {C.}~\bibnamefont {Porciani}}, \bibinfo
  {author} {\bibfnamefont {P.}~\bibnamefont {Monaco}}, \ and\ \bibinfo {author}
  {\bibfnamefont {A.~G.}\ \bibnamefont {S\'anchez}},\ }\href {\doibase
  10.1088/1475-7516/2020/03/056} {\bibfield  {journal} {\bibinfo  {journal}
  {JCAP}\ }\textbf {\bibinfo {volume} {03}},\ \bibinfo {pages} {056} (\bibinfo
  {year} {2020})},\ \Eprint {http://arxiv.org/abs/1908.01774} {arXiv:1908.01774
  [astro-ph.CO]} \BibitemShut {NoStop}%
\bibitem [{\citenamefont {Oddo}\ \emph {et~al.}(2021)\citenamefont {Oddo},
  \citenamefont {Rizzo}, \citenamefont {Sefusatti}, \citenamefont {Porciani},\
  and\ \citenamefont {Monaco}}]{Oddo:2021iwq}%
  \BibitemOpen
  \bibfield  {author} {\bibinfo {author} {\bibfnamefont {A.}~\bibnamefont
  {Oddo}}, \bibinfo {author} {\bibfnamefont {F.}~\bibnamefont {Rizzo}},
  \bibinfo {author} {\bibfnamefont {E.}~\bibnamefont {Sefusatti}}, \bibinfo
  {author} {\bibfnamefont {C.}~\bibnamefont {Porciani}}, \ and\ \bibinfo
  {author} {\bibfnamefont {P.}~\bibnamefont {Monaco}},\ }\href {\doibase
  10.1088/1475-7516/2021/11/038} {\bibfield  {journal} {\bibinfo  {journal}
  {JCAP}\ }\textbf {\bibinfo {volume} {11}},\ \bibinfo {pages} {038} (\bibinfo
  {year} {2021})},\ \Eprint {http://arxiv.org/abs/2108.03204} {arXiv:2108.03204
  [astro-ph.CO]} \BibitemShut {NoStop}%
\bibitem [{\citenamefont {Ivanov}\ \emph
  {et~al.}(2021{\natexlab{a}})\citenamefont {Ivanov}, \citenamefont {Philcox},
  \citenamefont {Nishimichi}, \citenamefont {Simonovi\'c}, \citenamefont
  {Takada},\ and\ \citenamefont {Zaldarriaga}}]{Ivanov:2021kcd}%
  \BibitemOpen
  \bibfield  {author} {\bibinfo {author} {\bibfnamefont {M.~M.}\ \bibnamefont
  {Ivanov}}, \bibinfo {author} {\bibfnamefont {O.~H.~E.}\ \bibnamefont
  {Philcox}}, \bibinfo {author} {\bibfnamefont {T.}~\bibnamefont {Nishimichi}},
  \bibinfo {author} {\bibfnamefont {M.}~\bibnamefont {Simonovi\'c}}, \bibinfo
  {author} {\bibfnamefont {M.}~\bibnamefont {Takada}}, \ and\ \bibinfo {author}
  {\bibfnamefont {M.}~\bibnamefont {Zaldarriaga}},\ }\href@noop {} {\
  (\bibinfo {year} {2021}{\natexlab{a}})},\ \Eprint
  {http://arxiv.org/abs/2110.10161} {arXiv:2110.10161 [astro-ph.CO]}
  \BibitemShut {NoStop}%
\bibitem [{\citenamefont {Philcox}\ and\ \citenamefont
  {Ivanov}(2021)}]{Philcox:2021kcw}%
  \BibitemOpen
  \bibfield  {author} {\bibinfo {author} {\bibfnamefont {O.~H.~E.}\
  \bibnamefont {Philcox}}\ and\ \bibinfo {author} {\bibfnamefont {M.~M.}\
  \bibnamefont {Ivanov}},\ }\href@noop {} {\  (\bibinfo {year} {2021})},\
  \Eprint {http://arxiv.org/abs/2112.04515} {arXiv:2112.04515 [astro-ph.CO]}
  \BibitemShut {NoStop}%
\bibitem [{\citenamefont {Alam}\ \emph {et~al.}(2017)\citenamefont {Alam} \emph
  {et~al.}}]{Alam:2016hwk}%
  \BibitemOpen
  \bibfield  {author} {\bibinfo {author} {\bibfnamefont {S.}~\bibnamefont
  {Alam}} \emph {et~al.} (\bibinfo {collaboration} {BOSS}),\ }\href {\doibase
  10.1093/mnras/stx721} {\bibfield  {journal} {\bibinfo  {journal} {Mon. Not.
  Roy. Astron. Soc.}\ }\textbf {\bibinfo {volume} {470}},\ \bibinfo {pages}
  {2617} (\bibinfo {year} {2017})},\ \Eprint {http://arxiv.org/abs/1607.03155}
  {arXiv:1607.03155 [astro-ph.CO]} \BibitemShut {NoStop}%
\bibitem [{\citenamefont {Cabass}\ \emph
  {et~al.}(2022{\natexlab{a}})\citenamefont {Cabass}, \citenamefont {Ivanov},
  \citenamefont {Philcox}, \citenamefont {Simonovi\'c},\ and\ \citenamefont
  {Zaldarriaga}}]{Cabass:2022wjy}%
  \BibitemOpen
  \bibfield  {author} {\bibinfo {author} {\bibfnamefont {G.}~\bibnamefont
  {Cabass}}, \bibinfo {author} {\bibfnamefont {M.~M.}\ \bibnamefont {Ivanov}},
  \bibinfo {author} {\bibfnamefont {O.~H.~E.}\ \bibnamefont {Philcox}},
  \bibinfo {author} {\bibfnamefont {M.}~\bibnamefont {Simonovi\'c}}, \ and\
  \bibinfo {author} {\bibfnamefont {M.}~\bibnamefont {Zaldarriaga}},\
  }\href@noop {} {\  (\bibinfo {year} {2022}{\natexlab{a}})},\ \Eprint
  {http://arxiv.org/abs/2201.07238} {arXiv:2201.07238 [astro-ph.CO]}
  \BibitemShut {NoStop}%
\bibitem [{\citenamefont {Perko}\ \emph {et~al.}(2016)\citenamefont {Perko},
  \citenamefont {Senatore}, \citenamefont {Jennings},\ and\ \citenamefont
  {Wechsler}}]{Perko:2016puo}%
  \BibitemOpen
  \bibfield  {author} {\bibinfo {author} {\bibfnamefont {A.}~\bibnamefont
  {Perko}}, \bibinfo {author} {\bibfnamefont {L.}~\bibnamefont {Senatore}},
  \bibinfo {author} {\bibfnamefont {E.}~\bibnamefont {Jennings}}, \ and\
  \bibinfo {author} {\bibfnamefont {R.~H.}\ \bibnamefont {Wechsler}},\
  }\href@noop {} {\  (\bibinfo {year} {2016})},\ \Eprint
  {http://arxiv.org/abs/1610.09321} {arXiv:1610.09321 [astro-ph.CO]}
  \BibitemShut {NoStop}%
\bibitem [{\citenamefont {Ivanov}\ \emph
  {et~al.}(2020{\natexlab{a}})\citenamefont {Ivanov}, \citenamefont
  {Simonovi\'c},\ and\ \citenamefont {Zaldarriaga}}]{Ivanov:2019pdj}%
  \BibitemOpen
  \bibfield  {author} {\bibinfo {author} {\bibfnamefont {M.~M.}\ \bibnamefont
  {Ivanov}}, \bibinfo {author} {\bibfnamefont {M.}~\bibnamefont {Simonovi\'c}},
  \ and\ \bibinfo {author} {\bibfnamefont {M.}~\bibnamefont {Zaldarriaga}},\
  }\href {\doibase 10.1088/1475-7516/2020/05/042} {\bibfield  {journal}
  {\bibinfo  {journal} {JCAP}\ }\textbf {\bibinfo {volume} {05}},\ \bibinfo
  {pages} {042} (\bibinfo {year} {2020}{\natexlab{a}})},\ \Eprint
  {http://arxiv.org/abs/1909.05277} {arXiv:1909.05277 [astro-ph.CO]}
  \BibitemShut {NoStop}%
\bibitem [{\citenamefont {D'Amico}\ \emph {et~al.}(2019)\citenamefont
  {D'Amico}, \citenamefont {Gleyzes}, \citenamefont {Kokron}, \citenamefont
  {Markovic}, \citenamefont {Senatore}, \citenamefont {Zhang}, \citenamefont
  {Beutler},\ and\ \citenamefont {Gil-Marín}}]{DAmico:2019fhj}%
  \BibitemOpen
  \bibfield  {author} {\bibinfo {author} {\bibfnamefont {G.}~\bibnamefont
  {D'Amico}}, \bibinfo {author} {\bibfnamefont {J.}~\bibnamefont {Gleyzes}},
  \bibinfo {author} {\bibfnamefont {N.}~\bibnamefont {Kokron}}, \bibinfo
  {author} {\bibfnamefont {D.}~\bibnamefont {Markovic}}, \bibinfo {author}
  {\bibfnamefont {L.}~\bibnamefont {Senatore}}, \bibinfo {author}
  {\bibfnamefont {P.}~\bibnamefont {Zhang}}, \bibinfo {author} {\bibfnamefont
  {F.}~\bibnamefont {Beutler}}, \ and\ \bibinfo {author} {\bibfnamefont
  {H.}~\bibnamefont {Gil-Marín}},\ }\href@noop {} {\  (\bibinfo {year}
  {2019})},\ \Eprint {http://arxiv.org/abs/1909.05271} {arXiv:1909.05271
  [astro-ph.CO]} \BibitemShut {NoStop}%
\bibitem [{\citenamefont {Ivanov}\ \emph
  {et~al.}(2020{\natexlab{b}})\citenamefont {Ivanov}, \citenamefont
  {Simonovi\'c},\ and\ \citenamefont {Zaldarriaga}}]{Ivanov:2019hqk}%
  \BibitemOpen
  \bibfield  {author} {\bibinfo {author} {\bibfnamefont {M.~M.}\ \bibnamefont
  {Ivanov}}, \bibinfo {author} {\bibfnamefont {M.}~\bibnamefont {Simonovi\'c}},
  \ and\ \bibinfo {author} {\bibfnamefont {M.}~\bibnamefont {Zaldarriaga}},\
  }\href {\doibase 10.1103/PhysRevD.101.083504} {\bibfield  {journal} {\bibinfo
   {journal} {Phys. Rev. D}\ }\textbf {\bibinfo {volume} {101}},\ \bibinfo
  {pages} {083504} (\bibinfo {year} {2020}{\natexlab{b}})},\ \Eprint
  {http://arxiv.org/abs/1912.08208} {arXiv:1912.08208 [astro-ph.CO]}
  \BibitemShut {NoStop}%
\bibitem [{\citenamefont {Nishimichi}\ \emph {et~al.}(2020)\citenamefont
  {Nishimichi}, \citenamefont {D'Amico}, \citenamefont {Ivanov}, \citenamefont
  {Senatore}, \citenamefont {Simonovi\'c}, \citenamefont {Takada},
  \citenamefont {Zaldarriaga},\ and\ \citenamefont
  {Zhang}}]{Nishimichi:2020tvu}%
  \BibitemOpen
  \bibfield  {author} {\bibinfo {author} {\bibfnamefont {T.}~\bibnamefont
  {Nishimichi}}, \bibinfo {author} {\bibfnamefont {G.}~\bibnamefont {D'Amico}},
  \bibinfo {author} {\bibfnamefont {M.~M.}\ \bibnamefont {Ivanov}}, \bibinfo
  {author} {\bibfnamefont {L.}~\bibnamefont {Senatore}}, \bibinfo {author}
  {\bibfnamefont {M.}~\bibnamefont {Simonovi\'c}}, \bibinfo {author}
  {\bibfnamefont {M.}~\bibnamefont {Takada}}, \bibinfo {author} {\bibfnamefont
  {M.}~\bibnamefont {Zaldarriaga}}, \ and\ \bibinfo {author} {\bibfnamefont
  {P.}~\bibnamefont {Zhang}},\ }\href {\doibase 10.1103/PhysRevD.102.123541}
  {\bibfield  {journal} {\bibinfo  {journal} {Phys. Rev. D}\ }\textbf {\bibinfo
  {volume} {102}},\ \bibinfo {pages} {123541} (\bibinfo {year} {2020})},\
  \Eprint {http://arxiv.org/abs/2003.08277} {arXiv:2003.08277 [astro-ph.CO]}
  \BibitemShut {NoStop}%
\bibitem [{\citenamefont {Cabass}\ \emph
  {et~al.}(2022{\natexlab{b}})\citenamefont {Cabass}, \citenamefont {Ivanov},
  \citenamefont {Lewandowski}, \citenamefont {Mirbabayi},\ and\ \citenamefont
  {Simonovi\'c}}]{Cabass:2022avo}%
  \BibitemOpen
  \bibfield  {author} {\bibinfo {author} {\bibfnamefont {G.}~\bibnamefont
  {Cabass}}, \bibinfo {author} {\bibfnamefont {M.~M.}\ \bibnamefont {Ivanov}},
  \bibinfo {author} {\bibfnamefont {M.}~\bibnamefont {Lewandowski}}, \bibinfo
  {author} {\bibfnamefont {M.}~\bibnamefont {Mirbabayi}}, \ and\ \bibinfo
  {author} {\bibfnamefont {M.}~\bibnamefont {Simonovi\'c}},\ }in\ \href@noop {}
  {\emph {\bibinfo {booktitle} {{2022 Snowmass Summer Study}}}}\ (\bibinfo
  {year} {2022})\ \Eprint {http://arxiv.org/abs/2203.08232} {arXiv:2203.08232
  [astro-ph.CO]} \BibitemShut {NoStop}%
\bibitem [{\citenamefont {Taruya}\ \emph {et~al.}(2008)\citenamefont {Taruya},
  \citenamefont {Koyama},\ and\ \citenamefont {Matsubara}}]{Taruya:2008pg}%
  \BibitemOpen
  \bibfield  {author} {\bibinfo {author} {\bibfnamefont {A.}~\bibnamefont
  {Taruya}}, \bibinfo {author} {\bibfnamefont {K.}~\bibnamefont {Koyama}}, \
  and\ \bibinfo {author} {\bibfnamefont {T.}~\bibnamefont {Matsubara}},\ }\href
  {\doibase 10.1103/PhysRevD.78.123534} {\bibfield  {journal} {\bibinfo
  {journal} {Phys. Rev. D}\ }\textbf {\bibinfo {volume} {78}},\ \bibinfo
  {pages} {123534} (\bibinfo {year} {2008})},\ \Eprint
  {http://arxiv.org/abs/0808.4085} {arXiv:0808.4085 [astro-ph]} \BibitemShut
  {NoStop}%
\bibitem [{\citenamefont {Assassi}\ \emph
  {et~al.}(2015{\natexlab{a}})\citenamefont {Assassi}, \citenamefont {Baumann},
  \citenamefont {Pajer}, \citenamefont {Welling},\ and\ \citenamefont {van~der
  Woude}}]{Assassi:2015jqa}%
  \BibitemOpen
  \bibfield  {author} {\bibinfo {author} {\bibfnamefont {V.}~\bibnamefont
  {Assassi}}, \bibinfo {author} {\bibfnamefont {D.}~\bibnamefont {Baumann}},
  \bibinfo {author} {\bibfnamefont {E.}~\bibnamefont {Pajer}}, \bibinfo
  {author} {\bibfnamefont {Y.}~\bibnamefont {Welling}}, \ and\ \bibinfo
  {author} {\bibfnamefont {D.}~\bibnamefont {van~der Woude}},\ }\href {\doibase
  10.1088/1475-7516/2015/11/024} {\bibfield  {journal} {\bibinfo  {journal}
  {JCAP}\ }\textbf {\bibinfo {volume} {11}},\ \bibinfo {pages} {024} (\bibinfo
  {year} {2015}{\natexlab{a}})},\ \Eprint {http://arxiv.org/abs/1505.06668}
  {arXiv:1505.06668 [astro-ph.CO]} \BibitemShut {NoStop}%
\bibitem [{\citenamefont {Assassi}\ \emph
  {et~al.}(2015{\natexlab{b}})\citenamefont {Assassi}, \citenamefont
  {Baumann},\ and\ \citenamefont {Schmidt}}]{Assassi:2015fma}%
  \BibitemOpen
  \bibfield  {author} {\bibinfo {author} {\bibfnamefont {V.}~\bibnamefont
  {Assassi}}, \bibinfo {author} {\bibfnamefont {D.}~\bibnamefont {Baumann}}, \
  and\ \bibinfo {author} {\bibfnamefont {F.}~\bibnamefont {Schmidt}},\ }\href
  {\doibase 10.1088/1475-7516/2015/12/043} {\bibfield  {journal} {\bibinfo
  {journal} {JCAP}\ }\textbf {\bibinfo {volume} {12}},\ \bibinfo {pages} {043}
  (\bibinfo {year} {2015}{\natexlab{b}})},\ \Eprint
  {http://arxiv.org/abs/1510.03723} {arXiv:1510.03723 [astro-ph.CO]}
  \BibitemShut {NoStop}%
\bibitem [{\citenamefont {Chudaykin}\ \emph {et~al.}(2020)\citenamefont
  {Chudaykin}, \citenamefont {Ivanov}, \citenamefont {Philcox},\ and\
  \citenamefont {Simonovi\'c}}]{Chudaykin:2020aoj}%
  \BibitemOpen
  \bibfield  {author} {\bibinfo {author} {\bibfnamefont {A.}~\bibnamefont
  {Chudaykin}}, \bibinfo {author} {\bibfnamefont {M.~M.}\ \bibnamefont
  {Ivanov}}, \bibinfo {author} {\bibfnamefont {O.~H.~E.}\ \bibnamefont
  {Philcox}}, \ and\ \bibinfo {author} {\bibfnamefont {M.}~\bibnamefont
  {Simonovi\'c}},\ }\href {\doibase 10.1103/PhysRevD.102.063533} {\bibfield
  {journal} {\bibinfo  {journal} {Phys. Rev. D}\ }\textbf {\bibinfo {volume}
  {102}},\ \bibinfo {pages} {063533} (\bibinfo {year} {2020})},\ \Eprint
  {http://arxiv.org/abs/2004.10607} {arXiv:2004.10607 [astro-ph.CO]}
  \BibitemShut {NoStop}%
\bibitem [{\citenamefont {Schmidt}\ \emph {et~al.}(2019)\citenamefont
  {Schmidt}, \citenamefont {Elsner}, \citenamefont {Jasche}, \citenamefont
  {Nguyen},\ and\ \citenamefont {Lavaux}}]{Schmidt:2018bkr}%
  \BibitemOpen
  \bibfield  {author} {\bibinfo {author} {\bibfnamefont {F.}~\bibnamefont
  {Schmidt}}, \bibinfo {author} {\bibfnamefont {F.}~\bibnamefont {Elsner}},
  \bibinfo {author} {\bibfnamefont {J.}~\bibnamefont {Jasche}}, \bibinfo
  {author} {\bibfnamefont {N.~M.}\ \bibnamefont {Nguyen}}, \ and\ \bibinfo
  {author} {\bibfnamefont {G.}~\bibnamefont {Lavaux}},\ }\href {\doibase
  10.1088/1475-7516/2019/01/042} {\bibfield  {journal} {\bibinfo  {journal}
  {JCAP}\ }\textbf {\bibinfo {volume} {01}},\ \bibinfo {pages} {042} (\bibinfo
  {year} {2019})},\ \Eprint {http://arxiv.org/abs/1808.02002} {arXiv:1808.02002
  [astro-ph.CO]} \BibitemShut {NoStop}%
\bibitem [{\citenamefont {Simonović}\ \emph {et~al.}(2018)\citenamefont
  {Simonović}, \citenamefont {Baldauf}, \citenamefont {Zaldarriaga},
  \citenamefont {Carrasco},\ and\ \citenamefont
  {Kollmeier}}]{Simonovic:2017mhp}%
  \BibitemOpen
  \bibfield  {author} {\bibinfo {author} {\bibfnamefont {M.}~\bibnamefont
  {Simonović}}, \bibinfo {author} {\bibfnamefont {T.}~\bibnamefont {Baldauf}},
  \bibinfo {author} {\bibfnamefont {M.}~\bibnamefont {Zaldarriaga}}, \bibinfo
  {author} {\bibfnamefont {J.~J.}\ \bibnamefont {Carrasco}}, \ and\ \bibinfo
  {author} {\bibfnamefont {J.~A.}\ \bibnamefont {Kollmeier}},\ }\href {\doibase
  10.1088/1475-7516/2018/04/030} {\bibfield  {journal} {\bibinfo  {journal}
  {JCAP}\ }\textbf {\bibinfo {volume} {1804}},\ \bibinfo {pages} {030}
  (\bibinfo {year} {2018})},\ \Eprint {http://arxiv.org/abs/1708.08130}
  {arXiv:1708.08130 [astro-ph.CO]} \BibitemShut {NoStop}%
\bibitem [{\citenamefont {Blas}\ \emph
  {et~al.}(2016{\natexlab{b}})\citenamefont {Blas}, \citenamefont {Garny},
  \citenamefont {Ivanov},\ and\ \citenamefont {Sibiryakov}}]{Blas:2015qsi}%
  \BibitemOpen
  \bibfield  {author} {\bibinfo {author} {\bibfnamefont {D.}~\bibnamefont
  {Blas}}, \bibinfo {author} {\bibfnamefont {M.}~\bibnamefont {Garny}},
  \bibinfo {author} {\bibfnamefont {M.~M.}\ \bibnamefont {Ivanov}}, \ and\
  \bibinfo {author} {\bibfnamefont {S.}~\bibnamefont {Sibiryakov}},\ }\href
  {\doibase 10.1088/1475-7516/2016/07/052} {\bibfield  {journal} {\bibinfo
  {journal} {JCAP}\ }\textbf {\bibinfo {volume} {1607}},\ \bibinfo {pages}
  {052} (\bibinfo {year} {2016}{\natexlab{b}})},\ \Eprint
  {http://arxiv.org/abs/1512.05807} {arXiv:1512.05807 [astro-ph.CO]}
  \BibitemShut {NoStop}%
\bibitem [{\citenamefont {Vasudevan}\ \emph {et~al.}(2019)\citenamefont
  {Vasudevan}, \citenamefont {Ivanov}, \citenamefont {Sibiryakov},\ and\
  \citenamefont {Lesgourgues}}]{Vasudevan:2019ewf}%
  \BibitemOpen
  \bibfield  {author} {\bibinfo {author} {\bibfnamefont {A.}~\bibnamefont
  {Vasudevan}}, \bibinfo {author} {\bibfnamefont {M.~M.}\ \bibnamefont
  {Ivanov}}, \bibinfo {author} {\bibfnamefont {S.}~\bibnamefont {Sibiryakov}},
  \ and\ \bibinfo {author} {\bibfnamefont {J.}~\bibnamefont {Lesgourgues}},\
  }\href {\doibase 10.1088/1475-7516/2019/09/037} {\bibfield  {journal}
  {\bibinfo  {journal} {JCAP}\ }\textbf {\bibinfo {volume} {09}},\ \bibinfo
  {pages} {037} (\bibinfo {year} {2019})},\ \Eprint
  {http://arxiv.org/abs/1906.08697} {arXiv:1906.08697 [astro-ph.CO]}
  \BibitemShut {NoStop}%
\bibitem [{\citenamefont {Alcock}\ and\ \citenamefont
  {Paczynski}(1979)}]{Alcock:1979mp}%
  \BibitemOpen
  \bibfield  {author} {\bibinfo {author} {\bibfnamefont {C.}~\bibnamefont
  {Alcock}}\ and\ \bibinfo {author} {\bibfnamefont {B.}~\bibnamefont
  {Paczynski}},\ }\href {\doibase 10.1038/281358a0} {\bibfield  {journal}
  {\bibinfo  {journal} {Nature}\ }\textbf {\bibinfo {volume} {281}},\ \bibinfo
  {pages} {358} (\bibinfo {year} {1979})}\BibitemShut {NoStop}%
\bibitem [{\citenamefont {Pajer}\ and\ \citenamefont
  {Zaldarriaga}(2013)}]{Pajer:2013jj}%
  \BibitemOpen
  \bibfield  {author} {\bibinfo {author} {\bibfnamefont {E.}~\bibnamefont
  {Pajer}}\ and\ \bibinfo {author} {\bibfnamefont {M.}~\bibnamefont
  {Zaldarriaga}},\ }\href {\doibase 10.1088/1475-7516/2013/08/037} {\bibfield
  {journal} {\bibinfo  {journal} {JCAP}\ }\textbf {\bibinfo {volume} {08}},\
  \bibinfo {pages} {037} (\bibinfo {year} {2013})},\ \Eprint
  {http://arxiv.org/abs/1301.7182} {arXiv:1301.7182 [astro-ph.CO]} \BibitemShut
  {NoStop}%
\bibitem [{\citenamefont {Barreira}\ \emph {et~al.}(2020)\citenamefont
  {Barreira}, \citenamefont {Cabass}, \citenamefont {Schmidt}, \citenamefont
  {Pillepich},\ and\ \citenamefont {Nelson}}]{Barreira:2020kvh}%
  \BibitemOpen
  \bibfield  {author} {\bibinfo {author} {\bibfnamefont {A.}~\bibnamefont
  {Barreira}}, \bibinfo {author} {\bibfnamefont {G.}~\bibnamefont {Cabass}},
  \bibinfo {author} {\bibfnamefont {F.}~\bibnamefont {Schmidt}}, \bibinfo
  {author} {\bibfnamefont {A.}~\bibnamefont {Pillepich}}, \ and\ \bibinfo
  {author} {\bibfnamefont {D.}~\bibnamefont {Nelson}},\ }\href {\doibase
  10.1088/1475-7516/2020/12/013} {\bibfield  {journal} {\bibinfo  {journal}
  {JCAP}\ }\textbf {\bibinfo {volume} {12}},\ \bibinfo {pages} {013} (\bibinfo
  {year} {2020})},\ \Eprint {http://arxiv.org/abs/2006.09368} {arXiv:2006.09368
  [astro-ph.CO]} \BibitemShut {NoStop}%
\bibitem [{\citenamefont {Barreira}(2022)}]{Barreira:2021ueb}%
  \BibitemOpen
  \bibfield  {author} {\bibinfo {author} {\bibfnamefont {A.}~\bibnamefont
  {Barreira}},\ }\href {\doibase 10.1088/1475-7516/2022/01/033} {\bibfield
  {journal} {\bibinfo  {journal} {JCAP}\ }\textbf {\bibinfo {volume} {01}},\
  \bibinfo {pages} {033} (\bibinfo {year} {2022})},\ \Eprint
  {http://arxiv.org/abs/2107.06887} {arXiv:2107.06887 [astro-ph.CO]}
  \BibitemShut {NoStop}%
\bibitem [{\citenamefont {Ivanov}\ \emph
  {et~al.}(2021{\natexlab{b}})\citenamefont {Ivanov}, \citenamefont {Philcox},
  \citenamefont {Simonovi\'c}, \citenamefont {Zaldarriaga}, \citenamefont
  {Nishimichi},\ and\ \citenamefont {Takada}}]{Ivanov:2021haa}%
  \BibitemOpen
  \bibfield  {author} {\bibinfo {author} {\bibfnamefont {M.~M.}\ \bibnamefont
  {Ivanov}}, \bibinfo {author} {\bibfnamefont {O.~H.~E.}\ \bibnamefont
  {Philcox}}, \bibinfo {author} {\bibfnamefont {M.}~\bibnamefont
  {Simonovi\'c}}, \bibinfo {author} {\bibfnamefont {M.}~\bibnamefont
  {Zaldarriaga}}, \bibinfo {author} {\bibfnamefont {T.}~\bibnamefont
  {Nishimichi}}, \ and\ \bibinfo {author} {\bibfnamefont {M.}~\bibnamefont
  {Takada}},\ }\href@noop {} {\  (\bibinfo {year} {2021}{\natexlab{b}})},\
  \Eprint {http://arxiv.org/abs/2110.00006} {arXiv:2110.00006 [astro-ph.CO]}
  \BibitemShut {NoStop}%
\bibitem [{\citenamefont {Philcox}\ \emph {et~al.}(2020)\citenamefont
  {Philcox}, \citenamefont {Ivanov}, \citenamefont {Simonovi\'c},\ and\
  \citenamefont {Zaldarriaga}}]{Philcox:2020vvt}%
  \BibitemOpen
  \bibfield  {author} {\bibinfo {author} {\bibfnamefont {O.~H.~E.}\
  \bibnamefont {Philcox}}, \bibinfo {author} {\bibfnamefont {M.~M.}\
  \bibnamefont {Ivanov}}, \bibinfo {author} {\bibfnamefont {M.}~\bibnamefont
  {Simonovi\'c}}, \ and\ \bibinfo {author} {\bibfnamefont {M.}~\bibnamefont
  {Zaldarriaga}},\ }\href {\doibase 10.1088/1475-7516/2020/05/032} {\bibfield
  {journal} {\bibinfo  {journal} {JCAP}\ }\textbf {\bibinfo {volume} {05}},\
  \bibinfo {pages} {032} (\bibinfo {year} {2020})},\ \Eprint
  {http://arxiv.org/abs/2002.04035} {arXiv:2002.04035 [astro-ph.CO]}
  \BibitemShut {NoStop}%
\bibitem [{\citenamefont {Philcox}\ \emph
  {et~al.}(2021{\natexlab{a}})\citenamefont {Philcox}, \citenamefont {Slepian},
  \citenamefont {Hou}, \citenamefont {Warner}, \citenamefont {Cahn},\ and\
  \citenamefont {Eisenstein}}]{Philcox:2021bwo}%
  \BibitemOpen
  \bibfield  {author} {\bibinfo {author} {\bibfnamefont {O.~H.~E.}\
  \bibnamefont {Philcox}}, \bibinfo {author} {\bibfnamefont {Z.}~\bibnamefont
  {Slepian}}, \bibinfo {author} {\bibfnamefont {J.}~\bibnamefont {Hou}},
  \bibinfo {author} {\bibfnamefont {C.}~\bibnamefont {Warner}}, \bibinfo
  {author} {\bibfnamefont {R.~N.}\ \bibnamefont {Cahn}}, \ and\ \bibinfo
  {author} {\bibfnamefont {D.~J.}\ \bibnamefont {Eisenstein}},\ }\href@noop {}
  {\  (\bibinfo {year} {2021}{\natexlab{a}})},\ \Eprint
  {http://arxiv.org/abs/2105.08722} {arXiv:2105.08722 [astro-ph.IM]}
  \BibitemShut {NoStop}%
\bibitem [{\citenamefont {Kalus}\ \emph {et~al.}(2019)\citenamefont {Kalus},
  \citenamefont {Percival}, \citenamefont {Bacon}, \citenamefont {Mueller},
  \citenamefont {Samushia}, \citenamefont {Verde}, \citenamefont {Ross},\ and\
  \citenamefont {Bernal}}]{Kalus:2018qsy}%
  \BibitemOpen
  \bibfield  {author} {\bibinfo {author} {\bibfnamefont {B.}~\bibnamefont
  {Kalus}}, \bibinfo {author} {\bibfnamefont {W.~J.}\ \bibnamefont {Percival}},
  \bibinfo {author} {\bibfnamefont {D.~J.}\ \bibnamefont {Bacon}}, \bibinfo
  {author} {\bibfnamefont {E.~M.}\ \bibnamefont {Mueller}}, \bibinfo {author}
  {\bibfnamefont {L.}~\bibnamefont {Samushia}}, \bibinfo {author}
  {\bibfnamefont {L.}~\bibnamefont {Verde}}, \bibinfo {author} {\bibfnamefont
  {A.~J.}\ \bibnamefont {Ross}}, \ and\ \bibinfo {author} {\bibfnamefont
  {J.~L.}\ \bibnamefont {Bernal}},\ }\href {\doibase 10.1093/mnras/sty2655}
  {\bibfield  {journal} {\bibinfo  {journal} {Mon. Not. Roy. Astron. Soc.}\
  }\textbf {\bibinfo {volume} {482}},\ \bibinfo {pages} {453} (\bibinfo {year}
  {2019})},\ \Eprint {http://arxiv.org/abs/1806.02789} {arXiv:1806.02789
  [astro-ph.CO]} \BibitemShut {NoStop}%
\bibitem [{\citenamefont {Ross}\ \emph {et~al.}(2017)\citenamefont {Ross} \emph
  {et~al.}}]{BOSS:2016apd}%
  \BibitemOpen
  \bibfield  {author} {\bibinfo {author} {\bibfnamefont {A.~J.}\ \bibnamefont
  {Ross}} \emph {et~al.} (\bibinfo {collaboration} {BOSS}),\ }\href {\doibase
  10.1093/mnras/stw2372} {\bibfield  {journal} {\bibinfo  {journal} {Mon. Not.
  Roy. Astron. Soc.}\ }\textbf {\bibinfo {volume} {464}},\ \bibinfo {pages}
  {1168} (\bibinfo {year} {2017})},\ \Eprint {http://arxiv.org/abs/1607.03145}
  {arXiv:1607.03145 [astro-ph.CO]} \BibitemShut {NoStop}%
\bibitem [{\citenamefont {Kitaura}\ \emph {et~al.}(2016)\citenamefont {Kitaura}
  \emph {et~al.}}]{Kitaura:2015uqa}%
  \BibitemOpen
  \bibfield  {author} {\bibinfo {author} {\bibfnamefont {F.-S.}\ \bibnamefont
  {Kitaura}} \emph {et~al.},\ }\href {\doibase 10.1093/mnras/stv2826}
  {\bibfield  {journal} {\bibinfo  {journal} {Mon. Not. Roy. Astron. Soc.}\
  }\textbf {\bibinfo {volume} {456}},\ \bibinfo {pages} {4156} (\bibinfo {year}
  {2016})},\ \Eprint {http://arxiv.org/abs/1509.06400} {arXiv:1509.06400
  [astro-ph.CO]} \BibitemShut {NoStop}%
\bibitem [{\citenamefont {Wadekar}\ and\ \citenamefont
  {Scoccimarro}(2019)}]{Wadekar:2019rdu}%
  \BibitemOpen
  \bibfield  {author} {\bibinfo {author} {\bibfnamefont {D.}~\bibnamefont
  {Wadekar}}\ and\ \bibinfo {author} {\bibfnamefont {R.}~\bibnamefont
  {Scoccimarro}},\ }\href@noop {} {\  (\bibinfo {year} {2019})},\ \Eprint
  {http://arxiv.org/abs/1910.02914} {arXiv:1910.02914 [astro-ph.CO]}
  \BibitemShut {NoStop}%
\bibitem [{\citenamefont {Wadekar}\ \emph {et~al.}(2020)\citenamefont
  {Wadekar}, \citenamefont {Ivanov},\ and\ \citenamefont
  {Scoccimarro}}]{Wadekar:2020hax}%
  \BibitemOpen
  \bibfield  {author} {\bibinfo {author} {\bibfnamefont {D.}~\bibnamefont
  {Wadekar}}, \bibinfo {author} {\bibfnamefont {M.~M.}\ \bibnamefont {Ivanov}},
  \ and\ \bibinfo {author} {\bibfnamefont {R.}~\bibnamefont {Scoccimarro}},\
  }\href {\doibase 10.1103/PhysRevD.102.123521} {\bibfield  {journal} {\bibinfo
   {journal} {Phys. Rev. D}\ }\textbf {\bibinfo {volume} {102}},\ \bibinfo
  {pages} {123521} (\bibinfo {year} {2020})},\ \Eprint
  {http://arxiv.org/abs/2009.00622} {arXiv:2009.00622 [astro-ph.CO]}
  \BibitemShut {NoStop}%
\bibitem [{\citenamefont {Philcox}\ \emph
  {et~al.}(2021{\natexlab{b}})\citenamefont {Philcox}, \citenamefont {Ivanov},
  \citenamefont {Zaldarriaga}, \citenamefont {Simonovic},\ and\ \citenamefont
  {Schmittfull}}]{Philcox:2020zyp}%
  \BibitemOpen
  \bibfield  {author} {\bibinfo {author} {\bibfnamefont {O.~H.~E.}\
  \bibnamefont {Philcox}}, \bibinfo {author} {\bibfnamefont {M.~M.}\
  \bibnamefont {Ivanov}}, \bibinfo {author} {\bibfnamefont {M.}~\bibnamefont
  {Zaldarriaga}}, \bibinfo {author} {\bibfnamefont {M.}~\bibnamefont
  {Simonovic}}, \ and\ \bibinfo {author} {\bibfnamefont {M.}~\bibnamefont
  {Schmittfull}},\ }\href {\doibase 10.1103/PhysRevD.103.043508} {\bibfield
  {journal} {\bibinfo  {journal} {Phys. Rev. D}\ }\textbf {\bibinfo {volume}
  {103}},\ \bibinfo {pages} {043508} (\bibinfo {year} {2021}{\natexlab{b}})},\
  \Eprint {http://arxiv.org/abs/2009.03311} {arXiv:2009.03311 [astro-ph.CO]}
  \BibitemShut {NoStop}%
\bibitem [{\citenamefont {Byun}\ \emph {et~al.}(2021)\citenamefont {Byun},
  \citenamefont {Oddo}, \citenamefont {Porciani},\ and\ \citenamefont
  {Sefusatti}}]{Byun:2020rgl}%
  \BibitemOpen
  \bibfield  {author} {\bibinfo {author} {\bibfnamefont {J.}~\bibnamefont
  {Byun}}, \bibinfo {author} {\bibfnamefont {A.}~\bibnamefont {Oddo}}, \bibinfo
  {author} {\bibfnamefont {C.}~\bibnamefont {Porciani}}, \ and\ \bibinfo
  {author} {\bibfnamefont {E.}~\bibnamefont {Sefusatti}},\ }\href {\doibase
  10.1088/1475-7516/2021/03/105} {\bibfield  {journal} {\bibinfo  {journal}
  {JCAP}\ }\textbf {\bibinfo {volume} {03}},\ \bibinfo {pages} {105} (\bibinfo
  {year} {2021})},\ \Eprint {http://arxiv.org/abs/2010.09579} {arXiv:2010.09579
  [astro-ph.CO]} \BibitemShut {NoStop}%
\bibitem [{\citenamefont {Biagetti}\ \emph {et~al.}(2021)\citenamefont
  {Biagetti}, \citenamefont {Castiblanco}, \citenamefont {Nore\~na},\ and\
  \citenamefont {Sefusatti}}]{Biagetti:2021tua}%
  \BibitemOpen
  \bibfield  {author} {\bibinfo {author} {\bibfnamefont {M.}~\bibnamefont
  {Biagetti}}, \bibinfo {author} {\bibfnamefont {L.}~\bibnamefont
  {Castiblanco}}, \bibinfo {author} {\bibfnamefont {J.}~\bibnamefont
  {Nore\~na}}, \ and\ \bibinfo {author} {\bibfnamefont {E.}~\bibnamefont
  {Sefusatti}},\ }\href@noop {} {\  (\bibinfo {year} {2021})},\ \Eprint
  {http://arxiv.org/abs/2111.05887} {arXiv:2111.05887 [astro-ph.CO]}
  \BibitemShut {NoStop}%
\bibitem [{\citenamefont {Brinckmann}\ and\ \citenamefont
  {Lesgourgues}(2019)}]{Brinckmann:2018cvx}%
  \BibitemOpen
  \bibfield  {author} {\bibinfo {author} {\bibfnamefont {T.}~\bibnamefont
  {Brinckmann}}\ and\ \bibinfo {author} {\bibfnamefont {J.}~\bibnamefont
  {Lesgourgues}},\ }\href {\doibase 10.1016/j.dark.2018.100260} {\bibfield
  {journal} {\bibinfo  {journal} {Phys. Dark Univ.}\ }\textbf {\bibinfo
  {volume} {24}},\ \bibinfo {pages} {100260} (\bibinfo {year} {2019})},\
  \Eprint {http://arxiv.org/abs/1804.07261} {arXiv:1804.07261 [astro-ph.CO]}
  \BibitemShut {NoStop}%
\bibitem [{\citenamefont {Chudaykin}\ \emph
  {et~al.}(2021{\natexlab{a}})\citenamefont {Chudaykin}, \citenamefont
  {Dolgikh},\ and\ \citenamefont {Ivanov}}]{Chudaykin:2020ghx}%
  \BibitemOpen
  \bibfield  {author} {\bibinfo {author} {\bibfnamefont {A.}~\bibnamefont
  {Chudaykin}}, \bibinfo {author} {\bibfnamefont {K.}~\bibnamefont {Dolgikh}},
  \ and\ \bibinfo {author} {\bibfnamefont {M.~M.}\ \bibnamefont {Ivanov}},\
  }\href {\doibase 10.1103/PhysRevD.103.023507} {\bibfield  {journal} {\bibinfo
   {journal} {Phys. Rev. D}\ }\textbf {\bibinfo {volume} {103}},\ \bibinfo
  {pages} {023507} (\bibinfo {year} {2021}{\natexlab{a}})},\ \Eprint
  {http://arxiv.org/abs/2009.10106} {arXiv:2009.10106 [astro-ph.CO]}
  \BibitemShut {NoStop}%
\bibitem [{\citenamefont {{Bennett}}\ \emph {et~al.}(2013)\citenamefont
  {{Bennett}}, \citenamefont {{Larson}}, \citenamefont {{Weiland}},
  \citenamefont {{Jarosik}}, \citenamefont {{Hinshaw}}, \citenamefont
  {{Odegard}}, \citenamefont {{Smith}}, \citenamefont {{Hill}}, \citenamefont
  {{Gold}}, \citenamefont {{Halpern}}, \citenamefont {{Komatsu}}, \citenamefont
  {{Nolta}}, \citenamefont {{Page}}, \citenamefont {{Spergel}}, \citenamefont
  {{Wollack}}, \citenamefont {{Dunkley}}, \citenamefont {{Kogut}},
  \citenamefont {{Limon}}, \citenamefont {{Meyer}}, \citenamefont {{Tucker}},\
  and\ \citenamefont {{Wright}}}]{2013ApJS..208...20B}%
  \BibitemOpen
  \bibfield  {author} {\bibinfo {author} {\bibfnamefont {C.~L.}\ \bibnamefont
  {{Bennett}}}, \bibinfo {author} {\bibfnamefont {D.}~\bibnamefont {{Larson}}},
  \bibinfo {author} {\bibfnamefont {J.~L.}\ \bibnamefont {{Weiland}}}, \bibinfo
  {author} {\bibfnamefont {N.}~\bibnamefont {{Jarosik}}}, \bibinfo {author}
  {\bibfnamefont {G.}~\bibnamefont {{Hinshaw}}}, \bibinfo {author}
  {\bibfnamefont {N.}~\bibnamefont {{Odegard}}}, \bibinfo {author}
  {\bibfnamefont {K.~M.}\ \bibnamefont {{Smith}}}, \bibinfo {author}
  {\bibfnamefont {R.~S.}\ \bibnamefont {{Hill}}}, \bibinfo {author}
  {\bibfnamefont {B.}~\bibnamefont {{Gold}}}, \bibinfo {author} {\bibfnamefont
  {M.}~\bibnamefont {{Halpern}}}, \bibinfo {author} {\bibfnamefont
  {E.}~\bibnamefont {{Komatsu}}}, \bibinfo {author} {\bibfnamefont {M.~R.}\
  \bibnamefont {{Nolta}}}, \bibinfo {author} {\bibfnamefont {L.}~\bibnamefont
  {{Page}}}, \bibinfo {author} {\bibfnamefont {D.~N.}\ \bibnamefont
  {{Spergel}}}, \bibinfo {author} {\bibfnamefont {E.}~\bibnamefont
  {{Wollack}}}, \bibinfo {author} {\bibfnamefont {J.}~\bibnamefont
  {{Dunkley}}}, \bibinfo {author} {\bibfnamefont {A.}~\bibnamefont {{Kogut}}},
  \bibinfo {author} {\bibfnamefont {M.}~\bibnamefont {{Limon}}}, \bibinfo
  {author} {\bibfnamefont {S.~S.}\ \bibnamefont {{Meyer}}}, \bibinfo {author}
  {\bibfnamefont {G.~S.}\ \bibnamefont {{Tucker}}}, \ and\ \bibinfo {author}
  {\bibfnamefont {E.~L.}\ \bibnamefont {{Wright}}},\ }\href {\doibase
  10.1088/0067-0049/208/2/20} {\bibfield  {journal} {\bibinfo  {journal}
  {"Astrophys. J."}\ }\textbf {\bibinfo {volume} {208}},\ \bibinfo {eid} {20}
  (\bibinfo {year} {2013})},\ \Eprint {http://arxiv.org/abs/1212.5225}
  {arXiv:1212.5225 [astro-ph.CO]} \BibitemShut {NoStop}%
\bibitem [{\citenamefont {Sabti}\ \emph
  {et~al.}(2021{\natexlab{a}})\citenamefont {Sabti}, \citenamefont {Mu\~noz},\
  and\ \citenamefont {Blas}}]{Sabti:2020ser}%
  \BibitemOpen
  \bibfield  {author} {\bibinfo {author} {\bibfnamefont {N.}~\bibnamefont
  {Sabti}}, \bibinfo {author} {\bibfnamefont {J.~B.}\ \bibnamefont {Mu\~noz}},
  \ and\ \bibinfo {author} {\bibfnamefont {D.}~\bibnamefont {Blas}},\ }\href
  {\doibase 10.1088/1475-7516/2021/01/010} {\bibfield  {journal} {\bibinfo
  {journal} {JCAP}\ }\textbf {\bibinfo {volume} {01}},\ \bibinfo {pages} {010}
  (\bibinfo {year} {2021}{\natexlab{a}})},\ \Eprint
  {http://arxiv.org/abs/2009.01245} {arXiv:2009.01245 [astro-ph.CO]}
  \BibitemShut {NoStop}%
\bibitem [{\citenamefont {Sabti}\ \emph
  {et~al.}(2021{\natexlab{b}})\citenamefont {Sabti}, \citenamefont {Mu\~noz},\
  and\ \citenamefont {Blas}}]{Sabti:2021unj}%
  \BibitemOpen
  \bibfield  {author} {\bibinfo {author} {\bibfnamefont {N.}~\bibnamefont
  {Sabti}}, \bibinfo {author} {\bibfnamefont {J.~B.}\ \bibnamefont {Mu\~noz}},
  \ and\ \bibinfo {author} {\bibfnamefont {D.}~\bibnamefont {Blas}},\
  }\href@noop {} {\  (\bibinfo {year} {2021}{\natexlab{b}})},\ \Eprint
  {http://arxiv.org/abs/2110.13161} {arXiv:2110.13161 [astro-ph.CO]}
  \BibitemShut {NoStop}%
\bibitem [{\citenamefont {Sabti}\ \emph {et~al.}(2022)\citenamefont {Sabti},
  \citenamefont {Mu\~noz},\ and\ \citenamefont {Blas}}]{Sabti:2021xvh}%
  \BibitemOpen
  \bibfield  {author} {\bibinfo {author} {\bibfnamefont {N.}~\bibnamefont
  {Sabti}}, \bibinfo {author} {\bibfnamefont {J.~B.}\ \bibnamefont {Mu\~noz}},
  \ and\ \bibinfo {author} {\bibfnamefont {D.}~\bibnamefont {Blas}},\ }\href
  {\doibase 10.1103/PhysRevD.105.043518} {\bibfield  {journal} {\bibinfo
  {journal} {Phys. Rev. D}\ }\textbf {\bibinfo {volume} {105}},\ \bibinfo
  {pages} {043518} (\bibinfo {year} {2022})},\ \Eprint
  {http://arxiv.org/abs/2110.13168} {arXiv:2110.13168 [astro-ph.CO]}
  \BibitemShut {NoStop}%
\bibitem [{\citenamefont {Bertolini}\ \emph {et~al.}(2016)\citenamefont
  {Bertolini}, \citenamefont {Schutz}, \citenamefont {Solon},\ and\
  \citenamefont {Zurek}}]{Bertolini:2016bmt}%
  \BibitemOpen
  \bibfield  {author} {\bibinfo {author} {\bibfnamefont {D.}~\bibnamefont
  {Bertolini}}, \bibinfo {author} {\bibfnamefont {K.}~\bibnamefont {Schutz}},
  \bibinfo {author} {\bibfnamefont {M.~P.}\ \bibnamefont {Solon}}, \ and\
  \bibinfo {author} {\bibfnamefont {K.~M.}\ \bibnamefont {Zurek}},\ }\href
  {\doibase 10.1088/1475-7516/2016/06/052} {\bibfield  {journal} {\bibinfo
  {journal} {JCAP}\ }\textbf {\bibinfo {volume} {06}},\ \bibinfo {pages} {052}
  (\bibinfo {year} {2016})},\ \Eprint {http://arxiv.org/abs/1604.01770}
  {arXiv:1604.01770 [astro-ph.CO]} \BibitemShut {NoStop}%
\bibitem [{\citenamefont {Konstandin}\ \emph {et~al.}(2019)\citenamefont
  {Konstandin}, \citenamefont {Porto},\ and\ \citenamefont
  {Rubira}}]{Konstandin:2019bay}%
  \BibitemOpen
  \bibfield  {author} {\bibinfo {author} {\bibfnamefont {T.}~\bibnamefont
  {Konstandin}}, \bibinfo {author} {\bibfnamefont {R.~A.}\ \bibnamefont
  {Porto}}, \ and\ \bibinfo {author} {\bibfnamefont {H.}~\bibnamefont
  {Rubira}},\ }\href {\doibase 10.1088/1475-7516/2019/11/027} {\bibfield
  {journal} {\bibinfo  {journal} {JCAP}\ }\textbf {\bibinfo {volume} {11}},\
  \bibinfo {pages} {027} (\bibinfo {year} {2019})},\ \Eprint
  {http://arxiv.org/abs/1906.00997} {arXiv:1906.00997 [astro-ph.CO]}
  \BibitemShut {NoStop}%
\bibitem [{\citenamefont {Creminelli}\ \emph {et~al.}(2013)\citenamefont
  {Creminelli}, \citenamefont {Noreña}, \citenamefont {Simonović},\ and\
  \citenamefont {Vernizzi}}]{Creminelli:2013mca}%
  \BibitemOpen
  \bibfield  {author} {\bibinfo {author} {\bibfnamefont {P.}~\bibnamefont
  {Creminelli}}, \bibinfo {author} {\bibfnamefont {J.}~\bibnamefont {Noreña}},
  \bibinfo {author} {\bibfnamefont {M.}~\bibnamefont {Simonović}}, \ and\
  \bibinfo {author} {\bibfnamefont {F.}~\bibnamefont {Vernizzi}},\ }\href
  {\doibase 10.1088/1475-7516/2013/12/025} {\bibfield  {journal} {\bibinfo
  {journal} {JCAP}\ }\textbf {\bibinfo {volume} {12}},\ \bibinfo {pages} {025}
  (\bibinfo {year} {2013})},\ \Eprint {http://arxiv.org/abs/1309.3557}
  {arXiv:1309.3557 [astro-ph.CO]} \BibitemShut {NoStop}%
\bibitem [{\citenamefont {Baldauf}\ \emph {et~al.}(2016)\citenamefont
  {Baldauf}, \citenamefont {Mirbabayi}, \citenamefont {Simonović},\ and\
  \citenamefont {Zaldarriaga}}]{Baldauf:2016sjb}%
  \BibitemOpen
  \bibfield  {author} {\bibinfo {author} {\bibfnamefont {T.}~\bibnamefont
  {Baldauf}}, \bibinfo {author} {\bibfnamefont {M.}~\bibnamefont {Mirbabayi}},
  \bibinfo {author} {\bibfnamefont {M.}~\bibnamefont {Simonović}}, \ and\
  \bibinfo {author} {\bibfnamefont {M.}~\bibnamefont {Zaldarriaga}},\
  }\href@noop {} {\  (\bibinfo {year} {2016})},\ \Eprint
  {http://arxiv.org/abs/1602.00674} {arXiv:1602.00674 [astro-ph.CO]}
  \BibitemShut {NoStop}%
\bibitem [{\citenamefont {Chudaykin}\ \emph
  {et~al.}(2021{\natexlab{b}})\citenamefont {Chudaykin}, \citenamefont
  {Ivanov},\ and\ \citenamefont {Simonovi\'c}}]{Chudaykin:2020hbf}%
  \BibitemOpen
  \bibfield  {author} {\bibinfo {author} {\bibfnamefont {A.}~\bibnamefont
  {Chudaykin}}, \bibinfo {author} {\bibfnamefont {M.~M.}\ \bibnamefont
  {Ivanov}}, \ and\ \bibinfo {author} {\bibfnamefont {M.}~\bibnamefont
  {Simonovi\'c}},\ }\href {\doibase 10.1103/PhysRevD.103.043525} {\bibfield
  {journal} {\bibinfo  {journal} {Phys. Rev. D}\ }\textbf {\bibinfo {volume}
  {103}},\ \bibinfo {pages} {043525} (\bibinfo {year} {2021}{\natexlab{b}})},\
  \Eprint {http://arxiv.org/abs/2009.10724} {arXiv:2009.10724 [astro-ph.CO]}
  \BibitemShut {NoStop}%
\bibitem [{\citenamefont {Wang}\ \emph {et~al.}(2019)\citenamefont {Wang},
  \citenamefont {Percival}, \citenamefont {Avila}, \citenamefont {Crittenden},\
  and\ \citenamefont {Bianchi}}]{Wang:2018xuy}%
  \BibitemOpen
  \bibfield  {author} {\bibinfo {author} {\bibfnamefont {M.~S.}\ \bibnamefont
  {Wang}}, \bibinfo {author} {\bibfnamefont {W.~J.}\ \bibnamefont {Percival}},
  \bibinfo {author} {\bibfnamefont {S.}~\bibnamefont {Avila}}, \bibinfo
  {author} {\bibfnamefont {R.}~\bibnamefont {Crittenden}}, \ and\ \bibinfo
  {author} {\bibfnamefont {D.}~\bibnamefont {Bianchi}},\ }\href {\doibase
  10.1093/mnras/stz829} {\bibfield  {journal} {\bibinfo  {journal} {Mon. Not.
  Roy. Astron. Soc.}\ }\textbf {\bibinfo {volume} {486}},\ \bibinfo {pages}
  {951} (\bibinfo {year} {2019})},\ \Eprint {http://arxiv.org/abs/1811.08155}
  {arXiv:1811.08155 [astro-ph.CO]} \BibitemShut {NoStop}%
\bibitem [{\citenamefont {Aghamousa}\ \emph {et~al.}(2016)\citenamefont
  {Aghamousa} \emph {et~al.}}]{Aghamousa:2016zmz}%
  \BibitemOpen
  \bibfield  {author} {\bibinfo {author} {\bibfnamefont {A.}~\bibnamefont
  {Aghamousa}} \emph {et~al.} (\bibinfo {collaboration} {DESI}),\ }\href@noop
  {} {\  (\bibinfo {year} {2016})},\ \Eprint {http://arxiv.org/abs/1611.00036}
  {arXiv:1611.00036 [astro-ph.IM]} \BibitemShut {NoStop}%
\bibitem [{\citenamefont {Amendola}\ \emph {et~al.}(2018)\citenamefont
  {Amendola} \emph {et~al.}}]{Amendola:2016saw}%
  \BibitemOpen
  \bibfield  {author} {\bibinfo {author} {\bibfnamefont {L.}~\bibnamefont
  {Amendola}} \emph {et~al.},\ }\href {\doibase 10.1007/s41114-017-0010-3}
  {\bibfield  {journal} {\bibinfo  {journal} {Living Rev. Rel.}\ }\textbf
  {\bibinfo {volume} {21}},\ \bibinfo {pages} {2} (\bibinfo {year} {2018})},\
  \Eprint {http://arxiv.org/abs/1606.00180} {arXiv:1606.00180 [astro-ph.CO]}
  \BibitemShut {NoStop}%
\bibitem [{\citenamefont {Schlegel}\ \emph {et~al.}(2019)\citenamefont
  {Schlegel} \emph {et~al.}}]{Schlegel:2019eqc}%
  \BibitemOpen
  \bibfield  {author} {\bibinfo {author} {\bibfnamefont {D.~J.}\ \bibnamefont
  {Schlegel}} \emph {et~al.},\ }\href@noop {} {\  (\bibinfo {year} {2019})},\
  \Eprint {http://arxiv.org/abs/1907.11171} {arXiv:1907.11171 [astro-ph.IM]}
  \BibitemShut {NoStop}%
\bibitem [{\citenamefont {Karagiannis}\ \emph {et~al.}(2018)\citenamefont
  {Karagiannis}, \citenamefont {Lazanu}, \citenamefont {Liguori}, \citenamefont
  {Raccanelli}, \citenamefont {Bartolo},\ and\ \citenamefont
  {Verde}}]{Karagiannis:2018jdt}%
  \BibitemOpen
  \bibfield  {author} {\bibinfo {author} {\bibfnamefont {D.}~\bibnamefont
  {Karagiannis}}, \bibinfo {author} {\bibfnamefont {A.}~\bibnamefont {Lazanu}},
  \bibinfo {author} {\bibfnamefont {M.}~\bibnamefont {Liguori}}, \bibinfo
  {author} {\bibfnamefont {A.}~\bibnamefont {Raccanelli}}, \bibinfo {author}
  {\bibfnamefont {N.}~\bibnamefont {Bartolo}}, \ and\ \bibinfo {author}
  {\bibfnamefont {L.}~\bibnamefont {Verde}},\ }\href {\doibase
  10.1093/mnras/sty1029} {\bibfield  {journal} {\bibinfo  {journal} {Mon. Not.
  Roy. Astron. Soc.}\ }\textbf {\bibinfo {volume} {478}},\ \bibinfo {pages}
  {1341} (\bibinfo {year} {2018})},\ \Eprint {http://arxiv.org/abs/1801.09280}
  {arXiv:1801.09280 [astro-ph.CO]} \BibitemShut {NoStop}%
\bibitem [{\citenamefont {Sailer}\ \emph {et~al.}(2021)\citenamefont {Sailer},
  \citenamefont {Castorina}, \citenamefont {Ferraro},\ and\ \citenamefont
  {White}}]{Sailer:2021yzm}%
  \BibitemOpen
  \bibfield  {author} {\bibinfo {author} {\bibfnamefont {N.}~\bibnamefont
  {Sailer}}, \bibinfo {author} {\bibfnamefont {E.}~\bibnamefont {Castorina}},
  \bibinfo {author} {\bibfnamefont {S.}~\bibnamefont {Ferraro}}, \ and\
  \bibinfo {author} {\bibfnamefont {M.}~\bibnamefont {White}},\ }\href@noop {}
  {\  (\bibinfo {year} {2021})},\ \Eprint {http://arxiv.org/abs/2106.09713}
  {arXiv:2106.09713 [astro-ph.CO]} \BibitemShut {NoStop}%
\bibitem [{\citenamefont {Chudaykin}\ and\ \citenamefont
  {Ivanov}(2019)}]{Chudaykin:2019ock}%
  \BibitemOpen
  \bibfield  {author} {\bibinfo {author} {\bibfnamefont {A.}~\bibnamefont
  {Chudaykin}}\ and\ \bibinfo {author} {\bibfnamefont {M.~M.}\ \bibnamefont
  {Ivanov}},\ }\href {\doibase 10.1088/1475-7516/2019/11/034} {\bibfield
  {journal} {\bibinfo  {journal} {JCAP}\ }\textbf {\bibinfo {volume} {11}},\
  \bibinfo {pages} {034} (\bibinfo {year} {2019})},\ \Eprint
  {http://arxiv.org/abs/1907.06666} {arXiv:1907.06666 [astro-ph.CO]}
  \BibitemShut {NoStop}%
\bibitem [{\citenamefont {Ivanov}(2021)}]{Ivanov:2021zmi}%
  \BibitemOpen
  \bibfield  {author} {\bibinfo {author} {\bibfnamefont {M.~M.}\ \bibnamefont
  {Ivanov}},\ }\href {\doibase 10.1103/PhysRevD.104.103514} {\bibfield
  {journal} {\bibinfo  {journal} {Phys. Rev. D}\ }\textbf {\bibinfo {volume}
  {104}},\ \bibinfo {pages} {103514} (\bibinfo {year} {2021})},\ \Eprint
  {http://arxiv.org/abs/2106.12580} {arXiv:2106.12580 [astro-ph.CO]}
  \BibitemShut {NoStop}%
\bibitem [{\citenamefont {Ivezi\'c}\ \emph {et~al.}(2019)\citenamefont
  {Ivezi\'c} \emph {et~al.}}]{LSST:2008ijt}%
  \BibitemOpen
  \bibfield  {author} {\bibinfo {author} {\bibfnamefont {v.}~\bibnamefont
  {Ivezi\'c}} \emph {et~al.} (\bibinfo {collaboration} {LSST}),\ }\href
  {\doibase 10.3847/1538-4357/ab042c} {\bibfield  {journal} {\bibinfo
  {journal} {Astrophys. J.}\ }\textbf {\bibinfo {volume} {873}},\ \bibinfo
  {pages} {111} (\bibinfo {year} {2019})},\ \Eprint
  {http://arxiv.org/abs/0805.2366} {arXiv:0805.2366 [astro-ph]} \BibitemShut
  {NoStop}%
\bibitem [{\citenamefont {Camera}\ \emph {et~al.}(2015)\citenamefont {Camera},
  \citenamefont {Santos},\ and\ \citenamefont {Maartens}}]{Camera:2014bwa}%
  \BibitemOpen
  \bibfield  {author} {\bibinfo {author} {\bibfnamefont {S.}~\bibnamefont
  {Camera}}, \bibinfo {author} {\bibfnamefont {M.~G.}\ \bibnamefont {Santos}},
  \ and\ \bibinfo {author} {\bibfnamefont {R.}~\bibnamefont {Maartens}},\
  }\href {\doibase 10.1093/mnras/stv040} {\bibfield  {journal} {\bibinfo
  {journal} {Mon. Not. Roy. Astron. Soc.}\ }\textbf {\bibinfo {volume} {448}},\
  \bibinfo {pages} {1035} (\bibinfo {year} {2015})},\ \bibinfo {note}
  {[Erratum: Mon.Not.Roy.Astron.Soc. 467, 1505--1506 (2017)]},\ \Eprint
  {http://arxiv.org/abs/1409.8286} {arXiv:1409.8286 [astro-ph.CO]} \BibitemShut
  {NoStop}%
\bibitem [{\citenamefont {Alonso}\ and\ \citenamefont
  {Ferreira}(2015)}]{Alonso:2015sfa}%
  \BibitemOpen
  \bibfield  {author} {\bibinfo {author} {\bibfnamefont {D.}~\bibnamefont
  {Alonso}}\ and\ \bibinfo {author} {\bibfnamefont {P.~G.}\ \bibnamefont
  {Ferreira}},\ }\href {\doibase 10.1103/PhysRevD.92.063525} {\bibfield
  {journal} {\bibinfo  {journal} {Phys. Rev. D}\ }\textbf {\bibinfo {volume}
  {92}},\ \bibinfo {pages} {063525} (\bibinfo {year} {2015})},\ \Eprint
  {http://arxiv.org/abs/1507.03550} {arXiv:1507.03550 [astro-ph.CO]}
  \BibitemShut {NoStop}%
\bibitem [{\citenamefont {Di~Dio}\ \emph {et~al.}(2017)\citenamefont {Di~Dio},
  \citenamefont {Perrier}, \citenamefont {Durrer}, \citenamefont {Marozzi},
  \citenamefont {Moradinezhad~Dizgah}, \citenamefont {Nore\~na},\ and\
  \citenamefont {Riotto}}]{DiDio:2016gpd}%
  \BibitemOpen
  \bibfield  {author} {\bibinfo {author} {\bibfnamefont {E.}~\bibnamefont
  {Di~Dio}}, \bibinfo {author} {\bibfnamefont {H.}~\bibnamefont {Perrier}},
  \bibinfo {author} {\bibfnamefont {R.}~\bibnamefont {Durrer}}, \bibinfo
  {author} {\bibfnamefont {G.}~\bibnamefont {Marozzi}}, \bibinfo {author}
  {\bibfnamefont {A.}~\bibnamefont {Moradinezhad~Dizgah}}, \bibinfo {author}
  {\bibfnamefont {J.}~\bibnamefont {Nore\~na}}, \ and\ \bibinfo {author}
  {\bibfnamefont {A.}~\bibnamefont {Riotto}},\ }\href {\doibase
  10.1088/1475-7516/2017/03/006} {\bibfield  {journal} {\bibinfo  {journal}
  {JCAP}\ }\textbf {\bibinfo {volume} {03}},\ \bibinfo {pages} {006} (\bibinfo
  {year} {2017})},\ \Eprint {http://arxiv.org/abs/1611.03720} {arXiv:1611.03720
  [astro-ph.CO]} \BibitemShut {NoStop}%
\bibitem [{\citenamefont {Castorina}\ and\ \citenamefont
  {di~Dio}(2022)}]{Castorina:2021xzs}%
  \BibitemOpen
  \bibfield  {author} {\bibinfo {author} {\bibfnamefont {E.}~\bibnamefont
  {Castorina}}\ and\ \bibinfo {author} {\bibfnamefont {E.}~\bibnamefont
  {di~Dio}},\ }\href {\doibase 10.1088/1475-7516/2022/01/061} {\bibfield
  {journal} {\bibinfo  {journal} {JCAP}\ }\textbf {\bibinfo {volume} {01}},\
  \bibinfo {pages} {061} (\bibinfo {year} {2022})},\ \Eprint
  {http://arxiv.org/abs/2106.08857} {arXiv:2106.08857 [astro-ph.CO]}
  \BibitemShut {NoStop}%
\bibitem [{\citenamefont {Andrews}\ \emph {et~al.}(2022)\citenamefont
  {Andrews}, \citenamefont {Jasche}, \citenamefont {Lavaux},\ and\
  \citenamefont {Schmidt}}]{Andrews:2022nvv}%
  \BibitemOpen
  \bibfield  {author} {\bibinfo {author} {\bibfnamefont {A.}~\bibnamefont
  {Andrews}}, \bibinfo {author} {\bibfnamefont {J.}~\bibnamefont {Jasche}},
  \bibinfo {author} {\bibfnamefont {G.}~\bibnamefont {Lavaux}}, \ and\ \bibinfo
  {author} {\bibfnamefont {F.}~\bibnamefont {Schmidt}},\ }\href@noop {} {\
  (\bibinfo {year} {2022})},\ \Eprint {http://arxiv.org/abs/2203.08838}
  {arXiv:2203.08838 [astro-ph.CO]} \BibitemShut {NoStop}%
\bibitem [{\citenamefont {Blas}\ \emph {et~al.}(2011)\citenamefont {Blas},
  \citenamefont {Lesgourgues},\ and\ \citenamefont {Tram}}]{Blas:2011rf}%
  \BibitemOpen
  \bibfield  {author} {\bibinfo {author} {\bibfnamefont {D.}~\bibnamefont
  {Blas}}, \bibinfo {author} {\bibfnamefont {J.}~\bibnamefont {Lesgourgues}}, \
  and\ \bibinfo {author} {\bibfnamefont {T.}~\bibnamefont {Tram}},\ }\href
  {\doibase 10.1088/1475-7516/2011/07/034} {\bibfield  {journal} {\bibinfo
  {journal} {JCAP}\ }\textbf {\bibinfo {volume} {1107}},\ \bibinfo {pages}
  {034} (\bibinfo {year} {2011})},\ \Eprint {http://arxiv.org/abs/1104.2933}
  {arXiv:1104.2933 [astro-ph.CO]} \BibitemShut {NoStop}%
\bibitem [{\citenamefont {Audren}\ \emph {et~al.}(2013)\citenamefont {Audren},
  \citenamefont {Lesgourgues}, \citenamefont {Benabed},\ and\ \citenamefont
  {Prunet}}]{Audren:2012wb}%
  \BibitemOpen
  \bibfield  {author} {\bibinfo {author} {\bibfnamefont {B.}~\bibnamefont
  {Audren}}, \bibinfo {author} {\bibfnamefont {J.}~\bibnamefont {Lesgourgues}},
  \bibinfo {author} {\bibfnamefont {K.}~\bibnamefont {Benabed}}, \ and\
  \bibinfo {author} {\bibfnamefont {S.}~\bibnamefont {Prunet}},\ }\href
  {\doibase 10.1088/1475-7516/2013/02/001} {\bibfield  {journal} {\bibinfo
  {journal} {JCAP}\ }\textbf {\bibinfo {volume} {1302}},\ \bibinfo {pages}
  {001} (\bibinfo {year} {2013})},\ \Eprint {http://arxiv.org/abs/1210.7183}
  {arXiv:1210.7183 [astro-ph.CO]} \BibitemShut {NoStop}%
\bibitem [{\citenamefont {Lewis}(2019)}]{Lewis:2019xzd}%
  \BibitemOpen
  \bibfield  {author} {\bibinfo {author} {\bibfnamefont {A.}~\bibnamefont
  {Lewis}},\ }\href@noop {} {\  (\bibinfo {year} {2019})},\ \Eprint
  {http://arxiv.org/abs/1910.13970} {arXiv:1910.13970 [astro-ph.IM]}
  \BibitemShut {NoStop}%
\end{thebibliography}%

\end{document}